# Physics-governed executable modelling of triboelectric nanogenerators

Hongfa Zhao[1,6], Baiqiao Wang[1,6], Tiancong Zhao[2], Chun Jin[1], Hanlin Zhou[1,3], Mingrui Shu[1], Minyi Xu[4], Liwei Lin[5], Wenbo Ding[1*], Zhong Lin Wang[3*]

[1]Shenzhen International Graduate School, Tsinghua University, Shenzhen 518055, China.

[2]Zhuhai Campus, Beijing Institute of Technology, Zhuhai 519088, China.

[3]Beijing Institute of Nanoenergy and Nanosystems, Chinese Academy of Sciences, Beijing 100085, China.

[4]Marine Engineering College, Dalian Maritime University, Dalian 116026, China.

[5]Department of Mechanical Engineering, University of California at Berkeley, Berkeley, CA 94720, USA.

[6]These authors contributed equally: Hongfa Zhao, Baiqiao Wang

*Email: dingwenbo@sz.tsinghua.edu.cn; zlwang@binn.cas.cn

**Abstract**

Predictive modelling of triboelectric nanogenerators (TENGs) remains fragmented across analytical theories, finite-geometry solvers and disconnected simulation workflows. These disparate approaches must be unified into an executable framework to advance quantitative TENG research. Here we introduce a charge-defined modelling framework and implement it as TENG-CLAW, a physics-governed platform for traceable TENG simulation. The framework establishes a self-consistent electrostatic hierarchy in which triboelectric charges, pre-charging charges and compensating electrode charges serve as defining state variables. This hierarchy connects the infinite-plate analytical limit for near-uniform fields with finite-geometry numerical formulations required for edge-dominated devices. Built on this basis, TENG-CLAW converts user-defined research requests into physically admissible simulation tasks, so that generated outputs are tied to explicit charge states, boundary conditions, solver routes and reusable artifacts across spatial, temporal, field-level, comparative and reporting workflows. This work establishes a rigorous computational basis for interpreting TENG mechanisms and provides reproducible research infrastructure for simulation and physics-guided device design.

## Introduction

Triboelectric nanogenerators (TENGs) have developed from energy-harvesting prototypes into electrostatic building blocks for self-powered sensing, human-machine interaction, wearable electronics, robotics and distributed intelligent systems[1-4]. Their operation couples contact electrification, dynamic charge redistribution, geometric modulation and electrical output generation[5-7]. As TENG research moves from device demonstration towards predictive design, the central challenge is no longer the absence of individual models. Rather, existing models are difficult to compose, execute and reproduce because their charge definitions, electrostatic assumptions, geometry treatments and simulation workflows are often separated from one another[8-11].

Substantial progress has been made in analytical modelling of contact-separation TENGs[12-15], generalized formulations for different working modes[16-19], quasi-electrostatic and three-dimensional charge models[20-22] and dynamic transducer models[23-26]. In particular, the initial-charge infinite-plate model provides an analytical description for large-aspect-ratio contact-separation devices[27, 28]. By incorporating triboelectric charges, pre-charging charges and compensating electrode charges, it quantitatively predicts transferred charge, current, voltage-charge relationships and load-dependent responses in near-uniform-field regimes. This model establishes a physically transparent analytical limit for TENG electrostatics.

This limit, however, does not fully capture finite-size charge redistribution, edge-localized accumulation, non-uniform fields or the transition from uniform-field behavior to finite-geometry behavior. More broadly, existing models often rely on different combinations of capacitor analogies, equivalent circuits, closed-form equations and numerical solvers[29-32]. Although each approach captures part of the device physics, the relationships among charge conditions, electrostatic constraints, observables and model validity are usually implicit[33-36]. This lack of standardization makes it difficult to compare models, select solver regimes and reuse previous simulations.

A second gap lies between theory and executable computation. Translating a TENG problem into a simulation still requires manual coordination of geometry, material parameters, charge states, motion laws, electrical boundary conditions, solver settings and post-processing[37-40]. These steps are often performed through disconnected scripts or manually configured numerical packages. Although such workflows can produce numerical fields or electrical outputs, the results are difficult to constrain and interpret within a unified physical framework. As a result, assumptions are hidden, outputs are hard to trace and simulations cannot be readily reused as computational evidence in later design studies.

Here we develop a charge-defined executable modelling framework for TENGs and implement it as TENG-CLAW, a physics-governed computational platform. The framework defines a self-consistent electrostatic closure in which triboelectric charges, pre-charging charges and compensating electrode charges constitute the initial charge ensemble of a TENG. Transferred charge, current, potential and field distributions are then derived under charge conservation, conductor equipotential conditions and geometry-dependent redistribution. This single physical formulation connects analytical and finite-geometry regimes, allowing solver selection to depend on device geometry and target observables rather than empirical preference. We further introduce an executable workflow that converts open-ended research requests into physically admissible TENG tasks. The workflow compiles user intent into a typed intermediate representation, applies deterministic physical and capability checks, selects the appropriate solver branch and stores outputs with trace metadata. Through charge-distribution mapping, dynamic output prediction, field visualization and browser-based reuse, TENG-CLAW makes TENG modelling inspectable, traceable and reproducible across analytical, numerical and workflow levels.

# Results

## Charge-defined electrostatic state and regime-adaptive solver hierarchy

A predictive TENG calculation must begin from a complete electrostatic state. In the present framework, a TENG is specified by charge conditions, material properties, geometry, mechanical excitation, electrical boundary conditions and target observables. Triboelectric surface charges, externally introduced pre-charging charges and compensating charges are treated as components of one initial charge-defined state rather than as separate fitting parameters for different outputs[27, 28, 41]. Under charge conservation, conductor equipotential conditions and electrostatic equilibrium, this state determines transferred charge, current, potential difference and electric-field distribution.

Mechanical actuation modifies the device geometry, alters the electrostatic domain and drives the system toward a new equilibrium under the same physical constraints. When the lateral dimension is much larger than the gap height, spatial variation is suppressed and the system approaches a near-uniform-field regime. In this limit, global observables can be obtained efficiently from an infinite-plate analytical model. When the lateral dimension and the gap height become comparable, boundary-induced non-uniformity becomes significant. Charge density, potential and electric field then become position-dependent, and edge localization, field distortion and nonlocal electrostatic coupling must be resolved explicitly by a finite-geometry solver.

This hierarchy defines the computational regime map of the framework (Fig. 1a,b). Large-aspect-ratio devices used for sensing often operate close to the near-uniform limit because their active area is much larger than the internal gap[42-44]. Geometrically confined or high-output devices, including vibration-, wind- and water-wave-driven harvesters[45-47], may require finite-geometry treatment because edge-induced redistribution can substantially change both transferred charge and field distribution. The modelling question is therefore not whether the analytical or numerical model is universally preferable, but how one charge-defined electrostatic problem can yield both limits without changing its physical assumptions.

We formulate this hierarchy as a charge-defined electrostatic closure. A TENG problem is specified by charge conditions established during fabrication, contact electrification and pre-charging, together with the electrostatic boundary conditions imposed by geometry and materials (Fig. 1c,d). Under quasi-electrostatic conditions, the admissible state is determined by Maxwell's equations, charge conservation, conductor equipotential constraints and electrostatic equilibrium. In compact form, the closure can be written as

$$\mathcal{C}_s[\rho_{\mathrm{s}},\rho_{\mathrm{b}}]=0,\quad \mathcal{C}_b[\phi;\rho_{\mathrm{s}},\rho_{\mathrm{b}},Q,\Omega(z)]=0, \tag{1}$$

where $\rho_{\mathrm{s}}$ denotes free source charges, $\rho_{\mathrm{b}}$ denotes polarization-induced bound charges, $Q$ is the transferred charge, $\phi$ is the electrostatic potential and $\Omega(z)$ represents the geometry-dependent domain. For clarity, the major symbols and their definitions are summarized in Supplementary Table 1.

For a contact-separation TENG with device area $S$, dielectric thickness $d_0$, relative permittivity $\varepsilon_r$, initial separation $z_0$, instantaneous separation $z(t)$ and effective source charge density $\sigma_{\mathrm{eff}}=\sigma_{\mathrm{te}}+\sigma_{\mathrm{tp}}$, the short-circuit transferred charge is

$$Q(t)=\int_{z_0}^{z(t)}\frac{\sigma_{\mathrm{eff}}S\varepsilon_r d_0}{\left(\varepsilon_r z+d_0\right)^2}\mathrm{d}z=\frac{\sigma_{\mathrm{eff}}S\varepsilon_r d_0\left[z(t)-z_0\right]}{\left[\varepsilon_r z(t)+d_0\right]\left[\varepsilon_r z_0+d_0\right]}. \tag{2}$$

The corresponding current is

$$I(t)=\frac{\mathrm{d}Q(t)}{\mathrm{d}t}=\frac{\sigma_{\mathrm{eff}}S\varepsilon_r d_0}{\left[\varepsilon_r z(t)+d_0\right]^2}\frac{\mathrm{d}z(t)}{\mathrm{d}t}. \tag{3}$$

The sign of $Q(t)$ and $I(t)$ depends on electrode convention. Compensating charges remain part of the complete electrostatic closure because they enforce equilibrium and charge conservation; in the standard two-electrode short-circuit branch, the measurable transferred charge is governed by the effective driving charge density $\sigma_{\mathrm{eff}}$.

The same admissible state determines field-level observables. The intrinsic potential difference and local electric fields can be expressed compactly as infinite-plate solution operators,

$$V(t)=\mathcal{C}_{\infty}^{V}\left(\sigma_{\text{te}},\sigma_{\text{tp}},\sigma_{c},d_0,\varepsilon_r,z(t)\right), \tag{4}$$

$$\mathbf{E}_k(t)=\mathcal{C}_{\infty}^{E_k}\left(\sigma_{\text{te}},\sigma_{\text{tp}},\sigma_{\text{c}},d_0,\varepsilon_r,z(t)\right), \quad k=1,\ldots,4. \tag{5}$$

Explicit expressions are provided in Supplementary Note 1. These operators represent the same charge-conservation and boundary-condition solution used to derive $Q(t)$ and $I(t)$, but expose potential and field observables rather than only lumped electrical outputs.

When finite-size effects are non-negligible, spatial structure becomes intrinsic. Charge density must then be treated as a position-dependent quantity, and electrostatic consistency must be enforced through conductor equipotential and charge-conservation constraints (Fig. 1e). The finite-geometry problem is written as the coupled integral-equation system

$$\sum_j \int_{\Gamma_j} K_{ij}\left(\mathbf{r},\mathbf{r}';z\right)\sigma_j\left(\mathbf{r}'\right)\mathrm{d}S' + \int_{\Gamma_b} K_{ib}\left(\mathbf{r},\mathbf{r}';z\right)\sigma_{\text{b}}\left(\mathbf{r}'\right)\mathrm{d}S' = V_i, \quad \mathbf{r}\in\Gamma_i, \tag{6}$$

together with global charge-conservation constraints,

$$\int_{\Gamma_i} \sigma_i(\mathbf{r})\mathrm{d}S = Q_i^{(0)} + s_i Q, \quad \sum_i Q_i = Q_{\text{tot}}. \tag{7}$$

Here $K_{\text{ij}}$ and $K_{\text{ib}}$ are geometry- and boundary-dependent electrostatic kernels, $\sigma_j\left(\mathbf{r}\right)$ is the free charge density on conducting surfaces, $\sigma_b\left(\mathbf{r}\right)$ is the bound charge density on dielectric surfaces, $V_i$ is the conductor potential, $\Gamma_i$ is the surface of conductor $i$, $Q_i^{(0)}$ is the initial charge assigned to that conductor and $s_i$ defines the direction of transferred charge relative to each electrode. Expanded equations are given in Supplementary Note 2.

Prescribed triboelectric and bound charges enter Eq. (7) as fixed source terms, whereas conductor charges are solved under equipotential and conservation constraints. The finite-geometry solution is therefore a redistribution of the same initial charge ensemble rather than an empirical finite-size correction. The kernel encodes nonlocal electrostatic coupling across the device, allowing edge localization, corner enhancement and field distortion to emerge self-consistently from geometry and boundary conditions.

The integral equations are solved numerically using the Method of Moments (Fig. 1f), which discretizes the continuous charge distribution into a finite-dimensional linear system,

$$\begin{bmatrix} \mathbf{K}_{cc}(z) & -\mathbf{P} \\ \mathbf{A} & \mathbf{0} \end{bmatrix} \begin{bmatrix} \boldsymbol{\sigma}_{\mathrm{c}} \\ \mathbf{V}_{\mathrm{c}} \end{bmatrix} = \begin{bmatrix} -\mathbf{K}_{cf}(z)\boldsymbol{\sigma}_{\mathrm{f}} \\ \mathbf{q}_{\mathrm{c}} \end{bmatrix}. \tag{8}$$

Here $\boldsymbol{\sigma}_{\mathrm{c}}$ is the vector of unknown conductor surface-charge densities, $\boldsymbol{\sigma}_{\mathrm{f}}$ collects prescribed source-charge distributions, $\mathbf{V}_{\mathrm{c}}$ is the vector of conductor potentials, $\mathbf{K}_{cc}(z)$ is the geometry-dependent conductor-conductor interaction matrix, and $\mathbf{K}_{cf}(z)\boldsymbol{\sigma}_{\mathrm{f}}$ gives the potential contribution from fixed source charges. The incidence matrix $\mathbf{P}$ maps conductor potentials to collocation points, and $\mathbf{A}$ integrates panel charges over each conductor to enforce charge conservation. For the short-circuit two-electrode branch used in the representative calculation, $\mathbf{V}_{\mathrm{c}}$ reduces to a common electrode potential and $\mathbf{q}_{\mathrm{c}}$ represents the conserved charge of the connected conductor system. Discretization, matrix construction and convergence are detailed in Supplementary Notes 3-7.

The finite-geometry solution reveals a clear separation between bulk and boundary behavior (Fig. 1g,h). In the interior, the charge density remains close to the uniform-field limit, showing that the leading-order response is consistent with the analytical branch. Near edges and corners, strong localization appears because of field concentration. The cross-sectional profile therefore gives a continuous transition from bulk-like response to edge-dominated redistribution.

The connection between analytical and spatially resolved descriptions is confirmed by comparing the finite-geometry solution with the infinite-plate limit and experimental measurements. In the asymptotic regime $l \gg z$, where $l$ is the characteristic lateral dimension, the finite-geometry solution converges to the infinite-plate prediction (Fig. 1i, experimental details are introduced in Supplementary Note 8, and the same TENG configuration [48, 49] is used as an experimental reference for both asymptotic convergence and executable TENG-CLAW output validation).

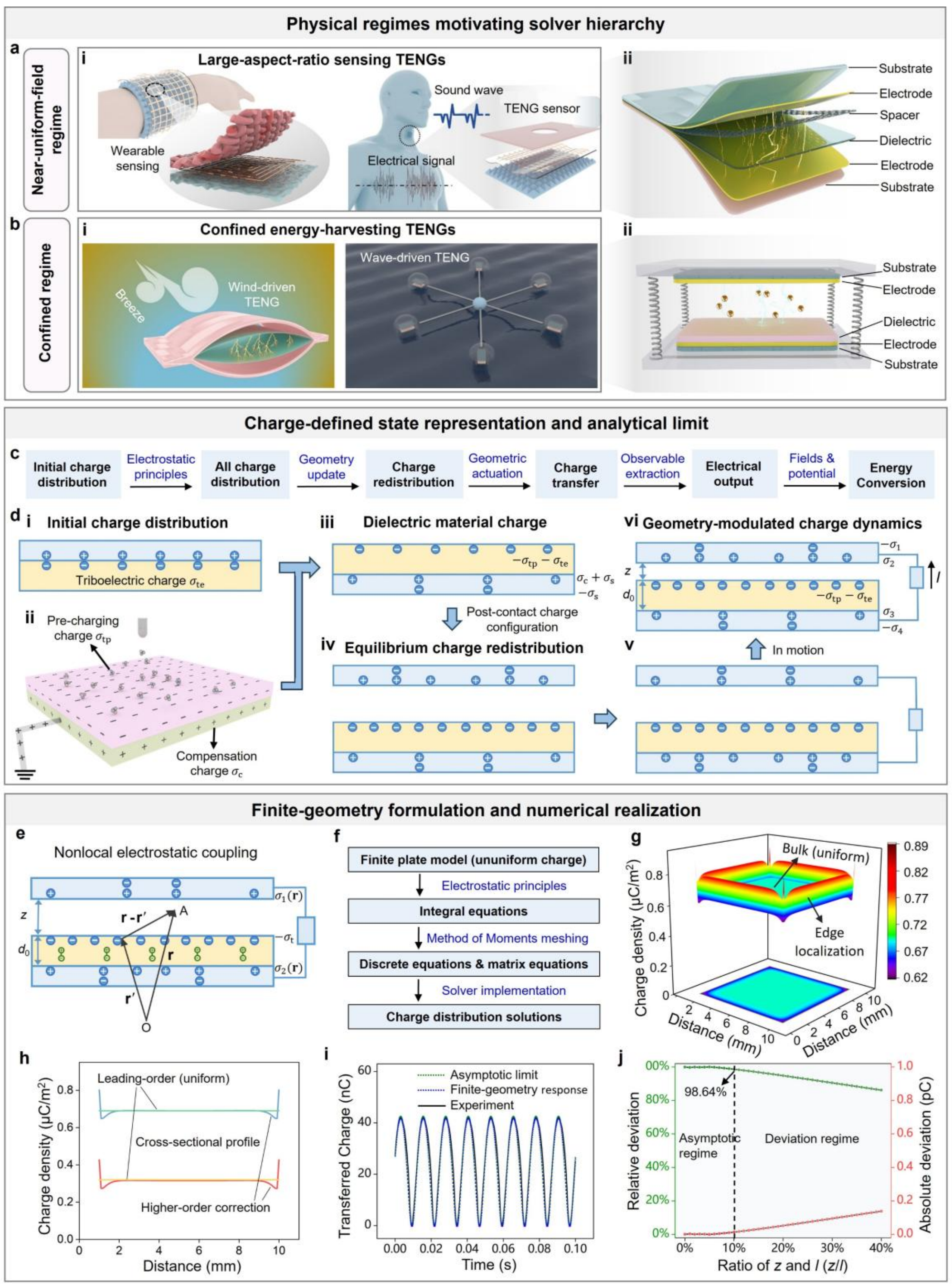

**Figure 1 Charge-defined electrostatic hierarchy and regime-adaptive TENG modelling. a,b,** Physical regimes motivating solver hierarchy, including near-uniform-field sensing TENGs and confined energy-harvesting TENGs. **c,** Conceptual workflow connecting initial charge distribution, electrostatic constraints, charge redistribution, geometric motion and measurable outputs within one modelling hierarchy. **d,** Formation and evolution of the initial charge ensemble composed of triboelectric

charges, pre-charging charges and compensating electrode charges. **e,** Finite-plate model resolving spatially non-uniform electrostatic distributions through nonlocal coupling. **f,** Numerical realization of the finite-geometry formulation using integral equations and the Method of Moments. **g,** Charge distribution on the independent electrode, showing bulk uniformity and edge localization. **h,** Cross-sectional charge-density profile comparing the leading-order uniform response with finite-geometry correction. **i,** Transferred charge predicted by the asymptotic analytical limit and finite-geometry model, compared with experiment. **j,** Relative deviation of the infinite-plate analytical prediction from the finite-geometry reference as a function of $z/$.

Using the finite-geometry solution as a spatially resolved reference, the relative deviation of the infinite-plate prediction remains below 1.4% for $z/l<0.1$ in the representative contact-separation configuration (Fig. 1j). This corresponds to agreement above 98.6%, indicating that the response is effectively near-uniform in this regime. As $z/l$ increases, deviations grow because edge-induced field distortion becomes more important. Even at $z\sim0.2l$, the infinite-plate model still captures the dominant response trend, showing that the approximation remains controlled beyond its strict asymptotic limit.

**From electrostatic formulation to physics-governed execution**

The charge-defined framework also specifies how TENG simulations should be posed, checked and executed. A meaningful calculation requires charge conditions, geometry, materials, motion law, electrical boundary conditions, model regime and target observables to be mutually consistent. TENG-CLAW operationalizes this requirement by converting open-ended research requests into executable TENG tasks whose assumptions, solver routes and outputs are explicitly recorded (Fig. 2).

The platform exposes a defined set of theory-faithful scientific actions, including charge-distribution simulation, time-series prediction, field snapshot generation, field animation, parameter optimization, comparative study and report generation. Each action corresponds to a physically interpretable traversal or aggregation of electrostatic states. For example, a time-series request requires a motion law, a sequence of instantaneous geometries, electrostatic solutions at each state, transferred-charge aggregation and temporal differentiation to obtain current. A field-animation request further generates state-indexed potential and electric-field maps along the same motion trajectory.

TENG-CLAW first compiles a user request into a typed intermediate representation, termed TENG-IR. A TENG-IR object records both physical inputs and research-level intent,

$$I_t = (\iota_t, g_t, \mu_t, \xi_t, \mathcal{O}_t, \mathcal{J}_t, \mathcal{K}_t, \mathcal{A}_t, \mathcal{R}_t, \beta_t), \tag{9}$$

where $\iota_t$ is the research intent, $g_t$ the geometry, $\mu_t$ the material parameters, $\xi_t$ the mechanical excitation, $\mathcal{O}_t$ the requested observables, $\mathcal{J}_t$ the scientific objective, $\mathcal{K}_t$ the constraints, $\mathcal{A}_t$ the requested artifacts, $\mathcal{R}_t$ the links to prior runs and $\beta_t$ the candidate backend plan. In implementation, the object also stores task identifiers, support status, missing information and capability gaps. TENG-IR is broader than a solver input file because executable TENG research may involve explanation, comparison, optimization, report generation or continuation from prior runs.

The compilation process is represented by the scientific state transition

$$(u_t, S_{t-1}) \xrightarrow{\Gamma} (I_t, C_t, \Pi_t, A_t, S_t), \tag{10}$$

where an open-ended request $u_t$ and the prior scientific state $S_{t-1}$ are compiled into structured intent $I_t$, physical context $C_t$, execution plan $\Pi_t$, generated artifacts $A_t$ and updated state $S_t$. This transition allows later requests to reuse an existing simulation state rather than restarting from isolated prompts.

Before execution, the task passes through a physical-governance operator,

$$G_{\text{phys}}(I_t, C_t) = (v_t, B_t, W_t, \pi_t, \eta_t), \quad v_t \in \{\text{pass}, \text{clarify}, \text{approximate}, \text{unsupported}\}. \tag{11}$$

The operator returns an admissibility verdict $v_t$, missing or ambiguous physical parameters $B_t$, model-assumption warnings $W_t$, selected computational branch $\pi_t$ and approximation level $\eta_t$. The deterministic preflight check uses the worker-side capability registry as the source of truth. It evaluates unit consistency, geometry compatibility, boundary-condition completeness, solver support, observable availability, safe default fields and whether an unsupported request is a candidate for controlled extension. Parameters that do not change the physical meaning of the task may receive controlled defaults; physically essential missing information leads to clarification.

Model routing follows the charge-defined hierarchy,

$$\pi_t = \begin{cases} \pi_\infty, & \chi < \chi_c \text{ and } \mathcal{O}_t \in \mathcal{O}_{\text{global}} \\ \pi_{\text{F}}, & \chi \geq \chi_c \text{ or } \mathcal{O}_t \in \mathcal{O}_{\text{spatial}} \end{cases}, \tag{12}$$

where $\pi_\infty$ is the infinite-plate analytical branch, $\pi_{\text{F}}$ is the finite-geometry numerical branch, $\chi$ is a geometry ratio, $\chi_c$ is an accuracy- and observable-dependent threshold, $\mathcal{O}_{\text{global}}$ includes lumped observables and $\mathcal{O}_{\text{spatial}}$ includes charge-density, potential and electric-field distributions. The threshold is not treated as a universal material constant. It depends on required accuracy, device geometry and the observable requested.

Each execution produces artifacts through

$$A_t = \mathcal{R}(I_t, C_t, \pi_t, \eta_t), \tag{13}$$

where $\mathcal{R}$ denotes solver execution and post-processing. Artifacts may include transferred-charge curves, current waveforms, charge-density maps, potential maps, electric-field maps, animations, raw numerical files or reports. Because each artifact is linked to its compiled intent, solver branch and approximation record, the output can be inspected, reproduced and compared across runs.

To preserve scientific continuity, TENG-CLAW stores executed studies in an experiment graph,

$$G_t = \left(V_t, E_t\right), \tag{14}$$

where each node contains the compiled intent, physical-critic outcome, execution plan, generated artifacts and scientific conclusion,

$$v_i = (I_i, C_i, \Pi_i, A_i, \Sigma_i). \tag{15}$$

Edges encode relations such as reuse, comparison, refinement and extension. This graph-backed memory allows later requests to recover previous baselines, compare new designs with earlier outputs, reopen field maps or generate reports without reconstructing the context from free-text dialogue.

This design keeps execution within an explicit physical and computational route, so that generated artifacts remain inspectable rather than merely plausible. When essential information is missing, TENG-CLAW requests clarification; when a simplification is required, the approximation is recorded; and when a request lies outside the supported backend, execution stops with an unsupported verdict before solver invocation. Controlled extension of unsupported but templatable tasks is isolated from stable execution and requires proposal, confirmation, validation and registration. In this way, TENG-CLAW converts the charge-defined electrostatic hierarchy into an executable research workflow without equating conversational fluency with physical validity.

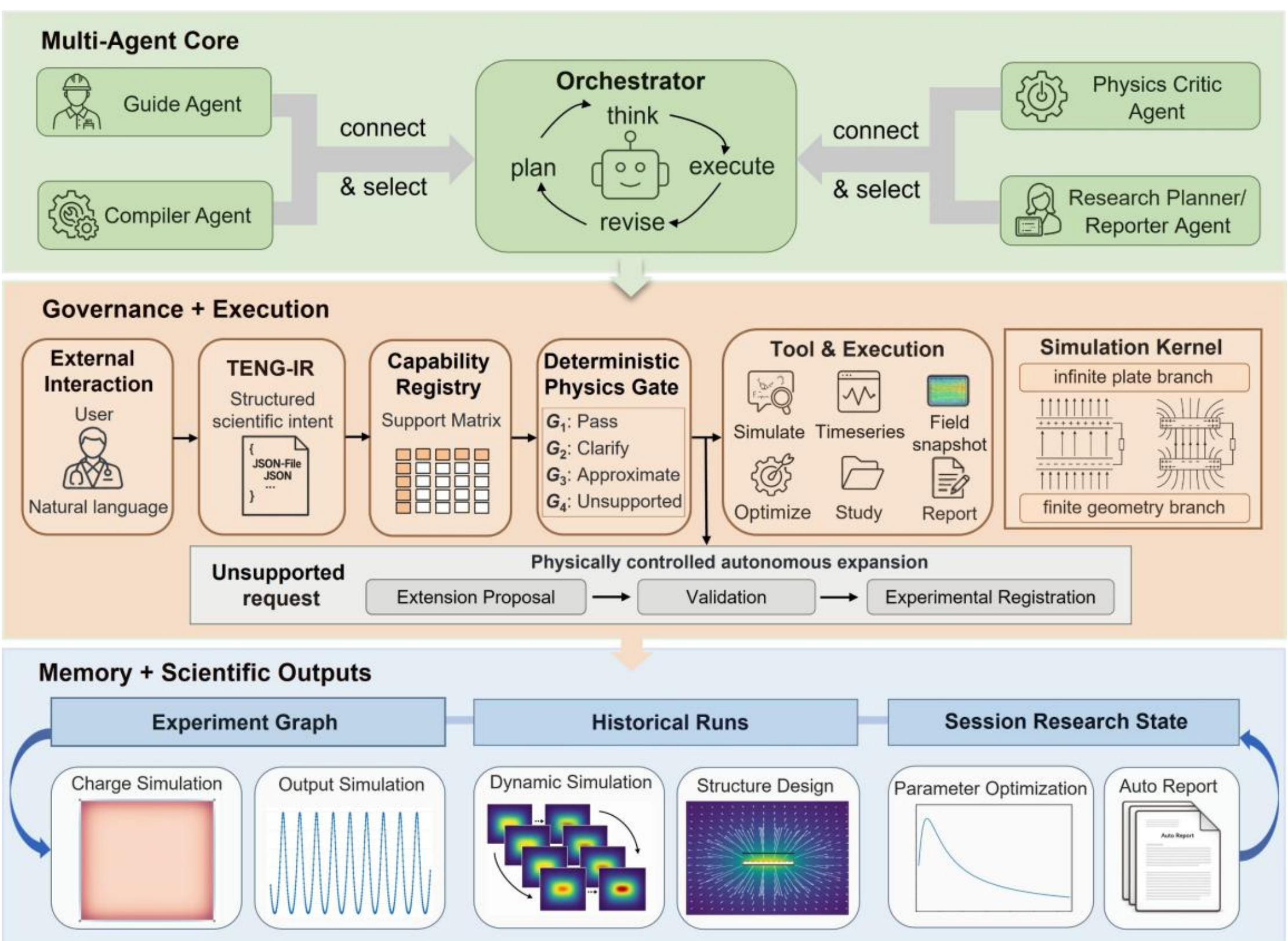

**Figure 2 Physics-governed executable workflow of TENG-CLAW.** An open-ended

research request is compiled into TENG-IR and checked by a physical-governance layer before solver execution. The task is then routed to the infinite-plate analytical branch or finite-geometry numerical branch according to device regime and requested observable. The resulting outputs are stored with trace metadata for inspection and reuse.

**Web-based realization and representative executable simulations**

We implemented TENG-CLAW as a web-based scientific workspace to demonstrate how the executable platform operates in practice (Fig. 3a). The workspace connects dialogue, model clarification, solver execution, result inspection and session-level reuse within one theory-governed environment. Browser-level operation, including model settings, trace-card inspection, result-card interpretation, results-browser reuse and additional cases, is described in Supplementary Note 10. Supplementary Video 1 shows representative browser interaction, including task specification, parameter clarification, executable simulation and artifact generation.

The representative cases test whether one platform can support three observable levels: spatial charge redistribution, lumped electrical output and field-level interpretation. In the first case, the user requests charge-distribution simulation. TENG-CLAW does not immediately return a nominal map. It identifies the required physical inputs, including grid size, surface charge density, dielectric thickness, relative permittivity, device shape and observation position (Fig. 3b). After these inputs are supplied, the task is compiled into TENG-IR and routed to the finite-geometry branch because spatial resolution is required. The charge maps are returned as artifacts linked to the interpreted physical state (Fig. 3c).

In the second case, the user defines a motion law, $z(t)=0.1+0.03\sin(160\pi t)$, and requests transferred charge and current (Fig. 4d). TENG-CLAW treats this request as a continuation of the same charge-evolution problem and returns the corresponding time-domain trajectories. The predicted charge and current are consistent with theoretical calculation and experimental measurements (Supplementary Note 8).

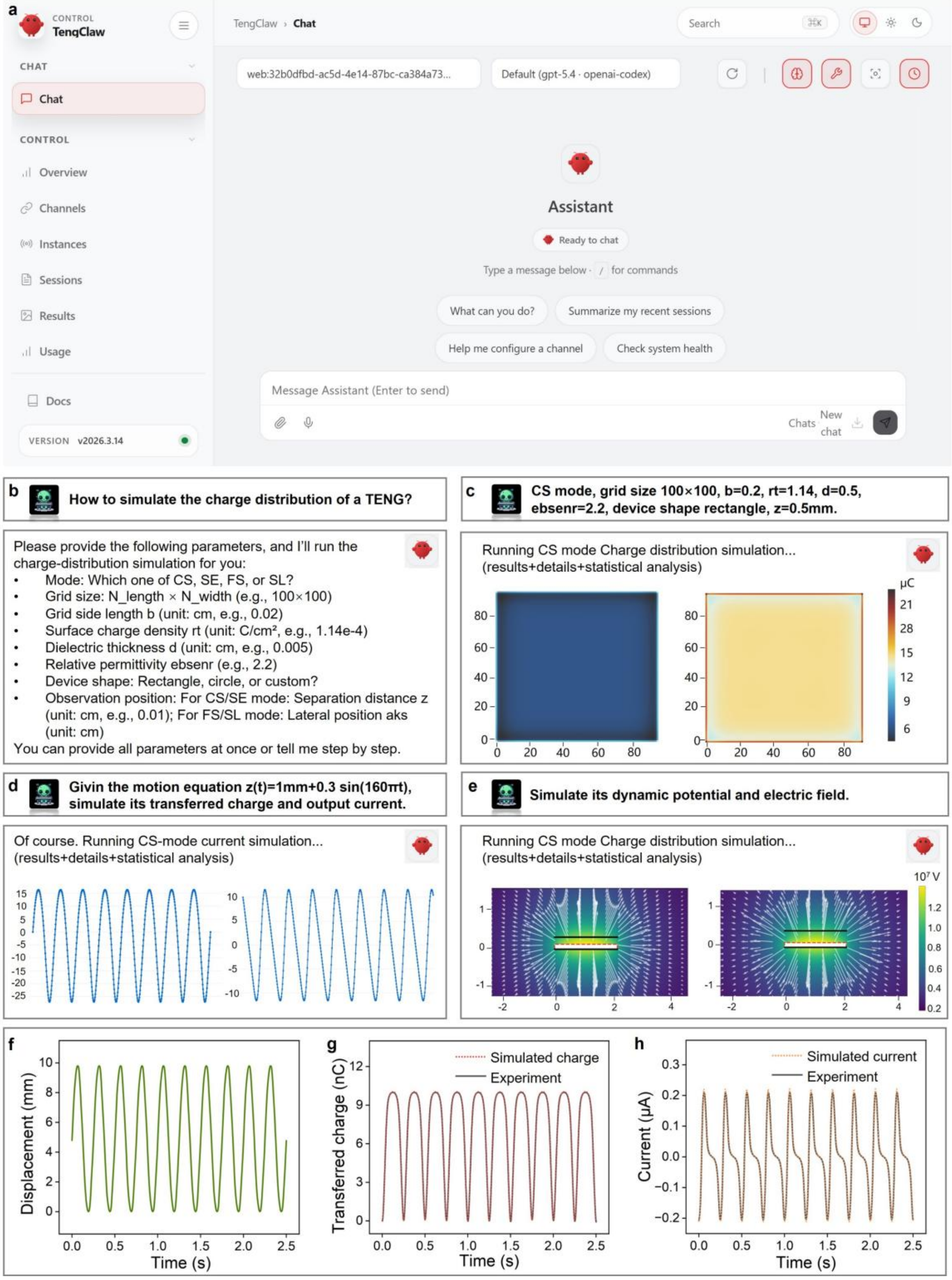

**Figure 3 Web-based realization and representative executable simulation cases.** a, Web interface of TENG-CLAW. b, Interactive clarification for charge-distribution simulation, where the system identifies the physical inputs required for an admissible finite-geometry calculation. c, Charge-distribution results generated after the task is compiled into TENG-IR and routed to the appropriate model branch. d, Dynamic transferred charge and current predicted from a user-defined motion law. e, Dynamic

potential and electric-field maps generated within the same executable simulation state. f, Linear-motor-driven displacement profile used for dynamic-output validation. g, Simulated and experimental transferred charge under the prescribed motion. h, Simulated and experimental current under the same motion.

The third case extends the same workspace from lumped outputs to spatial electrostatic interpretation. After charge transfer and current are obtained, the user requests dynamic potential and electric-field distributions (Fig. 3e). The system returns field maps in the same workspace, linking waveform-level observables to the underlying spatial electrostatic structure. TENG-CLAW therefore preserves the path from physical problem definition to spatial fields and measurable outputs, rather than generating isolated curves or images.

A linear-motor-driven experiment further validates dynamic-output prediction. The TENG was actuated with the prescribed displacement trajectory shown in Fig. 3f, and the simulated charge and current closely match the experimental trajectories (Fig. 3g,h). In addition to dialogue-based task specification, the platform also supports manual parameter input for direct simulation. Supplementary Video 2 demonstrates local self-deployment from the public code repository and includes a representative manually configured simulation of the linear-motor-driven TENG.

## Discussion

This work shifts TENG modelling from isolated post hoc calculations to physics-governed executable computation. The central contribution is not only a new solver or a web interface, but a charge-defined modelling framework in which physical state definition, solver-regime selection and workflow execution are connected. Each result originates from an admissible electrostatic problem in which charge states, geometry, boundary conditions and target observables are mutually consistent.

Within this formulation, analytical and finite-geometry descriptions become regime-dependent branches of one electrostatic closure. The infinite-plate branch provides transparency and efficiency for near-uniform fields, whereas the finite-geometry branch resolves boundary-induced redistribution when edge effects or spatial observables are important. The comparison between the two branches quantifies when the analytical approximation remains valid and when finite-geometry computation is required. This establishes a reproducible basis for model selection rather than relying on empirical solver preference.

The executable layer extends this physical hierarchy into a computational research workflow. TENG-CLAW compiles open-ended research goals into typed intermediate representations, checks physical and capability constraints before execution, records assumptions and approximations and stores results as reusable artifacts in an experiment graph. This structure addresses a central problem in device modelling: numerical solutions can be readily produced by simulation software, but physically meaningful solutions require an admissible model in which electrical conditions and solver routes are mutually constrained and explicitly preserved.

More broadly, the framework illustrates how domain theory can be transformed into reusable computational infrastructure. By making the charge state, solver route, assumptions and artifacts inspectable, TENG-CLAW provides a route towards reproducible simulation, automated design exploration and community-level comparison of TENG models. Although the present implementation is validated using TENGs, the same workflow principle—typed physical representation, physical-governance gating and experiment-graph memory—has considerable potential as an extensible basis for physics-constrained computation in other source-driven electrostatic or electromechanical systems with dynamic boundaries and spatially non-uniform fields.

## Methods

### User access to TENG-CLAW

TENG-CLAW was made available through a browser-based workspace in two deployment modes. In hosted mode, the OpenClaw gateway, TENG-CLAW plugin and TENG worker were deployed and maintained by our team, allowing users to access the workspace through a browser without local installation. In self-deployment mode, users installed the same gateway-plugin-worker stack in their own computing environment using the deployment instructions and accessed the workspace through a local or user-managed browser endpoint. Both modes provide the same workspace views, including Chat, Manual Sim, Results, Model Settings and Docs. Hosted mode supports immediate demonstration and evaluation. Self-deployment provides local control over model configuration, execution environment and data storage.

**Finite-plate numerical implementation**

Finite-plate results were obtained from the integral-equation formulation in the main text and the expanded equations in Supplementary Notes 2-7. The implementation used a point-matching Method of Moments scheme. Each electrode surface was divided into $N \times N$ square panels with side length $b$. The surface charge density on each panel was treated as a constant unknown, and the collocation point was placed at the panel center. Panels were numbered row by row, and the center coordinate of each panel was calculated from its row and column indices.

For each separation $z$, the interaction matrix was assembled from Coulomb potential contributions between source panels and collocation points. Off-diagonal entries were evaluated from the distance between distinct source and observation panels. Diagonal entries, where source and collocation panels coincide, were replaced by the finite-area self-potential of a uniformly charged square panel, as described in Supplementary Note 5. This treatment removes the point-source singularity while preserving the finite-size nature of the discretized surface.

The unknown conductor charge-density vectors were solved from the assembled linear system under charge-conservation and equipotential constraints. Prescribed triboelectric and bound-charge distributions were treated as fixed terms, whereas free charges on conducting electrodes were solved self-consistently. Transferred charge was obtained by integrating the solved charge density over the corresponding electrode surface. For time-dependent calculations, the prescribed motion law $z(t)$ was sampled at discrete time points, the electrostatic problem was solved independently at each separation, and current was calculated from the numerical derivative of the transferred-charge trajectory.

Spatial potential maps were calculated after the electrode charge densities were obtained. The potential at each point of the Cartesian field grid was evaluated by summing contributions from solved conductor charges and prescribed charge distributions. Electric-field maps were obtained from $\mathbf{E}(\mathbf{r}) = -\nabla\phi(\mathbf{r})$, with the gradient evaluated by finite differences. The same post-processing procedure was used for charge-density maps, potential maps and electric-field distributions in the main text.

Finite-plate charge-density calculations were performed using a custom Python implementation of the charge-defined model. Unless otherwise stated, the rectangular electrode was discretized with $N_{\text{length}}$=$N_{\text{width}}$=100 and the potential-field grid spacing was 0.2 mm. The dielectric polarization contribution was incorporated through the factor $-(\varepsilon_r - 1)/2\varepsilon_r$. The linear system was solved with PyTorch linear-algebra routines and can run on either CPU or GPU. Grid convergence was assessed by increasing $N_{\text{length}}$ and $N_{\text{width}}$ and confirming that the total transferred charge changed by less than the prescribed tolerance while the qualitative edge-localization pattern remained unchanged.

For dynamic simulations, the mechanical trajectory $z(t)$ was specified before electrostatic calculation. At each sampled time point, the instantaneous separation was passed to the analytical or finite-plate solver according to the model regime required by the requested observable. The transferred-charge sequence $Q(t)$ was assembled from individual electrostatic solutions, and the current was calculated by temporal differentiation.

**TENG-CLAW architecture and implementation**

TENG-CLAW was implemented as a web-based research workspace built on OpenClaw. OpenClaw provides the browser gateway, session runtime, workspace management and tool-routing infrastructure. The TENG worker contains the analytical and numerical simulation routines. The plugin layer registers TENG-CLAW tools, invokes orchestration, formats trace and result cards, exposes artifact routes and connects each session to short-horizon research memory.

The main executable entry point was tengclaw_orchestrate. A browser request passed through the OpenClaw session runtime and then entered the TENG-CLAW workflow. The Compiler converted the user request into a structured TENG research object. critic_preflight performed deterministic support and validity checks using the worker-side capability registry. The Physics Critic reported feasibility, missing parameters, warnings, approximations and fallback options. The Research Planner selected the executable route, and the Research Reporter converted numerical or visual outputs into result summaries and trace metadata. Execution was permitted only after the worker-side registry returned a supported action and a complete or safely defaultable parameter set.

The stable public worker exposed tengclaw_orchestrate, teng_simulate, teng_timeseries, teng_field_snapshot, teng_field_animation, teng_optimize, teng_study and teng_report. tengclaw_orchestrate was used as the default entry point for ordinary research interaction because it records intent interpretation, validity checks, solver selection and artifact generation in a single trace. Lower-level tools were used only when the required physical parameters had already been specified or when strict backend reproduction was required.

**Physical governance and reproducibility boundary**

The deterministic execution gate used the capability registry as the source of truth for supported modes, geometries, observables and solver actions. The preflight check covered unit consistency, mode and geometry compatibility, boundary-condition completeness, solver support, observable availability and safe defaultability of missing fields. The resulting verdict was one of four states: pass, clarify, approximate or unsupported. A pass verdict allowed stable execution. A clarify verdict required additional physical input. An approximate verdict allowed execution only with explicit assumptions. An unsupported verdict stopped stable execution before solver invocation.

The public deployment was restricted to registry-supported solver actions. Unsupported requests returned an unsupported verdict before solver invocation. Experimental extensions were isolated from stable execution and required proposal, confirmation, validation and registration; they were not used to generate the validated results reported in the main text.

**Artifacts, trace metadata and experiment graph**

Each execution produced result artifacts and trace metadata. Result artifacts included transferred-charge curves, current curves, charge-density maps, potential maps, electric-field maps, animations, raw data files or reports, depending on the requested action. Trace metadata recorded the interpreted task, supplied and missing parameters, preflight verdict, selected solver branch, warnings, approximations, artifact identifiers and links to prior runs.

A typical run directory contained a normalized request file, raw numerical output, summary file, report, static plot or poster, dynamic field animation when available and a per-run experiment-graph entry. The result card answered what a run produced, whereas the trace card recorded why a specific route was chosen. Experiment-graph entries stored task identifiers, TENG-IR objects, critic outputs, execution plans, run identifiers, metrics, conclusions, next actions and artifact paths. These records allow later requests to reuse, compare or extend previous calculations without reconstructing physical context from free-text chat history.

## Data availability

The data supporting the findings of this study are included in the main text, Supplementary Information and GitHub repository.

## Code availability

The source code, deployment instructions, representative configuration files and example workflows required to reproduce the analytical calculations, finite-geometry simulations and representative TENG-CLAW executions reported in this study are publicly available at GitHub: https://github.com/helloi1/tengclaw.

## Acknowledgements

This work was supported by National Key R&D Program of China (Grant No. 2024YFB3816000), Shenzhen Science and Technology Program (Grant Nos. KJZD20240903100905008 and JCYJ20240813111910014), Guangdong Innovative and Entrepreneurial Research Team Program (Grant No. 2021ZT09L197), and China Postdoctoral Science Foundation (Grant No. 2025M780094).

## Competing interests

The authors declare no competing interests.

# Supplementary Information for

# Physics-governed executable modelling of triboelectric nanogenerators

**Supplementary Table 1. Symbols and definitions**

| Symbol | Meaning | Notes / context |
|---|---|---|
| $\rho_s$ | Free source charge density | Appears in Eq. (1) |
| $\rho_b$ | Polarization-induced bound charge density | Appears in Eq. (1) |
| $Q$ | Transferred charge | Core macroscopic observable |
| $Q(t)$ | Time-dependent transferred charge | Used in Eqs. (3) and (4) |
| $I(t)$ | Output current | Defined as $I(t)=\mathrm{d}Q(t)/\mathrm{d}t$ |
| $\phi$ | Electrostatic potential | Used in the source-driven closure and field calculation |
| $V(t)$ | Intrinsic electrostatic potential difference between the two electrodes | Appears in Eq. (5) |
| $\mathbf{E}_k(t)$ | Piecewise electric field in region $k$ | Appears in Eq. (6) |
| $\mathbf{E}(\mathbf{r})$ | Electric field at position $\mathbf{r}$ | Calculated from $\mathbf{E}(\mathbf{r}) = -\nabla\phi(\mathbf{r})$ |
| $\Omega(z)$ | Geometry-dependent computational domain | Describes the device domain under separation ($z$) |
| $z_0$ | Initial electrode separation | Used in the contact-separation model |
| $z(t)$ | Instantaneous electrode separation | Describes mechanical modulation |
| $d_0$ | Dielectric thickness | Used in the infinite-plate model |
| $\varepsilon_0$ | Vacuum permittivity | Used in electrostatic field and potential calculations |
| $\varepsilon_r$ | Relative permittivity | Dielectric material parameter |
| $S$ | Effective device area | Used in the analytical transferred-charge expression |
| $\sigma_{\mathrm{te}}$ | Triboelectric charge density | One component of the initial charge ensemble |
| $\sigma_{\mathrm{tp}}$ | Pre-charging charge density | One component of the initial charge ensemble |
| $\sigma_{\mathrm{c}}$ | Compensating electrode charge density | Used to enforce electrostatic equilibrium and charge conservation |
| $\sigma_{\mathrm{eff}}$ | Effective driving source charge density | Defined as $\sigma_{\mathrm{eff}} = \sigma_{\mathrm{te}} + \sigma_{\mathrm{tp}}$ |
| $\mathcal{C}$ | Source-driven closure operator | Represents charge conservation, boundary conditions and electrostatic equilibrium |
| $\mathcal{P}_{\infty}$ | Infinite-plate electrostatic solution operator | Compact representation for the intrinsic potential difference |
| $\mathcal{E}_{k,\infty}$ | Infinite-plate electric-field solution operator for region (k) | Compact representation for piecewise electric fields |
| $k$ | Region index | Used for piecewise electric-field regions |
| $\mathbf{r}$ | Observation position vector | Used in the finite-geometry formulation |
| $\mathbf{r}'$ | Source position vector | Integration variable in the finite-geometry formulation |
| $\sigma_j(\mathbf{r})$ | Free charge density on conducting surface ($j$) | Position-dependent unknown in the finite-geometry model |
| $\sigma_{\mathrm{b}}(\mathbf{r})$ | Bound charge density on dielectric surfaces | Prescribed or induced source term in the finite-geometry model |

| $K_{ij}$ | Electrostatic kernel between conductor ($j$) and conductor ($i$) | Determined by geometry and material boundaries |
|---|---|---|
| $K_{ib}$ | Electrostatic kernel from bound/source charges to conductor ($i$) | Used in the coupled integral equation |
| $V_i$ | Electric potential of conductor ($i$) | Enforces conductor equipotential condition |
| $\Gamma_i$ | Surface of conductor ($i$) | Used in the charge-conservation constraint |
| $Q_i^{(0)}$ | Initial charge assigned to conductor ($i$) | Determined by the initial charge ensemble |
| $s_i$ | Direction factor for transferred charge relative to electrode ($i$) | Used in global charge conservation |
| $Q_i$ | Total charge on conductor ($i$) | Appears in the charge-conservation constraint |
| $Q_{tot}$ | Total conserved charge of the system | Global charge-conservation quantity |
| $\boldsymbol{\sigma}_{\mathrm{c}}$ | Vector of unknown conductor surface-charge densities | Solved in the discretized finite-plate system |
| $\boldsymbol{\sigma}_{\mathrm{f}}$ | Vector of prescribed fixed source-charge distributions | Includes triboelectric and bound charges |
| $\mathbf{V}_{\mathrm{c}}$ | Vector of conductor potentials | Unknown or constrained variable in the linear system |
| $\mathbf{q}_{\mathrm{c}}$ | Vector of conductor charge constraints | Used with $\mathbf{A}\boldsymbol{\sigma}_{\mathrm{c}} = \mathbf{q}_{\mathrm{c}}$ |
| $\mathbf{K}_{cc}(z)$ | Conductor-conductor interaction matrix | Geometry-dependent matrix in the finite-plate model |
| $\mathbf{K}_{cf}(z)$ | Matrix describing the contribution of fixed source charges to conductor collocation points | Appears in the discretized system |
| $\mathbf{P}$ | Potential-incidence matrix | Maps conductor potentials to collocation points |
| $\mathbf{A}$ | Charge-integration matrix | Integrates panel charges to enforce charge conservation |
| $l$ | Characteristic lateral dimension | Used to define the geometric regime |
| $\chi$ | Regime diagnostic parameter | Defined as $\chi = z / l$ |
| $b$ | Side length of a discretized panel | Used in the Method of Moments implementation |
| $N$ | Number of panels along each direction | Gives an $N \times N$ discretization |
| $N_{\text{length}}$ | Number of panels along the length direction | Used in grid convergence tests |
| $N_{\text{width}}$ | Number of panels along the width direction | Used in grid convergence tests |
| $Q_{\infty}$ | Transferred charge predicted by the infinite-plate model | Used in the deviation analysis |
| $Q_{\text{finite}}$ | Transferred charge predicted by the finite-plate model | Used as the spatially resolved reference |
| $\delta_Q$ | Relative transferred-charge deviation | Defined as $(Q_{\infty} - Q_{\text{finite}}) / Q_{\text{finite}} \times 100\%$ |
| $u_t$ | Open-ended user request at step ($t$) | Used in the TENGCLAW workflow |
| $S_{t-1}$ | Previous scientific state | Used in the state-transition equation |
| $S_t$ | Updated scientific state | Used in the state-transition equation |
| $I_t$ | Structured intent at step ($t$) | Workflow variable; not the same as current I($t$) |
| $C_t$ | Physical context at step ($t$) | Workflow variable |
| $\Pi_t$ | Execution plan at step ($t$) | Workflow variable |
| $A_t$ | Generated artifacts at step ($t$) | Workflow variable |
| $\Gamma$ | Scientific state-transition operator | Workflow variable; not the same as conductor surface $\Gamma_i$ |
| $G_{\text{phys}}$ | Physical governance operator | Evaluates task admissibility before execution |
| $v_t$ | Admissibility verdict | Values include pass, clarify, approximate and unsupported |
| $B_t$ | Missing or ambiguous physical parameters | Returned by the physical governance operator |
| $W_t$ | Warnings associated with model assumptions | Returned by the physical governance operator |
| $\pi_t$ | Selected computational branch | Infinite-plate analytical branch or finite-geometry numerical branch |
| $\eta_t$ | Approximation level | Records the level of approximation used in execution |
| $\nabla\phi(\mathbf{r})$ | Gradient of electrostatic potential | Used to obtain the electric field |
| $\sigma_1$ | Surface charge density on the first electrode charge layer | Supplementary Note 1; infinite-plate charge-layer solution |
| $\sigma_2$ | Surface charge density on the second electrode charge layer | Supplementary Note 1 |
| $\sigma_3$ | Surface charge density on the third electrode charge layer | Supplementary Note 1 |
| $\sigma_4$ | Surface charge density on the fourth electrode charge layer | Supplementary Note 1 |
| $V_t$ | Reference potential at the upper surface of the dielectric layer | Supplementary Note 1 |
| $\mathbf{e}_z$ | Unit vector along the vertical direction | Used for expressing one-dimensional electric fields |
| $\mathbf{e}_x$ | Unit vector along the $x$-direction | Used in Cartesian-coordinate representation |
| $\mathbf{e}_y$ | Unit vector along the $y$-direction | Used in Cartesian-coordinate representation |
| $q(z)$ | Net charge imbalance between the two electrodes as a function of separation | Used before deriving transferred charge |
| $\mathrm{d}z$ | Infinitesimal displacement of the dielectric layer or electrode separation | Used in differential charge-transfer derivation |
| $v$ | Relative velocity, $v = \mathrm{d}z/\mathrm{d}t$ | Used to express current under mechanical motion |

| $\mathbf{D}$ | Electric flux density | Used in the displacement-current description |
|---|---|---|
| $I_d$ | Displacement current inside the TENG | Obtained by integrating displacement-current density over area |
| $\mathbf{P}$ | Polarization vector | Used to express dipole-induced bound charge in Supplementary Note 2; distinct from the matrix P in the main text |
| $\mathbf{n}$ | Surface normal vector | Used in $\sigma_b = \mathbf{P} \cdot \mathbf{n}$ |
| $\sigma_t$ | Total prescribed triboelectric-related source charge density | Defined as $\sigma_t = \sigma_{te} + \sigma_{tp}$ in Supplementary Note 2 |
| $\sigma_0$ | Combined charge-density parameter | Defined as $\sigma_0 = \sigma_{te} + \sigma_c$ |
| $S_1$ | Surface domain of the independent electrode | Used in finite-plate integral equations |
| $S_2$ | Surface domain of the base electrode | Used in finite-plate integral equations |
| $S_t$ | Surface domain of the dielectric triboelectric layer | Used as an integration domain; not the workflow state $S_t$ in the main text |
| $S_1'$ | Source-point area of the independent electrode | Used in integral equations |
| $S_2'$ | Source-point area of the base electrode | Used in integral equations |
| $S_t'$ | Source-point area of the dielectric surface | Used in integral equations |
| $\mathbf{r}_1$ | Field-point coordinate on the first electrode | Supplementary Note 2 |
| $\mathbf{r}_2$ | Field-point coordinate on the second electrode | Supplementary Note 2 |
| $x$ | Cartesian coordinate along the $x$-axis | Used in explicit finite-plate equations |
| $y$ | Cartesian coordinate along the $y$-axis | Used in explicit finite-plate equations |
| $x'$ | Source-point $x$-coordinate | Used in surface integrals |
| $y'$ | Source-point $y$-coordinate | Used in surface integrals |
| $x_1$ | $x$-coordinate of a field point on electrode 1 | Supplementary Note 2 |
| $y_1$ | $y$-coordinate of a field point on electrode 1 | Supplementary Note 2 |
| $x_2$ | $x$-coordinate of a field point on electrode 2 | Supplementary Note 2 |
| $y_2$ | $y$-coordinate of a field point on electrode 2 | Supplementary Note 2 |
| $n$ | Number of basis functions or weighting functions | Supplementary Note 3; distinct from surface normal n |
| $f_m$ | Subdomain basis function for the $m$-th segment | Used in Method of Moments discretization |
| $\sigma_{2m}$ | Expansion coefficient of the charge density on the $m$-th subdomain of electrode 2 | Supplementary Note 3 |
| $S_m$ | Area of the $m$-th subdomain segment | Used in subdomain basis definition |
| $L(\sigma_2)$ | Representative integral operator acting on $\sigma_2$ | Used to illustrate discretization |
| $w_m$ | Weighting function in the Method of Moments | Used in point-matching, Galerkin or least-squares schemes |
| $\delta(\mathbf{r}-\mathbf{r}_{\mathrm{m}})$ | Dirac delta weighting function | Used in the point-matching method |
| $r_m$ | Collocation point for the m-th weighting function | Supplementary Note 3 |
| $l_{mn}$ | Matrix element obtained by testing $L(f_n)$ with $w_m$ | Supplementary Note 3 |
| $g_m$ | Tested right-hand-side term associated with $w_m$ | Supplementary Note 3 |
| $m$ | Source-segment or basis-function index | Used in discretized equations |
| $i$ | Collocation-point or column index, depending on context | Used in discretization and surface numbering |
| $j$ | Row index of a discretized electrode segment | Supplementary Note 4 |
| $x_k$ | Center x-coordinate of the $k$-th segment | Supplementary Note 4 |
| $y_k$ | Center y-coordinate of the $k$-th segment | Supplementary Note 4 |
| $\boldsymbol{\sigma}_1$ | Discretized charge-density vector on the independent electrode | Supplementary Notes 5 and 6 |
| $\boldsymbol{\sigma}_2$ | Discretized charge-density vector on the base electrode | Supplementary Notes 5 and 6 |
| $\boldsymbol{\sigma}_{\mathrm{t}}$ | Discretized prescribed triboelectric-charge vector | Supplementary Note 5 |
| $\boldsymbol{\sigma}_{\mathrm{e}}$ | Discretized prescribed bound-charge vector | Appears in the vector form on page 10; likely corresponds to σ_b vector |
| $\sigma_{1m}$ | Charge-density element of the $\sigma_1$ vector on the $m$-th segment | Supplementary Note 5 |
| $\sigma_{2m}$ | Charge-density element of the $\sigma_2$ vector on the m-th segment | Supplementary Note 5 |
| $a$ | Auxiliary vector used to express the electrode charge-density solution | Supplementary Note 6 |
| $u$ | Auxiliary vector used to simplify the relation between the $\boldsymbol{\sigma}_1$ vector and the $\boldsymbol{\sigma}_2$ vector | Supplementary Note 6 |
| $h$ | Vertical coordinate or height of a field-evaluation point | Used in the finite-plate potential equation |
| $V(x,y,z)$ | Spatial potential distribution in Cartesian coordinates | Supplementary Note 7 |
| $\mathrm{V}(x_i, y_i, z_i)$ | Discretized potential at the $i$-th field point | Supplementary Note 7 |
| $z_i$ | $z$-coordinate of the $i$-th field point | Supplementary Note 7 |
| $N \times N \times (d + z)$ | Three-dimensional discretization of the computational domain | Supplementary Note 7 |

## Supplementary Note 1. Charge distribution and electrical response in the infinite-plate model

We consider a TENG consisting of a dielectric layer with thickness $d_0$ and relative permittivity $\varepsilon_r$. The initial internal separation between the dielectric layer and the opposite electrode is denoted by $z_0$. Within the infinite-plate approximation, edge effects are neglected and the electrostatic field is treated as one-dimensional[1, 2]. By applying charge conservation, the zero-field condition and the equipotential condition of the electrodes, the governing equations are obtained as follows:

$$\begin{cases} -\sigma_1+\sigma_2+\sigma_3-\sigma_4=\sigma_{\mathrm{te}}+\sigma_{\mathrm{c}} \\ -\dfrac{\sigma_{\mathrm{tp}}+\sigma_{\mathrm{te}}}{2\varepsilon_0}+\dfrac{\sigma_1}{2\varepsilon_0}+\dfrac{\sigma_2}{2\varepsilon_0}+\dfrac{\sigma_3}{2\varepsilon_0}-\dfrac{\sigma_4}{2\varepsilon_0}=0 \\ \dfrac{\sigma_{\mathrm{tp}}+\sigma_{\mathrm{te}}}{2\varepsilon_0}+\dfrac{\sigma_1}{2\varepsilon_0}-\dfrac{\sigma_2}{2\varepsilon_0}-\dfrac{\sigma_3}{2\varepsilon_0}-\dfrac{\sigma_4}{2\varepsilon_0}=0 \\ V_{\mathrm{t}}-\dfrac{\sigma_2}{\varepsilon_0}z_0=V_{\mathrm{t}}+\dfrac{\sigma_3}{\varepsilon_0\varepsilon_{\mathrm{r}}}d_0 \end{cases}$$

Here, $\sigma_1$, $\sigma_2$, $\sigma_3$ and $\sigma_4$ denote the surface charge densities on the corresponding electrode charge layers. The parameters $\sigma_{\mathrm{te}}$, $\sigma_{\mathrm{tp}}$ and $\sigma_c$ represent the triboelectric charge density, pre-charging charge density and compensating electrode charge density, respectively. $\varepsilon_0$ is the vacuum permittivity, and $V_{\mathrm{t}}$ is the reference potential at the upper surface of the dielectric layer. Solving the above electrostatic system gives the charge densities on the four electrode charge layers:

$$\begin{cases} \sigma_1=\dfrac{\sigma_{\mathrm{tp}}-\sigma_{\mathrm{c}}}{2} \\ \sigma_2=\dfrac{\left(\sigma_{\mathrm{tp}}+\sigma_{\mathrm{te}}\right)d_0}{z_0\varepsilon_{\mathrm{r}}+d_0} \\ \sigma_3=\dfrac{(\sigma_{\mathrm{tp}}+\sigma_{\mathrm{te}})z_0\varepsilon_{\mathrm{r}}}{z_0\varepsilon_{\mathrm{r}}+d_0} \\ \sigma_4=\dfrac{\sigma_{\mathrm{tp}}-\sigma_{\mathrm{c}}}{2} \end{cases}$$

Once the charge densities are determined, the electric-field distribution in each region of the TENG can be derived. Taking $\mathbf{e}_z$ as the unit vector along the vertical direction, the piecewise electric fields are expressed as:

$$\mathbf{E}_1 = \frac{\sigma_{\mathrm{tp}} - \sigma_{\mathrm{c}}}{2\varepsilon_0}\mathbf{e}_{\mathrm{z}}$$

$$\mathbf{E}_2 = \frac{(\sigma_{\mathrm{tp}} + \sigma_{\mathrm{te}})z_0}{\varepsilon_0\left(z_0\varepsilon_{\mathrm{r}} + d_0\right)}\mathbf{e}_{\mathrm{z}}$$

$$\mathbf{E}_3 = -\frac{(\sigma_{\mathrm{tp}} + \sigma_{\mathrm{te}})d_0}{\varepsilon_0\left(z_0\varepsilon_{\mathrm{r}} + d_0\right)}\mathbf{e}_{\mathrm{z}}$$

$$\mathbf{E}_4 = \frac{\sigma_{\mathrm{tp}} - \sigma_{\mathrm{c}}}{2\varepsilon_0}\mathbf{e}_{\mathrm{z}}$$

The net charge imbalance between the two electrodes is obtained from the difference between the total charges carried by the corresponding electrode charge layers:

$$q\left(z\right) = \left[\sigma_3 - \sigma_4 - \left(\sigma_2 - \sigma_1\right)\right]S = \frac{(\sigma_{\mathrm{tp}} + \sigma_{\mathrm{te}})\left(z\varepsilon_{\mathrm{r}} - d_0\right)S}{z\varepsilon_{\mathrm{r}} + d_0}$$

When the separation between the dielectric layer and the independent electrode increases by an infinitesimal displacement d$z$, the corresponding transferred charge is determined by the variation of this net charge imbalance:

$$\mathrm{d}Q = \frac{1}{2}\left[q\left(z + \mathrm{d}z\right) - q\left(z\right)\right]$$

Taking the differential limit gives the transferred charge per unit displacement:

$$\frac{\mathrm{d}Q}{\mathrm{d}z} = \frac{1}{2}\frac{q\left(z + \mathrm{d}z\right) - q\left(z\right)}{\mathrm{d}z} = \frac{1}{2}\frac{\mathrm{d}q}{\mathrm{d}z} = \frac{(\sigma_{\mathrm{tp}} + \sigma_{\mathrm{te}})\varepsilon_{\mathrm{r}}Sd_0}{\left(z\varepsilon_{\mathrm{r}} + d_0\right)^2}$$

Integrating this expression from the initial separation $z_0$ to an arbitrary separation $z$ gives the transferred charge as a function of separation:

$$Q = \int_{z_0}^{z}\mathrm{d}Q = \int_{z_0}^{z}\frac{(\sigma_{\mathrm{tp}} + \sigma_{\mathrm{te}})\varepsilon_{\mathrm{r}}Sd_0}{\left(z\varepsilon_{\mathrm{r}} + d_0\right)^2}\mathrm{d}z = (\sigma_{\mathrm{tp}} + \sigma_{\mathrm{te}})\varepsilon_{\mathrm{r}}Sd_0\int_{z_0}^{z}\frac{\mathrm{d}z}{\left(z\varepsilon_{\mathrm{r}} + d_0\right)^2} = \frac{(\sigma_{\mathrm{tp}} + \sigma_{\mathrm{te}})\varepsilon_{\mathrm{r}}Sd_0\left(z - z_0\right)}{\left(z\varepsilon_{\mathrm{r}} + d_0\right)\left(z_0\varepsilon_{\mathrm{r}} + d_0\right)}$$

For a time-dependent separation $z(t)$, the short-circuit current is obtained by differentiating the transferred charge with respect to time. If the relative velocity is denoted by $v$=d$z$/d$t$, the current can be written as:

$$I(t)=\frac{\mathrm{d}Q}{\mathrm{d}t}=\frac{\mathrm{d}Q}{\mathrm{d}z}v=\frac{(\sigma_{\mathrm{tp}}+\sigma_{\mathrm{te}})\varepsilon_{\mathrm{r}}Sd_0v}{\left(z\varepsilon_{\mathrm{r}}+d_0\right)^2}$$

The same electrical response can also be described from the displacement-current perspective[3, 4]. The displacement-current density is determined by the temporal variation of the electric flux density $D$:

$$\frac{\partial D}{\partial t}=\varepsilon_0\frac{\partial E_3}{\partial t}=\varepsilon_0\frac{\partial E_3}{\partial z}\frac{\partial z}{\partial t}=\frac{(\sigma_{\mathrm{tp}}+\sigma_{\mathrm{te}})d_0\varepsilon_{\mathrm{r}}v}{\left(z\varepsilon_{\mathrm{r}}+d_0\right)^2}$$

where $D$ is the electric flux density. Integrating the displacement-current density over the active area $S$ gives the total displacement current inside the TENG:

$$I_{\mathrm{d}}=\int_S\frac{\partial D}{\partial t}dS=\frac{(\sigma_{\mathrm{tp}}+\sigma_{\mathrm{te}})d_0\varepsilon_{\mathrm{r}}Sv}{\left(z\varepsilon_{\mathrm{r}}+d_0\right)^2}$$

Therefore, the displacement current inside the TENG is equal to the conduction current in the external circuit. This equivalence confirms the consistency between the charge-transfer description and Maxwell's displacement-current description.

The intrinsic potential difference between the two electrodes is obtained by integrating the electric field across the dielectric layer and the air gap:

$$V=E_2d_0+E_3z_0=\frac{\left(\sigma_{\mathrm{tp}}+\sigma_{\mathrm{c}}\right)d_0+\left(\sigma_{\mathrm{tp}}-\sigma_{\mathrm{c}}+2\sigma_{\mathrm{te}}\right)\varepsilon_{\mathrm{r}}z_0}{2\varepsilon_0\varepsilon_{\mathrm{r}}}$$

This derivation shows that, in the infinite-plate limit, the charge distribution, electric field, potential difference, transferred charge and current are all determined by the same initial charge ensemble and electrostatic boundary conditions. The measurable electrical response is therefore a consequence of geometry-driven charge redistribution rather than an independently imposed circuit variable.

**Supplementary Note 2. Governing equation for the finite plate model**

The electric-dipole contribution can be represented as the superposition of two layers of bound charges[5], denoted by $\sigma_{\mathrm{b}}$. In the infinite-plate model, this dipole contribution has a negligible influence on the overall electrostatic response. In the finite-plate model, however, the dipole-induced bound charges can affect the electric field in the space outside the dielectric layer. Because this contribution remains much smaller than that of the triboelectric charges, the bound charge density can be evaluated using the infinite-plate solution and subsequently introduced into the finite-plate formulation with negligible error. The bound charge density is given by:

$$\sigma_{\mathrm{b}} = \mathbf{P} \cdot \mathbf{n} = \frac{z(\varepsilon_{\mathrm{r}} - 1)\left(\sigma_{\mathrm{tp}} + \sigma_{\mathrm{te}}\right)}{\varepsilon_{\mathrm{r}} z + d_0}$$

Unlike the infinite-plate model, the finite-plate model does not assume a spatially uniform electrode charge distribution. Instead, the charge densities vary with position on the electrode surfaces. Specifically, $\sigma_1$ and $\sigma_2$ are treated as functions of the spatial coordinate $\mathbf{r}$. Here, $\mathbf{r}$ denotes the position vector from the origin O to the field point, where the potential is evaluated, whereas $\mathbf{r}'$ denotes the position vector from the origin O to the source point, where the source charge is located. The quantity $\mathbf{r} - \mathbf{r}'$ therefore represents the separation vector between the field point and the source point. In the Cartesian coordinate system, the position vector can be written as $\mathbf{r} = x\mathbf{e}_{\mathrm{x}} + y\mathbf{e}_{\mathrm{y}} + z\mathbf{e}_{\mathrm{z}}$. By applying charge conservation and the equipotential condition of the two electrodes, the governing integral equations of the finite-plate model are obtained as:

$$\begin{cases} \int_{S_1'} \sigma_1(\mathbf{r})\mathrm{d}S_1 + \int_{S_2} \sigma_1(\mathbf{r})\mathrm{d}S_2 = \sigma_0 S \\ \dfrac{1}{4\pi\varepsilon_0}\left[\int_{S_1} \dfrac{\sigma_1(\mathbf{r}')}{|\mathbf{r}_1 - \mathbf{r}'|}\mathrm{d}S_1' + \int_{S_{\mathrm{t}}} \dfrac{-\sigma_{\mathrm{t}}(\mathbf{r}') + \sigma_{\mathrm{b}}(\mathbf{r}')}{|\mathbf{r}_1 - \mathbf{r}'|}\mathrm{d}S_{\mathrm{t}}' + \int_{S_2} \dfrac{\sigma_2(\mathbf{r}') - \sigma_{\mathrm{b}}(\mathbf{r}')}{|\mathbf{r}_1 - \mathbf{r}'|}\mathrm{d}S_2'\right] = V \\ \dfrac{1}{4\pi\varepsilon_0}\left[\int_{S_2} \dfrac{\sigma_2(\mathbf{r}') - \sigma_{\mathrm{b}}(\mathbf{r}')}{|\mathbf{r}_2 - \mathbf{r}'|}\mathrm{d}S_2' + \int_{S_{\mathrm{t}}} \dfrac{-\sigma_{\mathrm{t}}(\mathbf{r}') + \sigma_{\mathrm{b}}(\mathbf{r}')}{|\mathbf{r}_2 - \mathbf{r}'|}\mathrm{d}S_{\mathrm{t}}' + \int_{S_1} \dfrac{\sigma_1(\mathbf{r}')}{|\mathbf{r}_2 - \mathbf{r}'|}\mathrm{d}S_1'\right] = V \end{cases}$$

where $\sigma_t = \sigma_{\mathrm{tp}} + \sigma_{\mathrm{te}}$ and $\sigma_0 = \sigma_{\mathrm{te}} + \sigma_c$. The integration domains $S_1'$, $S_2'$ and $S_t'$ correspond to the source-point areas of the independent electrode, the base electrode and the dielectric surface, respectively. The vectors $\mathbf{r}_1$ and $\mathbf{r}_2$ denote the field-point coordinates on the two electrodes.

The first equation enforces global charge conservation. The second and third equations impose the equipotential conditions of the two electrodes, indicating that the electrode potential $V$ is generated by the combined contribution of all charges distributed within the TENG, including the electrode charges, triboelectric charges and bound charges. Expressed explicitly in the Cartesian coordinate system, the finite-plate governing equations become:

$$
\begin{cases}
\iint_{S_1} \sigma_1(x,y)dxdy + \iint_{S_2} \sigma_2(x,y)dxdy = \sigma_0 S \\
\frac{1}{4\pi\varepsilon_0}\iint_{S_1} \frac{\sigma_1(x',y')}{\sqrt{(x_1-x')^2+(y_1-y')^2}}dx'dy' + \frac{1}{4\pi\varepsilon_0}\iint_{S_t} \frac{-\sigma_t(x',y')}{\sqrt{(x_1-x')^2+(y_1-y')^2+z_0^2}}dx'dy' \\
+\frac{1}{4\pi\varepsilon_0}\iint_{S_t} \frac{\sigma_b(x',y')}{\sqrt{(x_1-x')^2+(y_1-y')^2+z_0^2}}dx'dy' + \frac{1}{4\pi\varepsilon_0}\iint_{S_2} \frac{\sigma_2(x',y')}{\sqrt{(x_1-x')^2+(y_1-y')^2+(d_0+z_0)^2}}dx'dy' \\
+\frac{1}{4\pi\varepsilon_0}\iint_{S_2} \frac{-\sigma_b(x',y')}{\sqrt{(x_1-x')^2+(y_1-y')^2+(d_0+z_0)^2}}dx'dy' = V \\
\frac{1}{4\pi\varepsilon_0}\iint_{S_2} \frac{\sigma_2(x',y')}{\sqrt{(x_2-x')^2+(y_2-y')^2}}dx'dy' + \frac{1}{4\pi\varepsilon_0}\iint_{S_2} \frac{-\sigma_b(x',y')}{\sqrt{(x_2-x')^2+(y_2-y')^2+z_0^2}}dx'dy' \\
+\frac{1}{4\pi\varepsilon_0}\iint_{S_t} \frac{-\sigma_t(x',y')}{\sqrt{(x_2-x')^2+(y_2-y')^2+z_0^2}}dx'dy' + \frac{1}{4\pi\varepsilon_0}\iint_{S_t} \frac{\sigma_b(x',y')}{\sqrt{(x_2-x')^2+(y_2-y')^2+(d_0+z_0)^2}}dx'dy' \\
+\frac{1}{4\pi\varepsilon_0}\iint_{S_1} \frac{\sigma_1(x',y')}{\sqrt{(x_2-x')^2+(y_2-y')^2+(d_0+z_0)^2}}dx'dy' = V
\end{cases}
$$

These equations form the continuous integral-equation description of the finite-plate TENG. They retain the same charge-source basis as the infinite-plate model, but allow the electrode charge densities to redistribute spatially under the influence of finite geometry, nonlocal electrostatic coupling and boundary-induced field distortion.

**Supplementary Note 3. Discretization of the integral equations**

To solve the finite-plate model of the TENG, the unknown continuous charge distribution is approximated by a linear combination of $n$ basis functions[6]. The basis functions are prescribed, whereas the corresponding expansion coefficients remain unknown. These coefficients are determined by projecting the governing equation onto a set of $n$ weighting functions, thereby converting the original functional equation into a finite system of algebraic equations. In this work, a point-matching implementation of the Method of Moments is adopted, in which the integral equation is enforced at selected collocation points within the computational domain.

Basis functions can generally be divided into global basis functions and subdomain basis functions[7]. Global basis functions can provide faster convergence when the qualitative form of the unknown solution is known in advance. However, their construction usually requires prior knowledge of the spatial characteristics of the solution. Subdomain basis functions are more flexible and easier to implement because they do not assume a specific global form of the unknown function. Although they typically require a larger number of subdivisions to achieve convergence, they are well suited for resolving spatially non-uniform charge distributions in finite-geometry TENGs.

The discretization procedure is illustrated using the following representative integral term:

$$L(\sigma_2) = \frac{1}{4\pi\varepsilon_0} \iint_{S_2'} \frac{\sigma_2(x', y')}{\sqrt{(x_2 - x')^2 + (y_2 - y')^2}} dx'dy'$$

where the subdomain basis function is defined as:

$$\sigma_2 = \sum_m f_{2m} \sigma_{2m}$$

and the basis function:

$$f_{2m} = \begin{cases} 1, (x, y) \in S_m \\ 0, (x, y) \notin S_m \end{cases}$$

Substituting this expansion into the integral expression gives:

$$L(\sigma_2) = \sum_m \frac{1}{4\pi\varepsilon_0} \frac{\sigma_{2m}}{\sqrt{(x_{2m} - x')^2 + (y_{2m} - y')^2}}$$

The remaining terms in the governing equations are discretized in the same manner.

The choice of weighting function determines the specific numerical scheme used to transform the integral equation. In the point-matching method, the Dirac delta function is used as the weighting function, so that the residual equation is enforced directly at selected collocation points. In the Galerkin method, the basis functions themselves are used as the weighting functions. In the least-squares method, the residual is used to construct the weighting operation. Here, the Dirac delta function is selected, which simplifies the inner product to the evaluation of the corresponding function at a specified point:

$$l_{mn} =< w_m ,\ L(f_n) >= \int_V \delta(\mathbf{r} - \mathbf{r}_\mathrm{m}) L(f_n) \mathrm{d}V = L(f_n)\big|_{\mathbf{r}=\mathbf{r}_\mathrm{m}}$$

Similarly, for the right-hand side of the equation, the inner product becomes:

$$g_m =< w_m ,\ g >= \int_V \delta(\mathbf{r} - \mathbf{r}_\mathrm{m}) g \mathrm{d}V = g(\mathbf{r}_\mathrm{m})$$

Because $V$ is a constant in the original equation, its inner product with the weighting function remains $V$. By applying the above discretization and point-matching procedure to the finite-plate integral equations, the continuous problem is transformed into a system of $2n$+1 algebraic equations:

$$\sum_{m=1}^{n} S_m (\sigma_{1m} + \sigma_{1m} - \sigma_0) = 0$$

$$\sum_{m=1}^{n} \frac{S_m}{4\pi\varepsilon_0} \left( \frac{\sigma_{1m}}{\sqrt{(x_{1m} - x_i)^2 + (y_{1m} - y_i)^2}} - \frac{\sigma_\mathrm{t} - \sigma_\mathrm{b}}{\sqrt{(x_{1m} - x_i)^2 + (y_{1m} - y_i)^2 + z^2}} + \frac{\sigma_{2m} + \sigma_\mathrm{b}}{\sqrt{(x_{1m} - x_i)^2 + (y_{1m} - y_i)^2 + (d_0 + z)^2}} \right) = V (\forall i \in [1, n], i \in \mathbb{Z})$$

$$\sum_{m=1}^{n} \frac{S_m}{4\pi\varepsilon_0} \left( \frac{\sigma_{2m} - \sigma_\mathrm{b}}{\sqrt{(x_{2m} - x_i)^2 + (y_{2m} - y_i)^2}} - \frac{\sigma_\mathrm{t} - \sigma_\mathrm{b}}{\sqrt{(x_{2m} - x_i)^2 + (y_{2m} - y_i)^2 + {d_0}^2}} + \frac{\sigma_{1m}}{\sqrt{(x_{2m} - x_i)^2 + (y_{2m} - y_i)^2 + (d_0 + z)^2}} \right) = V (\forall i \in [1, n], i \in \mathbb{Z})$$

This discrete system provides the numerical basis for solving the spatial charge distributions on the two electrodes in the finite-plate model.

## Supplementary Note 4. Discretization of the electrode surface

The electrode surface is discretized into $N \times N$ square segments. The top-left corner of the electrode is defined as the origin of the coordinate system. The segments in the first row are numbered sequentially from left to right, and the numbering then continues row by row from top to bottom. Under this numbering convention, the center coordinates of the $k$-th segment are given by:

$$x_k = (i - \frac{1}{2})b$$

$$y_k = (j - \frac{1}{2})b$$

$$k = (i-1)N + j$$

where $b$ is the side length of each segment, and $i$ and $j$ denote the column and row indices of the segment, respectively. This discretization provides the geometric basis for the point-matching implementation of the Method of Moments.

| 1 | 2 | 3 | ······ | $N$ |
|---|---|---|---|---|
| $N$+1 | $N$+2 | $N$+3 | ······ | 2$N$ |
| 2$N$+1 | 2$N$+2 | 2$N$+3 | | 3$N$ |
| ⋮ | ⋮ | | ⋱ | ⋮ |
| $N^2$-$N$ | $N^2$-$N$+1 | $N^2$-$N$+2 | ······ | $N^2$ |

**Supplementary Figure 1**. Discretization of the electrode surface into $N \times N$ segments.

**Supplementary Note 5. Matrix-equation formulation**

The discretized finite-plate equations can be written as follows:

$$\sum_{m=1}^{N^2}(\sigma_{1m}+\sigma_{2m})=N^2\sigma_0$$

$$\sum_{m=1}^{n}\frac{b^2}{4\pi\varepsilon_0}\left(\frac{\sigma_{1m}}{\sqrt{(x_{1m}-x_i)^2+(y_{1m}-y_i)^2}}-\frac{\sigma_{\mathrm{t}}-\sigma_{\mathrm{b}}}{\sqrt{(x_{1m}-x_i)^2+(y_{1m}-y_i)^2+z^2}}+\frac{\sigma_{2m}+\sigma_{\mathrm{b}}}{\sqrt{(x_{1m}-x_i)^2+(y_{1m}-y_i)^2+(d_0+z)^2}}\right)=V(\forall i\in[1,n],i\in\mathbb{Z})$$

$$\sum_{m=1}^{n}\frac{b^2}{4\pi\varepsilon_0}\left(\frac{\sigma_{2m}-\sigma_{\mathrm{b}}}{\sqrt{(x_{2m}-x_i)^2+(y_{2m}-y_i)^2}}-\frac{\sigma_{\mathrm{t}}-\sigma_{\mathrm{b}}}{\sqrt{(x_{2m}-x_i)^2+(y_{2m}-y_i)^2+d_0^{\,2}}}+\frac{\sigma_{1m}}{\sqrt{(x_{2m}-x_i)^2+(y_{2m}-y_i)^2+(d_0+z)^2}}\right)=V(\forall i\in[1,n],i\in\mathbb{Z})$$

In these equations, the first relation enforces charge conservation, whereas the remaining relations impose the equipotential conditions at the selected collocation points on the two electrodes.

When the source segment and the collocation segment coincide, that is, when $m = i$, the corresponding term represents the potential contribution of a uniformly charged segment to itself. In this self-interaction case, the direct point-source expression becomes singular because the denominator tends to zero. Therefore, the self-potential term is evaluated separately by integrating the contribution over the finite area of the segment:

$$\int_{-b/2}^{b/2}dx\int_{-b/2}^{b/2}dy\frac{1}{4\pi\varepsilon_0\sqrt{x^2+y^2}}=\frac{b\mathit{Arc}\,\mathrm{Cosh}(17)}{\pi\varepsilon_0}=\frac{b\cdot 0.8814}{\pi\varepsilon_0}$$

Using this self-interaction correction, the discretized equations can be assembled into the following matrix form:

$$\sum_m \sigma_{1m}+\sigma_{2m}=N^2\sigma_0$$

$$H\boldsymbol{\sigma}_1+H_{d_0+z}\boldsymbol{\sigma}_2=H_z\boldsymbol{\sigma}_{\mathrm{t}}-(H_z-H_{z+d_0})\boldsymbol{\sigma}_{\mathrm{b}}+\mathbf{V}$$

$$H\boldsymbol{\sigma}_2+H_{d_0+z}\boldsymbol{\sigma}_1=(H-H_{d_0})\boldsymbol{\sigma}_{\mathrm{b}}+H_{d_0}\boldsymbol{\sigma}_{\mathrm{t}}+\mathbf{V}$$

where $H$ is the interaction matrix, and $\boldsymbol{\sigma}_1$, $\boldsymbol{\sigma}_2$, $\boldsymbol{\sigma}_{\mathrm{t}}$, $\boldsymbol{\sigma}_{\mathrm{b}}$ and $\mathbf{V}$ are the corresponding discretized vectors:

$$\boldsymbol{\sigma}_1=\begin{pmatrix}\sigma_{11}\\ \sigma_{12}\\ \vdots\\ \sigma_{1m}\\ \vdots\\ \sigma_{1N^2}\end{pmatrix}_{N^2},\quad \boldsymbol{\sigma}_2=\begin{pmatrix}\sigma_{21}\\ \sigma_{22}\\ \vdots\\ \sigma_{2m}\\ \vdots\\ \sigma_{2N^2}\end{pmatrix}_{N^2},\quad \boldsymbol{\sigma}_\mathrm{t}=\begin{pmatrix}\sigma_\mathrm{t}\\ \sigma_\mathrm{t}\\ \vdots\\ \sigma_\mathrm{t}\\ \vdots\\ \sigma_\mathrm{t}\end{pmatrix}_{N^2},\quad \boldsymbol{\sigma}_\mathrm{e}=\begin{pmatrix}\sigma_\mathrm{e}\\ \sigma_\mathrm{e}\\ \vdots\\ \sigma_\mathrm{e}\\ \vdots\\ \sigma_\mathrm{e}\end{pmatrix}_{N^2},\quad \mathbf{V}=\begin{pmatrix}V\\ V\\ \vdots\\ V\\ \vdots\\ V\end{pmatrix}_{N^2}$$

Here, $\boldsymbol{\sigma}_1$ and $\boldsymbol{\sigma}_2$ denote the unknown charge-density vectors on the independent electrode and the base electrode, respectively. The vectors $\boldsymbol{\sigma}_\mathrm{t}$ and $\boldsymbol{\sigma}_\mathrm{b}$ represent the prescribed triboelectric-charge and bound-charge distributions, and $\mathbf{V}$ denotes the electrode-potential vector evaluated at the collocation points.

The matrix $H$ contains the electrostatic interaction coefficients between all pairs of source and collocation segments. Its diagonal elements correspond to the self-interaction terms, whereas the off-diagonal elements describe the Coulomb interaction between different segments:

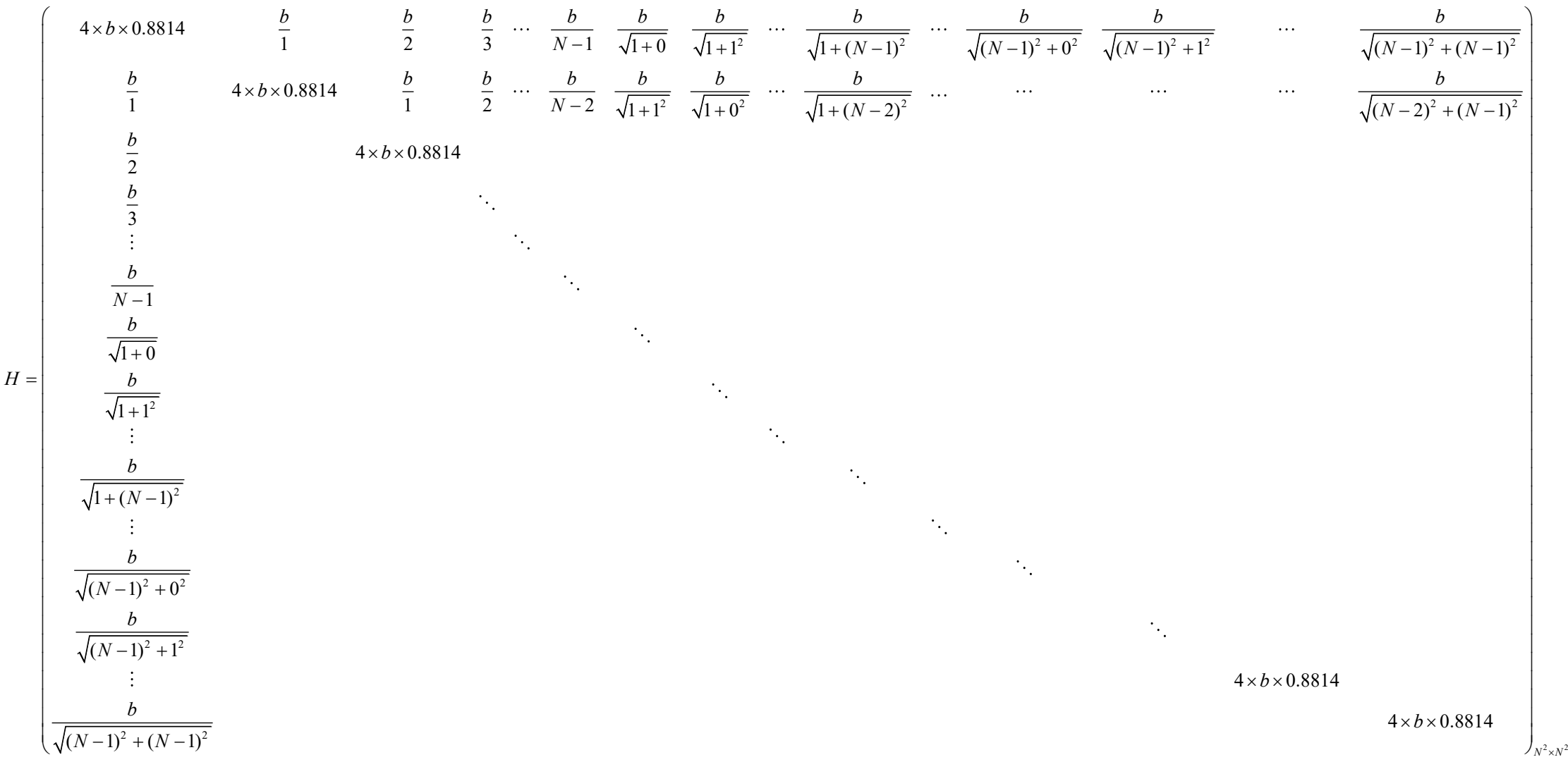

This matrix formulation converts the finite-plate integral equations into a finite-dimensional linear algebra problem, allowing the spatial charge distributions on the two electrodes to be solved numerically.

**Supplementary Note 6. Electrode charge-distribution vectors**

Starting from the matrix equation derived above, the two electrode charge-density vectors can be related by subtracting the equipotential equations for the two electrodes:

$$H\boldsymbol{\sigma}_2 + H_{d_0+z}\boldsymbol{\sigma}_1 - (H - H_{d_0})\boldsymbol{\sigma}_{\mathrm{b}} - H_{d_0}\boldsymbol{\sigma}_t = H\boldsymbol{\sigma}_1 + H_{d_0+z}\boldsymbol{\sigma}_2 - H_z\boldsymbol{\sigma}_t + (H_z - H_{z+d_0})\boldsymbol{\sigma}_{\mathrm{b}}$$

Rearranging this relation gives:

$$\boldsymbol{\sigma}_2 - \boldsymbol{\sigma}_1 = (H - H_{d_0+z})^{-1}((H_{d_0} - H_z)\boldsymbol{\sigma}_{\mathrm{t}} + (H - H_{d_0} + H_z - H_{z+d_0})\boldsymbol{\sigma}_{\mathrm{b}})$$

For compactness, we define the auxiliary vector **u** as:

$$\mathbf{u} = (H_{d_0} - H_z)\boldsymbol{\sigma}_{\mathrm{t}} + (H - H_{d_0} + H_z - H_{z+d_0})\boldsymbol{\sigma}_{\mathrm{b}}$$

Substituting this definition into the first matrix equation gives:

$$H\boldsymbol{\sigma}_1 + H\mathbf{u} + H_{d_0+z}\boldsymbol{\sigma}_1 = (H - H_{d_0})\boldsymbol{\sigma}_{\mathrm{b}} + H_{d_0}\boldsymbol{\sigma}_{\mathrm{t}} + \mathbf{V}$$

The charge-density vector $\boldsymbol{\sigma}_1$ can then be expressed as:

$$\boldsymbol{\sigma}_1 = (H + H_{d_0+z})^{-1}((H - H_{d_0})\boldsymbol{\sigma}_{\mathrm{b}} + H_{d_0}(\boldsymbol{\sigma}_{\mathrm{t}} - H\mathbf{u}) + (H - H_{d_0+z})^{-1}\mathbf{V}$$

We further define:

$$\mathbf{a} = (H + H_{d_0+z})^{-1}((H - H_{d_0})\boldsymbol{\sigma}_{\mathrm{b}} + H_{d_0}\boldsymbol{\sigma}_{\mathrm{t}} - H\mathbf{u})$$

so that the first electrode charge-density vector can be written in terms of **a**, **u** and the electrode potential vector **V**. Substituting this expression into the charge-conservation equation gives the following relation for $V$:

$$\sum_{m=1}^{N^2}(2a_m + u_m) + 2V \cdot (H + H_{d_0+z})^{-1}\Big|_{sum \quad all} = N^2\sigma_0$$

Thus, the electrode potential is obtained as:

$$V = \frac{\dfrac{N^2\sigma_0 - \sum_{m=1}^{N^2}(2a_m + u_m)}{2}}{(H + H_{d_0+z})^{-1}\Big|_{sum \quad all}}$$

Substituting this result back into the matrix equation gives the charge-density vector on the independent electrode:

$$\boldsymbol{\sigma}_1 = (H + H_{d_0+z})^{-1}((H - H_{d_0})\boldsymbol{\sigma}_{\mathrm{b}} + H_{d_0}\boldsymbol{\sigma}_{\mathrm{t}} - H((H_{d_0} - H_z)\boldsymbol{\sigma}_{\mathrm{t}} + (H - H_{d_0} + H_z - H_{z+d_0})\boldsymbol{\sigma}_{\mathrm{b}}))$$

$$+(H - H_{d_0+z})^{-1}\overrightarrow{\frac{N^2\sigma_0 - \sum_{m=1}^{N^2}(2a_m + u_m)}{2(H + H_{d_0+z})^{-1}\Big|_{sum \quad all}}}$$

After $\boldsymbol{\sigma_2}$ is obtained, the charge-density vector on the first electrode can be calculated from the previously derived relation between $\boldsymbol{\sigma_1}$ and $\boldsymbol{\sigma_2}$. This procedure provides the complete numerical solution for the spatial charge-density distributions on both electrodes. The resulting vectors $\boldsymbol{\sigma_1}$ and $\boldsymbol{\sigma_2}$ are solved by Python and then used to calculate the potential distribution and electric-field distribution in the finite plate model.

**Supplementary Note 7. Potential and electric-field calculation in the finite-plate model.**

In the finite-plate model, the electric field can be calculated directly from the spatial charge distribution:

$$\mathbf{E}(\mathbf{r})=\frac{1}{4\pi\varepsilon_0}[\int_{S_1}\frac{\sigma_1(\mathbf{r}')(\mathbf{r}_1-\mathbf{r}')}{|\mathbf{r}-\mathbf{r}'|^3}dS_1'+\int_{S_t}\frac{[-\sigma_t(\mathbf{r}')+\sigma_b(\mathbf{r}')](\mathbf{r}_1-\mathbf{r}')}{|\mathbf{r}-\mathbf{r}'|^3}dS_t'+\int_{S_2}\frac{[\sigma_2(\mathbf{r}')-\sigma_b(\mathbf{r}')](\mathbf{r}_1-\mathbf{r}')}{|\mathbf{r}-\mathbf{r}'|^3}dS_2']$$

However, this direct approach requires vector integration over the charged surfaces, which is computationally demanding. To simplify the calculation, we first evaluate the spatial potential distribution $V$, and then obtain the electric field by taking the negative gradient of the potential:

$$V(\mathbf{r})=\frac{1}{4\pi\varepsilon_0}[\int_{S_1}\frac{\sigma_1(\mathbf{r}')}{|\mathbf{r}-\mathbf{r}'|}dS_1'+\int_{S_t}\frac{-\sigma_t(\mathbf{r}')+\sigma_b(\mathbf{r}')}{|\mathbf{r}-\mathbf{r}'|}dS_t'+\int_{S_2}\frac{\sigma_2(\mathbf{r}')-\sigma_b(\mathbf{r}')}{|\mathbf{r}-\mathbf{r}'|}dS_2']$$

$$\mathbf{E}(\mathbf{r})=-\nabla V(\mathbf{r})=-\frac{\partial V(\mathbf{r})}{\partial x}\mathbf{e_x}-\frac{\partial V(\mathbf{r})}{\partial y}\mathbf{e_y}-\frac{\partial V(\mathbf{r})}{\partial z}\mathbf{e_z}$$

The Method of Moments is then applied to obtain a numerical solution for the potential equation. After the potential distribution is determined, the electric field is calculated using the finite-difference method[8]. Written in the Cartesian coordinate system, the potential equation is expressed as:

$$V(x,y,z)=\frac{1}{4\pi\varepsilon_0}[\iint_{S_1}\frac{\sigma_1(x',y')}{\sqrt{(x-x')^2+(y-y')^2+(h-d-z')^2}}dx'dy'$$
$$+\iint_{S_t}\frac{-\sigma_t(x',y')+\sigma_b(x',y')}{\sqrt{(x-x')^2+(y-y')^2+(h-d)^2}}dx'dy'+\iint_{S_2}\frac{\sigma_2(x',y')-\sigma_b(x',y')}{\sqrt{(x-x')^2+(y-y')^2+h^2}}dx'dy']$$

Similar to the discretization procedure used for solving the charge-distribution equations, the computational domain is divided into $N\times N\times(d+z)$ discrete elements. This gives the following discretized form of the potential equation:

$$V(i)=\sum_{m=1}^{N^2}\frac{S_m}{4\pi\varepsilon_0}(\frac{\sigma_{1m}}{\sqrt{(x_m-x_i)^2+(y_m-y_i)^2+(d+z-h_i)^2}}-\frac{\sigma_t-\sigma_b}{\sqrt{(x_m-x_i)^2+(y_m-y_i)^2+(d-h_i)^2}})$$
$$+\sum_{m=1}^{N^2}\frac{S_m}{4\pi\varepsilon_0}\frac{\sigma_{2m}-\sigma_b}{\sqrt{(x_m-x_i)^2+(y_m-y_i)^2+h_i^2}}(\forall i\in[1,(d+z)\cdot N^2],i\in\mathbb{Z})$$

The numerical solution of the discretized equations is obtained using the same custom Python implementation used for the finite-plate calculations in the main text. Linear systems are solved with PyTorch linear-algebra routines, so the implementation can run on either CPU or GPU. The resulting potential distribution within the internal space of the TENG can then be visualized by slicing the three-dimensional potential field, as shown in Supplementary Figure 2.

After the potential distribution is obtained, the electric-field distribution is calculated from the potential gradient using the finite-difference method. The calculated field lines extend from the electrodes toward the dielectric layer. In the central region of the device, the electric field remains nearly uniform, whereas near the edges the field lines bend outward because of finite-size effects. This behavior is consistent with the theoretical expectation that finite geometry preserves a near-uniform field in the bulk region while producing field distortion near the boundaries.

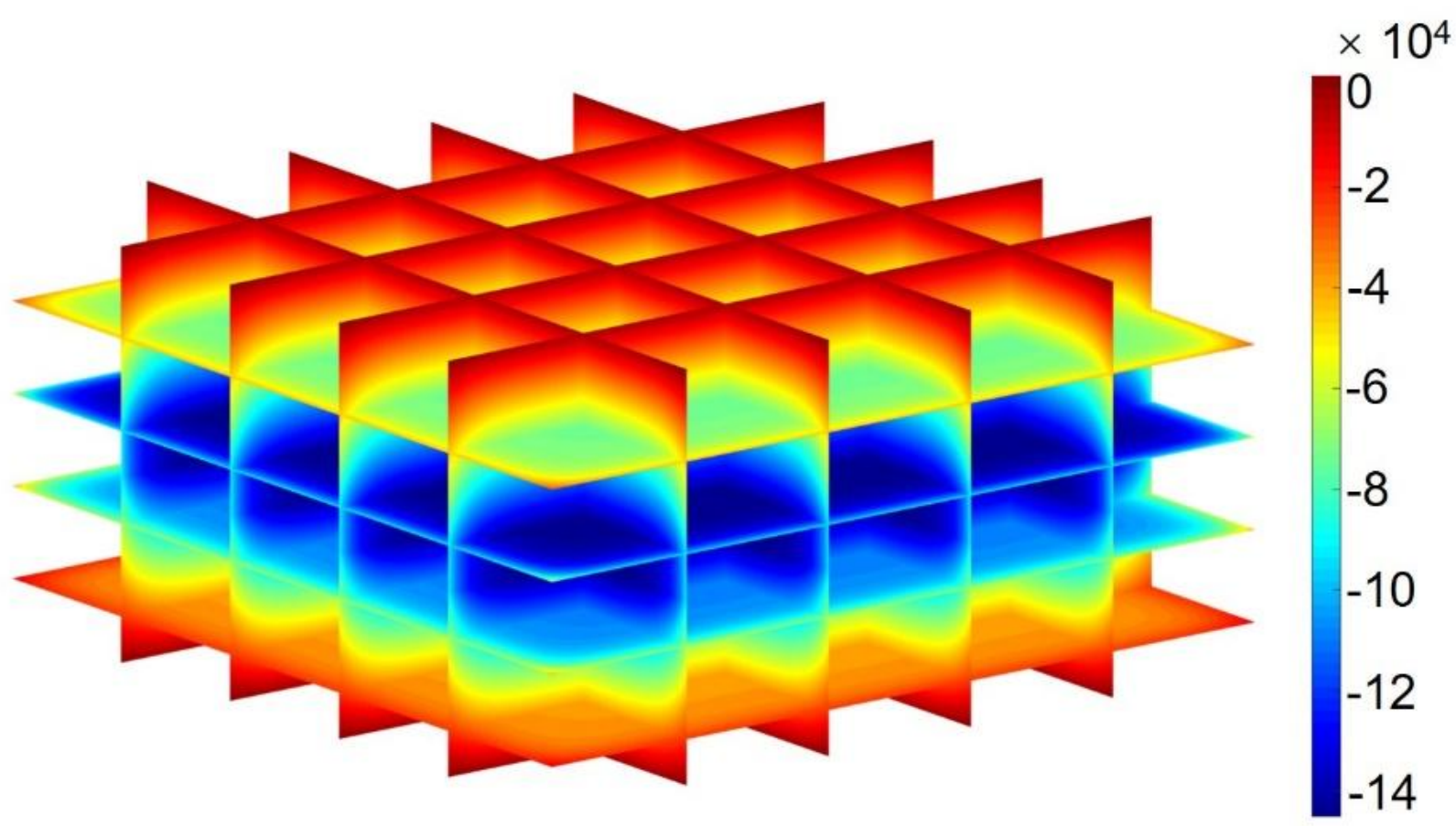

**Supplementary Figure 2. Potential distribution of the TENG.** The sliced potential map shows the spatial potential distribution calculated from the finite-plate model. The result illustrates the nearly uniform potential variation in the central region and the boundary-induced distortion near the edges.

**Supplementary Note 8. Experimental device and TENGCLAW validation**

The experimental validation in Fig. 2g was performed using a Helmholtz-resonator-based triboelectric nanogenerator (HR-TENG), whose device structure and fabrication procedure have been reported in our previous studies[9, 10]. Briefly, the HR-TENG consists of a Helmholtz resonant cavity, an aluminium film with uniformly distributed acoustic holes, and a fluorinated ethylene propylene (FEP) film with a conductive ink-printed back electrode. The resonant cavity concentrates and amplifies the incident acoustic wave, thereby driving periodic contact–separation motion between the aluminium and FEP triboelectric layers.

The resonant cavity used in the experiment had dimensions of 73 mm × 73 mm × 40 mm. One or two acoustic tubes with an inner diameter of 5.0 mm and a length of 32 mm were fixed to the cavity. The aluminium film served as the electropositive triboelectric layer and contained 440 uniformly distributed acoustic holes. Its length, width and thickness were 45 mm, 45 mm and 0.1 mm, respectively, and the diameter of each hole was 0.5 mm. The FEP film served as the electronegative triboelectric layer because of its strong electronegativity and mechanical flexibility. The FEP film had a thickness of 50 μm and an effective working area of 45 mm × 45 mm. Since FEP is electrically insulating, a micrometre-thick conductive ink electrode was printed on the back side of the FEP film for charge collection.

For the comparison between the finite-geometry solution and the infinite-plate prediction, the characteristic lateral dimension was taken as the working length of the triboelectric interface, $l$=45mm. The separation distance $z$during operation was much smaller than $l$, so the device operated in the asymptotic regime $l \gg z$. Under this condition, edge-induced charge redistribution becomes a higher-order correction, and the finite-geometry solution is expected to converge to the infinite-plate analytical prediction.

In addition to validating the asymptotic convergence between the finite-geometry and infinite-plate descriptions, this HR-TENG was also used to evaluate the executable simulation capability of TENGCLAW. After the device geometry, material parameters, charge conditions and mechanical excitation were provided through the dialogue interface, TENGCLAW generated the corresponding electrical response without additional manual model reconstruction. The simulated transferred charge, current and voltage show high consistency with the experimental outputs of the HR-TENG (Supplementary Figure 3), further confirming that the source-driven hierarchy can be translated into quantitatively reliable device-level simulation.

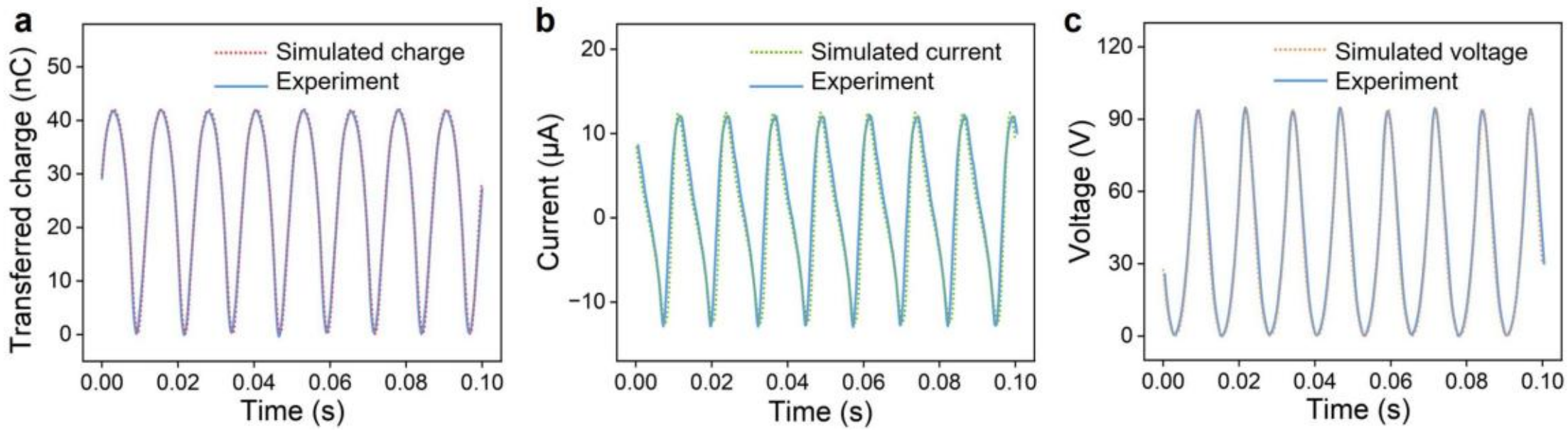

**Supplementary Figure 3. Comparison between TENGCLAW simulations and experimental measurements.** Comparison between simulated and experimental **a** transferred charge, **b** current and **c** voltage of the HR-TENG.

**Supplementary Note 9. Additional TENGCLAW implementation details**

TENGCLAW implements the source-driven electrostatic hierarchy described in the main text as a theory-governed, web-based research workspace. The system is built on OpenClaw, which provides the browser gateway, session runtime, workspace management and tool-routing infrastructure[11]. The TENG worker encapsulates the analytical and numerical simulation routines, while the plugin layer registers the TENGCLAW tools, invokes the orchestration workflow, formats trace and result cards, exposes artifact routes and connects each session to short-horizon research memory.

This implementation is designed to preserve the physical structure of the theory during computation. A user request is not treated as an isolated software command, but as a research task that must be interpreted, checked for physical admissibility, routed to an appropriate model branch and returned as a reusable scientific artifact.

**Execution stack and public tools**

The practical execution path proceeds from a browser request to the OpenClaw session runtime and then to tengclaw_orchestrate, which coordinates the Compiler, deterministic critic_preflight, Physics Critic, Research Planner and Research Reporter. The user-facing dialogue and guidance functions are provided by the surrounding OpenClaw dialogue shell, whereas physical execution is performed through the TENGCLAW worker and its registered tools.

Within this workflow, the Compiler converts a natural-language request into a structured TENG research object. The deterministic critic_preflight layer performs support and validity checks using the worker-side capability registry. The Physics Critic explains the resulting verdict, assumptions, missing parameters or required clarifications. The Research Planner selects the stable or experimental execution route and links the task to prior runs when relevant. The Research Reporter converts numerical and visual outputs into reusable conclusions, result summaries and trace metadata. This role separation ensures that language-model explanation remains distinct from permission to execute a physical computation.

The public TENGCLAW tools include:

*tengclaw_orchestrate*: the recommended research entry point. It compiles TENG-IR, invokes deterministic preflight, reuses recent scientific state and writes trace metadata.

*teng_simulate*: direct single-case simulation for users who have already specified the required physical parameters.

*teng_timeseries*: time-domain traversal under a specified motion law, returning transferred-charge and current trajectories.

*teng_optimize*: parameter sweep or optimization under a declared objective and constraints.

*teng_field_snapshot*: static charge, potential or electric-field analysis at a specified device state.

*teng_field_animation*: state-indexed visualization of field evolution along a motion trajectory.

*teng_study*: multi-case comparative study under shared physical conditions.

*teng_report*: report generation from completed runs or comparative studies.

For most scientific interactions, tengclaw_orchestrate is the preferred entry point because it preserves the complete chain from intent interpretation to physical governance, model routing, solver execution and artifact generation. The lower-level tools are primarily useful for reproducibility, debugging or direct simulation when the user has already specified the relevant physical parameters.

**TENG-IR, deterministic physical governance and model routing**

The central representation used by TENGCLAW is TENG-IR, a typed intermediate representation that preserves the physical content of a research request. TENG-IR records the research intent, working mode, geometry, material parameters, source-charge specification, mechanical excitation, requested observables, objective, constraints, artifact requirements, links to prior runs and candidate backend plan. In the context of the source-driven hierarchy, TENG-IR is more than a solver input file. It records why the calculation is being performed, which model regime is admissible, which assumptions have been made and which artifacts are required for interpretation or follow-up.

After compilation, the deterministic preflight layer maps TENG-IR to an admissibility verdict. The implemented verdicts are pass, clarify, approximate and unsupported. A pass verdict indicates that the task is sufficiently specified and supported by the stable backend. A clarify verdict indicates that one or more missing variables would change the scientific meaning of the task and must therefore be provided before execution. An approximate verdict indicates that execution is possible only under an explicit simplification or fallback. An unsupported verdict indicates that the request lies outside current stable support and should not be executed silently.

The preflight checks include the following categories. Mode and geometry support determines whether the requested TENG working mode and geometry class are supported by the current backend. Observable support determines whether transferred charge, current, voltage, charge maps, potential maps, electric-field maps, animations or reports can be computed by the selected branch. Boundary completeness checks whether source-charge density, dielectric thickness, relative permittivity, separation distance, device size, observation plane and circuit condition have been specified. Default safety determines whether missing parameters can be assigned controlled implementation defaults without changing the physical meaning of the task. Extension candidacy determines whether an unsupported task is non-extensible, templatable or internally eligible for controlled extension.

Model routing follows the same source-driven hierarchy used in the main theory. Near-uniform configurations are routed to the infinite-plate analytical branch when the device geometry satisfies the corresponding regime assumptions. Requests requiring spatial charge redistribution, potential maps, electric-field distributions or finite-size edge effects are routed to the finite-geometry numerical branch. In this way, model selection is not left to a manual user choice, but is determined by the admissible physical regime of the task.

**Solver branches, artifact generation and traceable outputs**

The stable solver layer exposes a compact vocabulary of scientific actions: simulate, timeseries, field_snapshot, field_animation, optimize, study and report. These actions are defined in terms of TENG research practice rather than generic software operations. For example, a timeseries request corresponds to sampling a prescribed motion trajectory, solving the admissible electrostatic state at each sampled configuration, aggregating the transferred charge and calculating the current by temporal differentiation. A field_animation request corresponds to generating a state-indexed visualization of potential or electric-field evolution under the same interpreted physical assumptions.

The infinite-plate branch implements the near-uniform analytical limit of the source-driven electrostatic hierarchy. The finite-geometry branch resolves position-dependent charge density and field structure using the numerical formulation described in Supplementary Notes 2–7. These two branches are treated as different regimes of the same charge-source problem. Consequently, the computed observables remain linked to the initial charge ensemble, geometry, material configuration and motion law specified in TENG-IR.

Each execution returns both machine-readable and human-readable artifacts. Depending on the selected action, these artifacts may include transferred-charge trajectories, current curves, charge-density maps, potential maps, electric-field maps, comparison plots, MP4 animations, raw data files and summary reports. The interface presents these outputs as result cards rather than as detached files. A separate trace card records the interpreted intent, preflight verdict, selected model branch, warnings, approximations, reuse links and execution plan. The pairing of result card and trace card makes each computational result scientifically inspectable.

**Persistence, scientific memory and artifact addressing**

TENGCLAW stores each completed computation as explicit scientific state rather than relying only on conversational history. A stable worker run writes a run bundle containing at least summary.json, raw.json, report.md when a report is produced, optional preview figures or MP4 animations, and a per-run experiment_graph.json snapshot. At the graph level, the system also maintains an append-only

experiment_graph.jsonl log. The implementation therefore functions as a lightweight scientific ledger rather than as a separate graph database.

Lineage is carried through fields such as parent_graph_ids, parent_run_ids, reuse_run_ids and extension metadata. These links allow the system to resolve follow-up requests such as continuing from a baseline, comparing a new candidate with an earlier study, reopening a report or generating the field animation associated with a previous selected run. This memory model is distinct from ordinary chat history because it binds each scientific conclusion to the interpreted task, critic outcome, execution path and produced artifacts.

Session memory is maintained separately from the run graph. The context engine keeps a session-scoped research_state.json file containing recent goals, conclusions, run identifiers, graph identifiers, pending extension plans and other short-horizon research memory. This state is injected into later turns as structured context. Artifacts are served through the TENGCLAW plugin route after reconstruction from each run artifact map and validation of a derived token. The artifact route therefore does not function as an unstructured file browser; plots, reports and animations remain tied to the runs that generated them.

**Public browser deployment, identity isolation and extension boundary**

The public browser deployment differs from a single-user local debugging setup. On first visit, the gateway issues a signed, long-lived tenantId cookie and maps it to a tenant-specific agentId. Each tenant stores its own model configuration, encrypted model token, sessions, recent-memory state, experiment graph, runs and artifacts. The precise server root is deployment-specific; in the deployment described here, these states are written under a tenant-specific TENGAgent output tree rather than a shared global workspace.

Two security and reproducibility details are relevant to the manuscript description. First, the model token saved by a public user is encrypted with AES-256-GCM before being written to the tenant store, and the plaintext token is not returned to the browser after saving[12]. Second, the gateway rewrites and validates session keys so that a browser can access only sessions belonging to its own tenant agent. In public mode, the browser-visible interface is intentionally restricted to Model Settings, the TENGCLAW chat workspace and the Results browser. Administrative configuration, pairing, cron, debugging and other control-plane functions are hidden or rejected at the server boundary.

Controlled self-extension is deployment-sensitive. In internal research mode, unsupported but templatable requests may enter a controlled lifecycle consisting of proposal, confirmation, validation and isolated registration. In the public browser deployment used for ordinary users, experimental solver-extension generation is disabled. Unsupported requests are therefore surfaced as support boundaries rather than being converted into executable experimental modules. This distinction is essential for interpreting the extension-related elements in the workflow diagram and prevents the public system from appearing more autonomous than it is.

**Supplementary Note 10. Practical browser workflow and representative use cases**

This note illustrates how the TENGCLAW implementation described in Supplementary Note 9 is used in a practical browser session. The purpose is not to introduce an additional physical model, but to clarify how the source-driven hierarchy is exposed to users as a traceable research workflow. In this workflow, natural-language requests, physical admissibility checks, model routing, solver execution and artifact reuse are connected within the same browser environment.

A typical TENGCLAW session begins in the browser workspace. When the user opens the web page, the gateway initializes the browser identity state. The user then specifies the model provider, optional base URL, model name and API key in the Model Settings panel. Scientific interaction proceeds mainly through the TENGCLAW chat workspace, where the user can describe a research task in natural language rather than manually constructing a solver script.

After a request is submitted, TENGCLAW returns both a trace card and a result card. The trace card should be inspected first because it records how the request was interpreted, which physical parameters were recognized, which parameters were missing, what verdict was returned by the deterministic critic, which model branch was selected and whether previous runs were reused. The result card then presents the numerical or visual outputs, together with links to the generated artifacts. The Results browser provides access to previous runs, reports, plots, raw data and animations, allowing earlier calculations to be reopened, compared or extended.

For most research tasks, users should begin with an orchestrated request rather than directly invoking a low-level tool. For example, a request such as "simulate the transferred charge and current of this contact–separation device under sinusoidal motion" allows TENGCLAW to compile the task, check physical completeness, select an admissible model branch and store the resulting run as scientific state. Direct tool calls are more appropriate when the user has already specified the working mode, geometry, material constants, charge density, motion law and desired observable, and wants to execute a reproducible single-step calculation. This distinction is important because the central function of TENGCLAW is not merely numerical execution, but the governed transition from a scientific question to a physically admissible computation.

**Representative case 1: completing an under-specified charge-distribution request**

A user may begin with a broad question such as “How do I simulate the charge distribution of a TENG?” This request is scientifically meaningful, but it is not yet physically complete. TENGCLAW therefore identifies the missing physical commitments instead of returning an unconstrained charge map. Depending on the requested working mode and geometry, the system may ask for the electrode or dielectric shape, grid resolution, device size, surface charge density, dielectric thickness, relative permittivity, separation distance and observation position.

Once these quantities are supplied, the request is compiled into TENG-IR and routed to the finite-geometry branch if spatial resolution is required. The returned charge map is therefore tied to a specified source ensemble, geometry and boundary condition. This prevents the visualization from becoming a nominal image generated from an incomplete prompt.

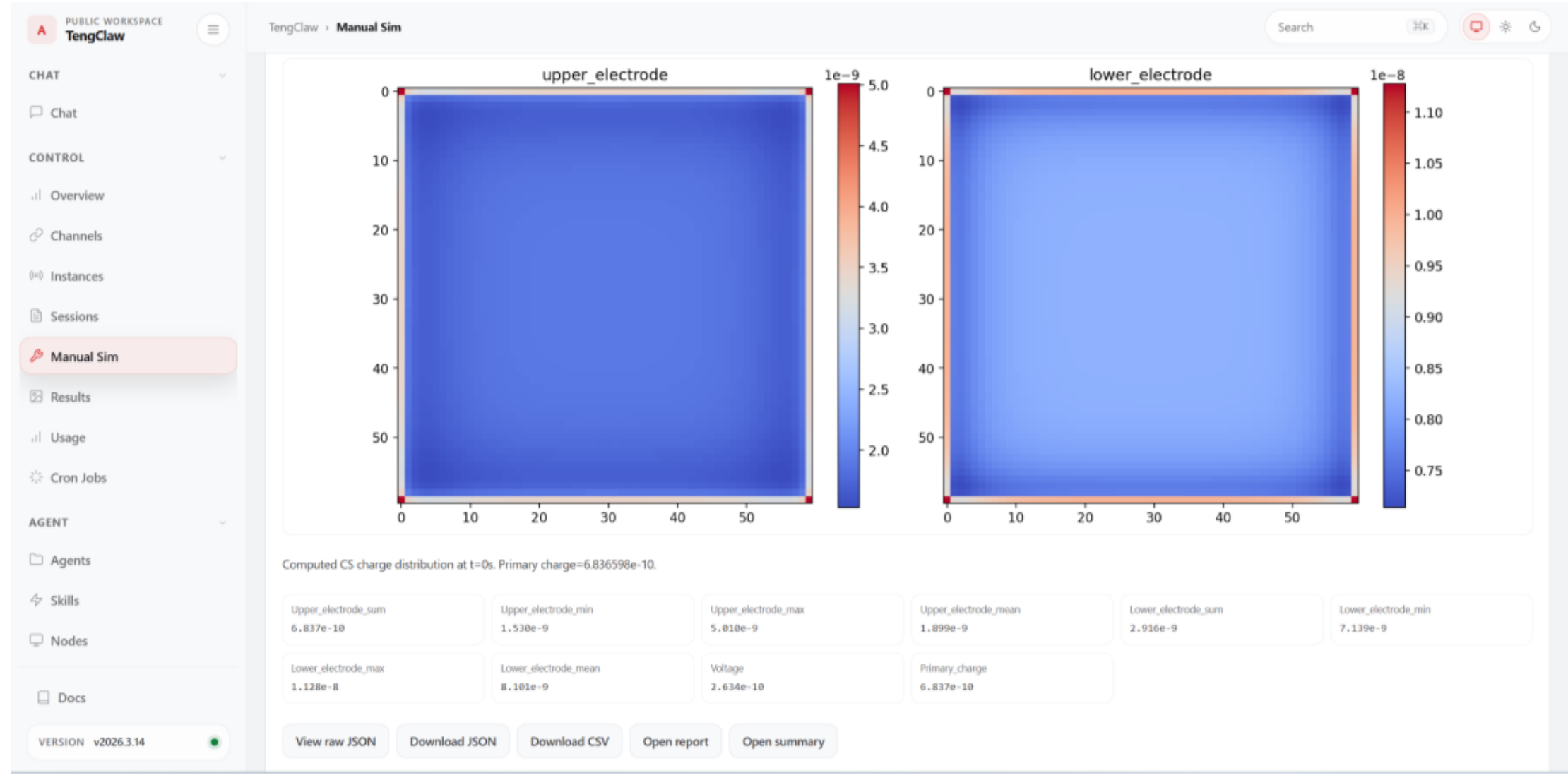

**Representative case 2: predicting dynamic output from a defined motion law**

After a contact–separation configuration has been specified, the user can request transferred charge and current under a defined motion law, such as $z(t)=z_0+A\sin(\omega t)$. TENGCLAW treats this request as geometric modulation of the same source-driven electrostatic state. The workflow samples the prescribed motion trajectory, evaluates the admissible electrostatic state at each sampled separation, calculates the transferred charge and obtains the current from its time derivative.

This case corresponds to the dynamic-output workflow discussed in the main text, but the same procedure can be applied to other supported periodic or user-defined trajectories. The important point is that the time-domain electrical response is not treated as an isolated curve-fitting task. It is computed as a derived observable of the evolving charge ensemble.

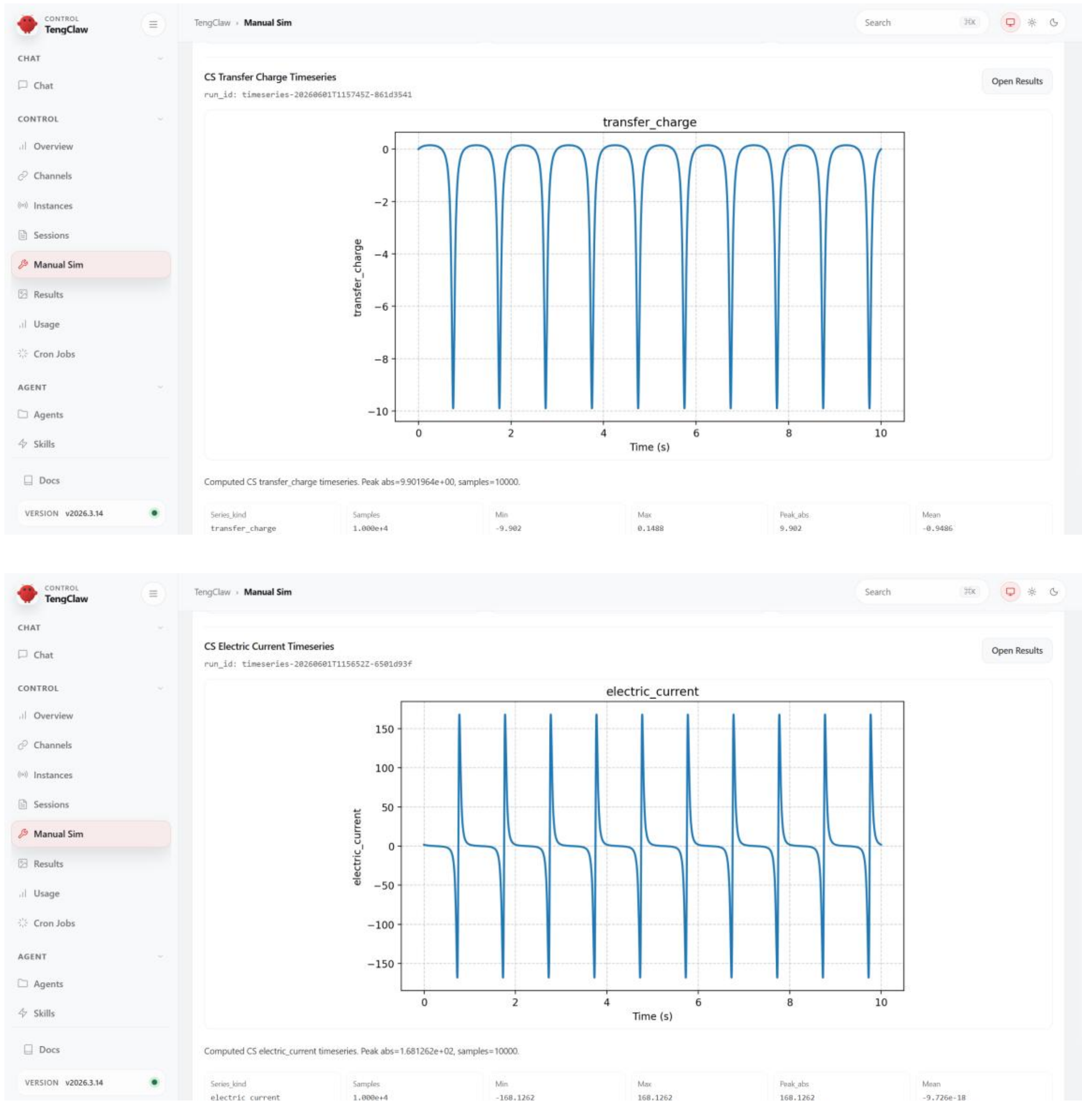

**Representative case 3: connecting lumped electrical output with field-level interpretation**

After a dynamic-output run, the user may ask why the current waveform has a particular shape, where the electric field becomes concentrated during the cycle or how the potential distribution evolves with separation. Because TENGCLAW stores the interpreted device and motion context, the system can reuse the existing physical state and generate potential snapshots, electric-field maps or a field animation along the same trajectory.

In this case, field visualization is treated as a continuation of the same source-driven calculation rather than as a separate post-processing image. The resulting maps help connect transferred charge and current to spatial electrostatic redistribution, edge concentration and the evolving potential landscape. This preserves the chain from source charges to field structure and then to measurable electrical output.

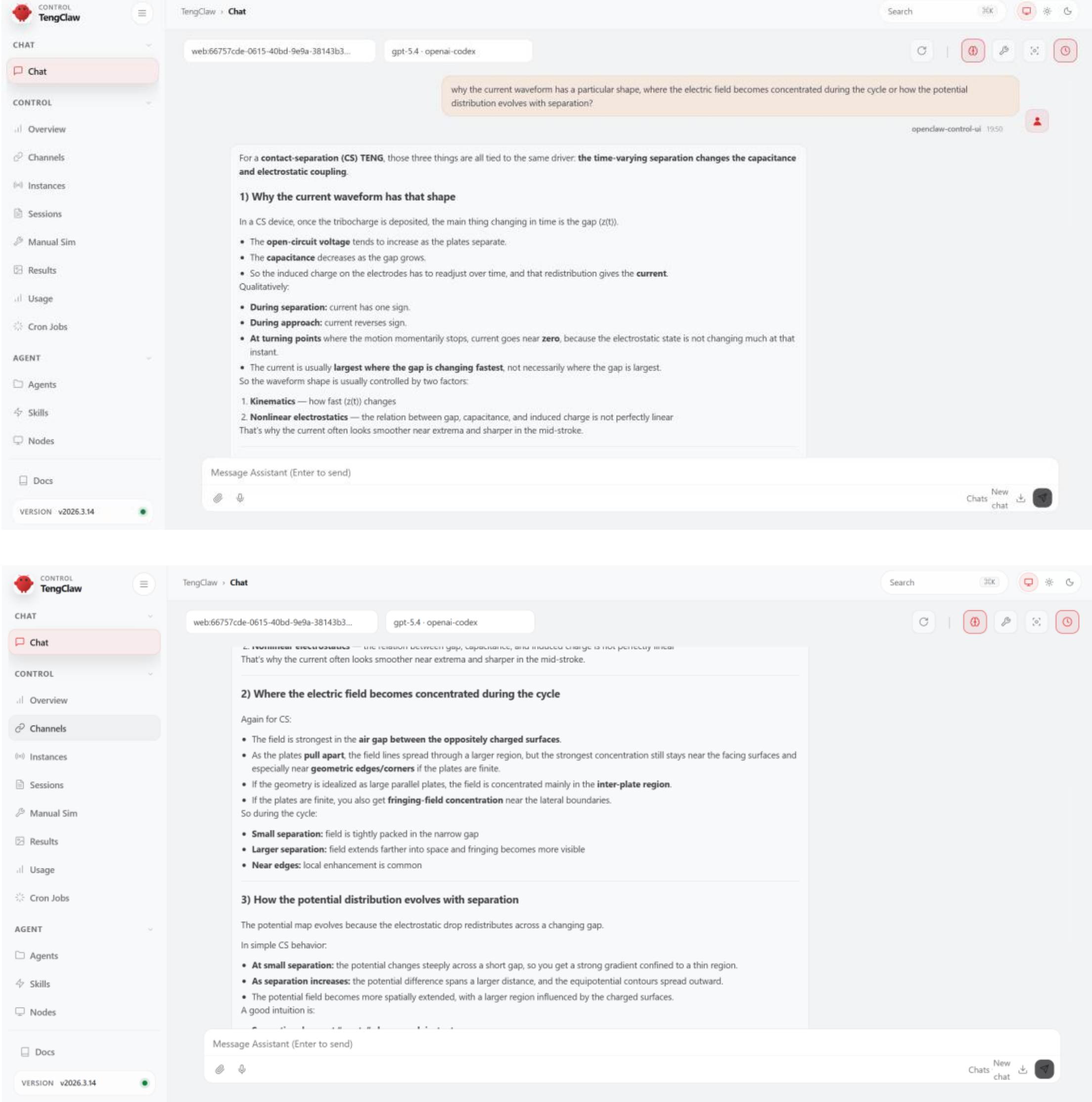

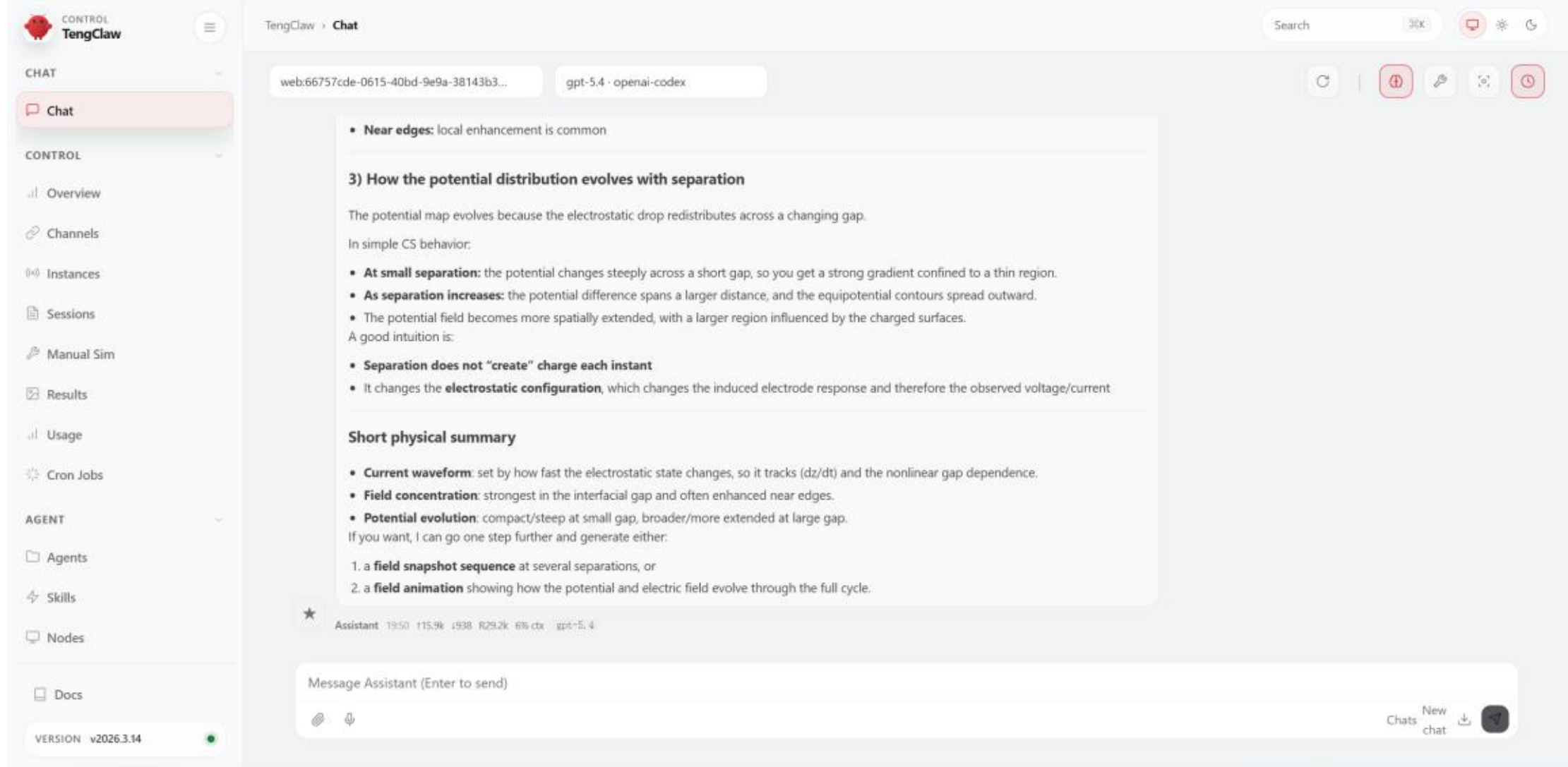

**Representative case 4: comparative design study under shared conditions**

A common research task is to compare multiple electrode sizes, dielectric thicknesses, gap distances or candidate geometries under the same motion law. TENGCLAW can organize this task as a study-level request rather than as a set of unrelated simulations. The user specifies the shared physical conditions, the design variables to be compared and the ranking metric, such as peak current, transferred charge, average output or a normalized performance indicator.

If the ranking metric is ambiguous, the deterministic critic requests clarification before execution. The returned study includes ranked candidates, linked run identifiers and reusable artifacts. A follow-up request can then ask for the field distribution, report or raw data of the best-performing candidate without restating the full physical setup. This allows comparative design exploration to remain traceable across multiple runs.

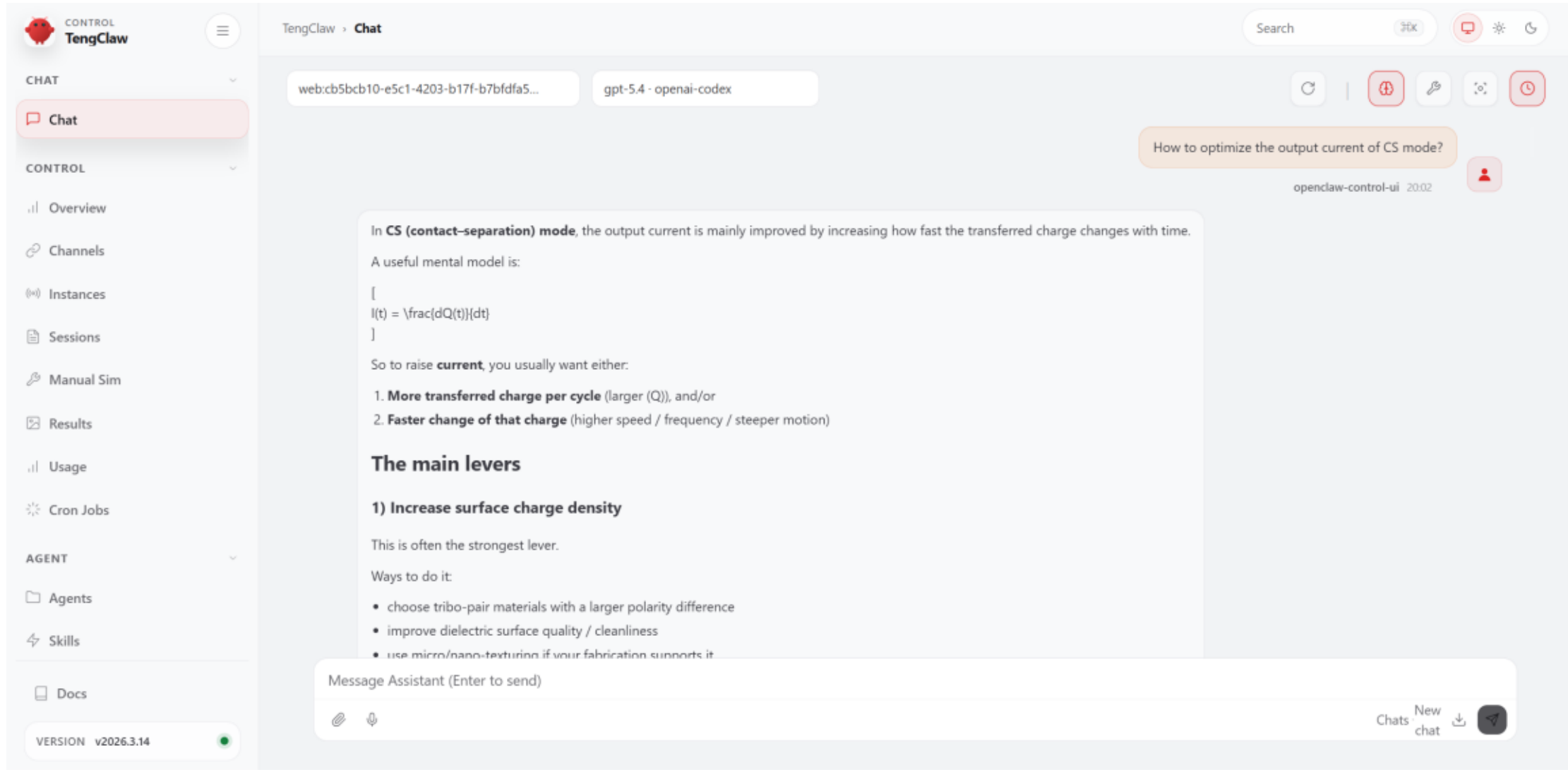
CONTROL
TengClaw
CHAT
Chat
CONTROL
Overview
Channels
Instances
Sessions
Manual Sim
Results
Usage
Cron Jobs
AGENT
Agents
Skills
Docs
VERSION v2026.3.14
TengClaw › Chat
Search
web:cb5bcb10-e5c1-4203-b17f-b7bfdfa5...
gpt-5.4 · openai-codex
How to optimize the output current of CS mode?
openclaw-control-ui 20:02
In CS (contact–separation) mode, the output current is mainly improved by increasing how fast the transferred charge changes with time.
A useful mental model is:
[
I(t) = \frac{dQ(t)}{dt}
]
So to raise current, you usually want either:
1. More transferred charge per cycle (larger (Q)), and/or
2. Faster change of that charge (higher speed / frequency / steeper motion)
The main levers
1) Increase surface charge density
This is often the strongest lever.
Ways to do it:
• choose tribo-pair materials with a larger polarity difference
• improve dielectric surface quality / cleanliness
Message Assistant (Enter to send)
Chats
New chat

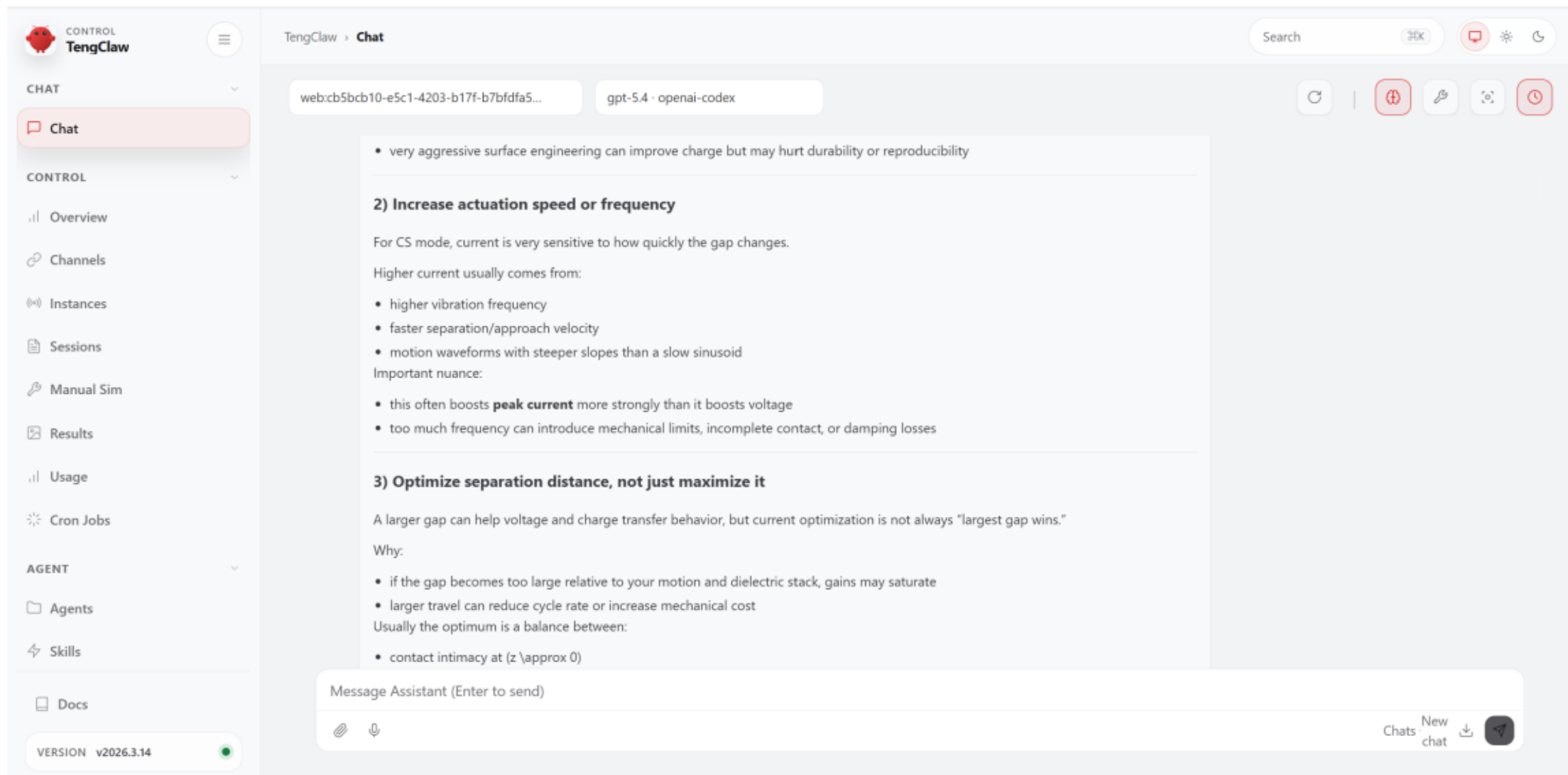
CONTROL
TengClaw
CHAT
Chat
CONTROL
Overview
Channels
Instances
Sessions
Manual Sim
Results
Usage
Cron Jobs
AGENT
Agents
Skills
Docs
VERSION v2026.3.14
TengClaw › Chat
Search
web:cb5bcb10-e5c1-4203-b17f-b7bfdfa5...
gpt-5.4 · openai-codex
• very aggressive surface engineering can improve charge but may hurt durability or reproducibility
2) Increase actuation speed or frequency
For CS mode, current is very sensitive to how quickly the gap changes.
Higher current usually comes from:
• higher vibration frequency
• faster separation/approach velocity
• motion waveforms with steeper slopes than a slow sinusoid
Important nuance:
• this often boosts peak current more strongly than it boosts voltage
• too much frequency can introduce mechanical limits, incomplete contact, or damping losses
3) Optimize separation distance, not just maximize it
A larger gap can help voltage and charge transfer behavior, but current optimization is not always "largest gap wins."
Why:
• if the gap becomes too large relative to your motion and dielectric stack, gains may saturate
• larger travel can reduce cycle rate or increase mechanical cost
Usually the optimum is a balance between:
• contact intimacy at (z \approx 0)
Message Assistant (Enter to send)
Chats
New chat

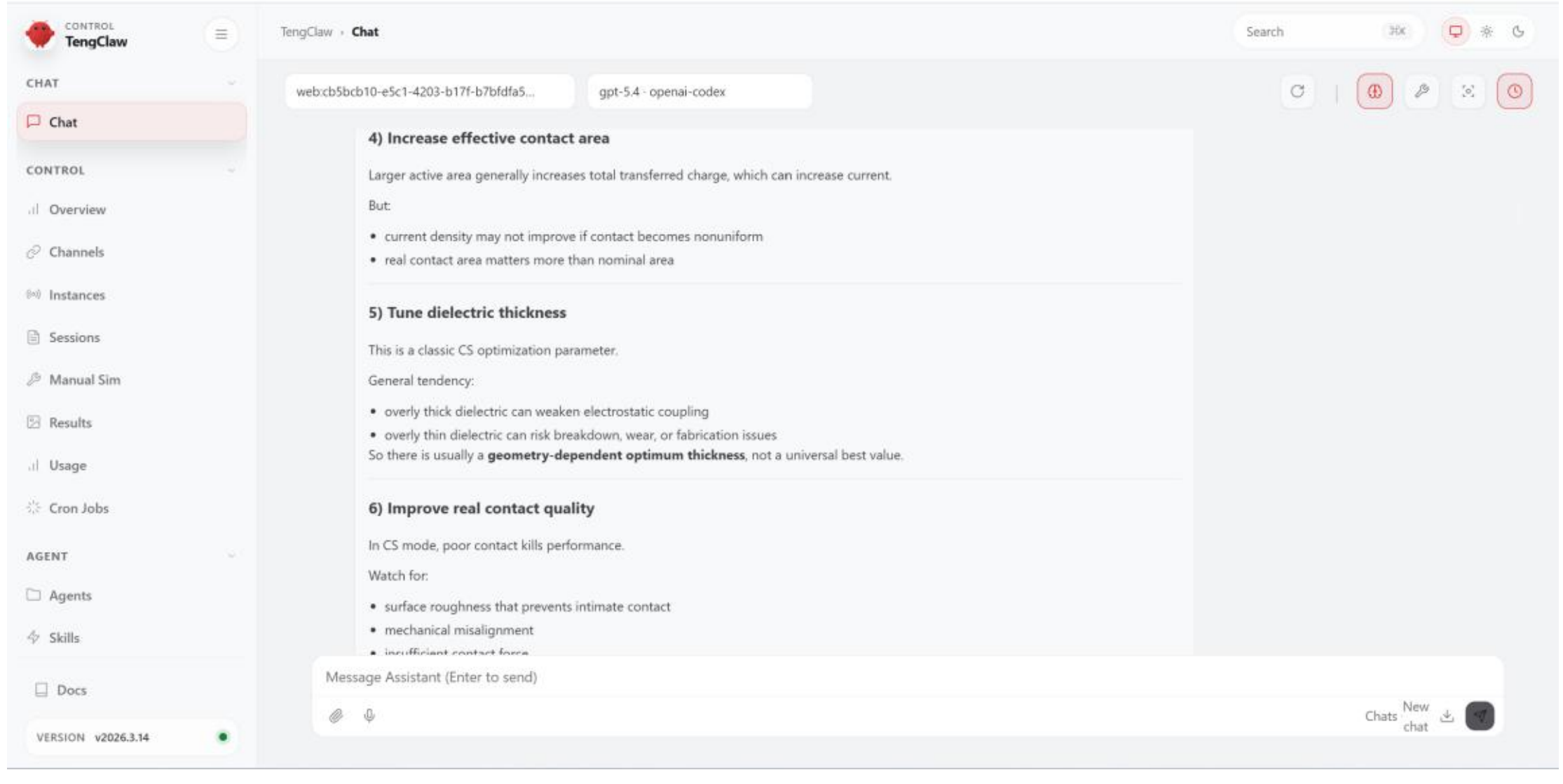

TengClaw › Chat
4) Increase effective contact area
Larger active area generally increases total transferred charge, which can increase current.
But:
• current density may not improve if contact becomes nonuniform
• real contact area matters more than nominal area
5) Tune dielectric thickness
This is a classic CS optimization parameter.
General tendency:
• overly thick dielectric can weaken electrostatic coupling
• overly thin dielectric can risk breakdown, wear, or fabrication issues
So there is usually a geometry-dependent optimum thickness, not a universal best value.
6) Improve real contact quality
In CS mode, poor contact kills performance.
Watch for:
• surface roughness that prevents intimate contact
• mechanical misalignment
Message Assistant (Enter to send)

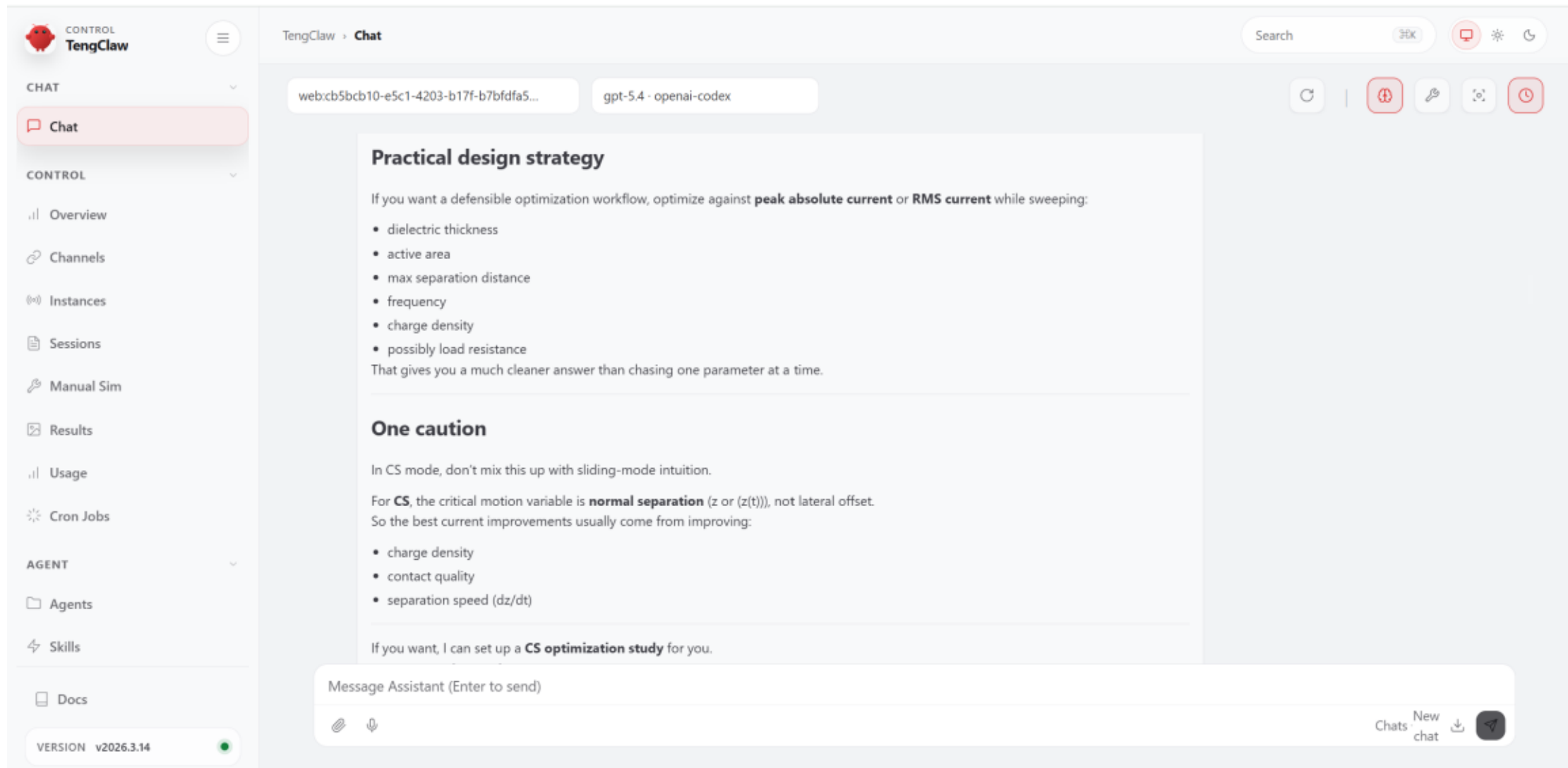

TengClaw › Chat
Practical design strategy
If you want a defensible optimization workflow, optimize against peak absolute current or RMS current while sweeping:
• dielectric thickness
• active area
• max separation distance
• frequency
• charge density
• possibly load resistance
That gives you a much cleaner answer than chasing one parameter at a time.
One caution
In CS mode, don't mix this up with sliding-mode intuition.
For CS, the critical motion variable is normal separation (z or (z(t)), not lateral offset.
So the best current improvements usually come from improving:
• charge density
• contact quality
• separation speed (dz/dt)
If you want, I can set up a CS optimization study for you.
Message Assistant (Enter to send)

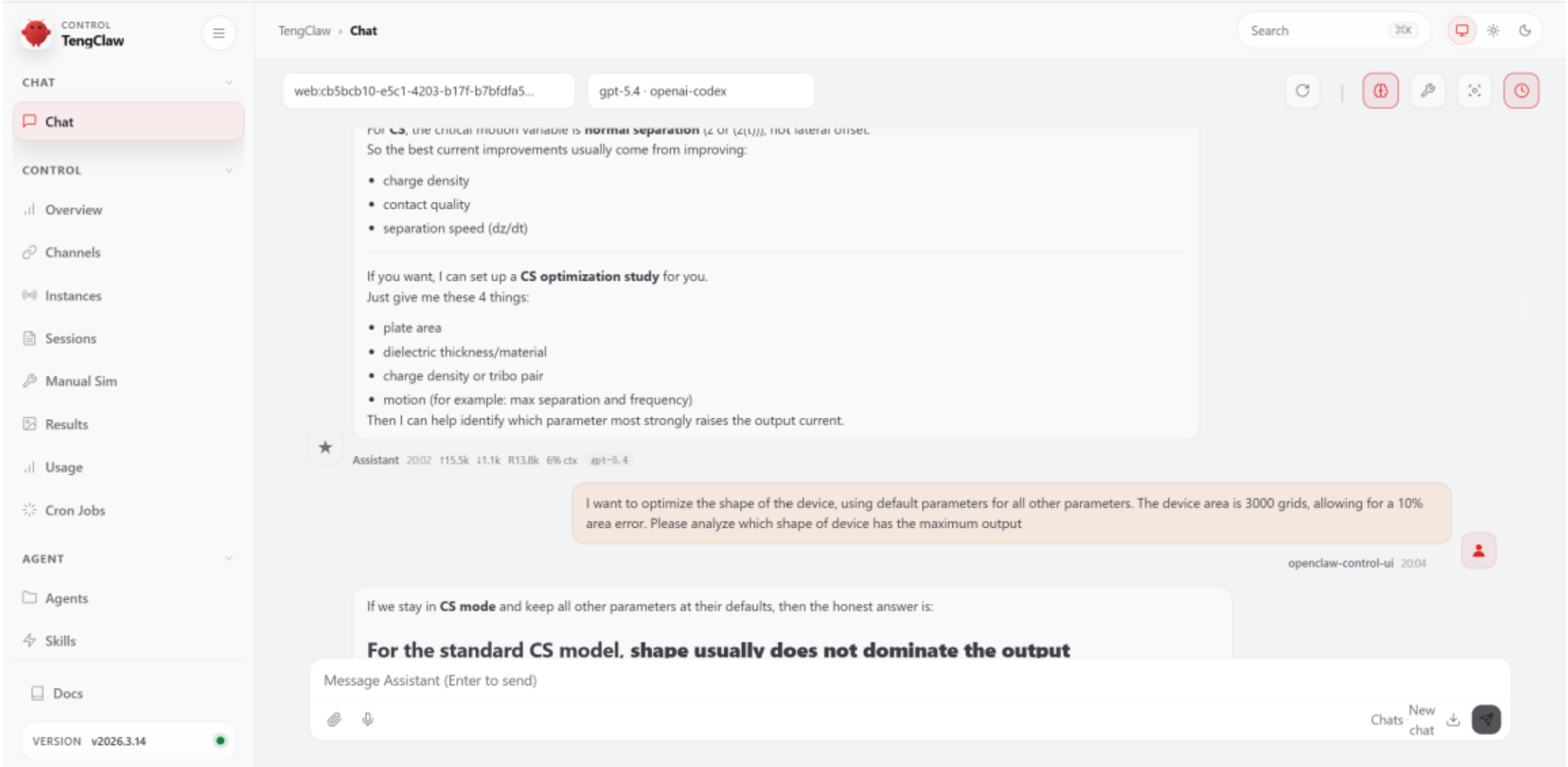

TengClaw › Chat
So the best current improvements usually come from improving:
• charge density
• contact quality
• separation speed (dz/dt)
If you want, I can set up a CS optimization study for you.
Just give me these 4 things:
• plate area
• dielectric thickness/material
• charge density or tribo pair
• motion (for example: max separation and frequency)
Then I can help identify which parameter most strongly raises the output current.
Assistant
I want to optimize the shape of the device, using default parameters for all other parameters. The device area is 3000 grids, allowing for a 10% area error. Please analyze which shape of device has the maximum output
openclaw-control-ui
If we stay in CS mode and keep all other parameters at their defaults, then the honest answer is:
For the standard CS model, shape usually does not dominate the output
Message Assistant (Enter to send)

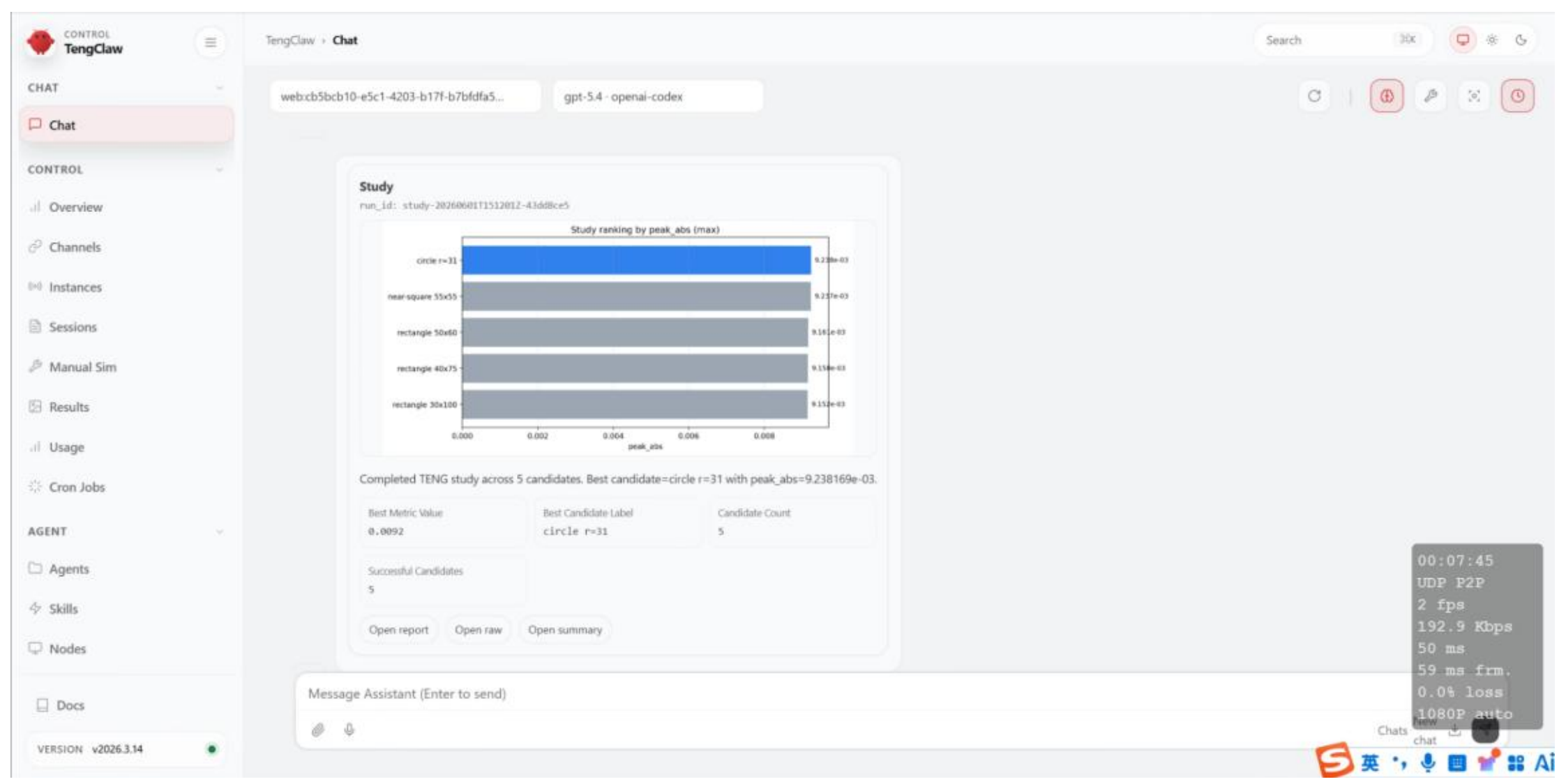

CONTROL
TengClaw
CHAT
Chat
CONTROL
Overview
Channels
Instances
Sessions
Manual Sim
Results
Usage
Cron Jobs
AGENT
Agents
Skills
Nodes
Docs
VERSION v2026.3.14
TengClaw › Chat
Search
web:cb5bcb10-e5c1-4203-b17f-b7bfdfa5...
gpt-5.4 · openai-codex
Study
Study ranking by peak_abs (max)
circle r=31
near-square 55x55
rectangle 50x60
rectangle 40x75
rectangle 30x100
0.000
0.002
0.004
0.006
0.008
peak_abs
Completed TENG study across 5 candidates. Best candidate=circle r=31 with peak_abs=9.238169e-03.
Best Metric Value
0.0092
Best Candidate Label
circle r=31
Candidate Count
5
Successful Candidates
5
Open report
Open raw
Open summary
Message Assistant (Enter to send)
00:07:45
UDP P2P
2 fps
192.9 Kbps
50 ms
59 ms frm.
0.0% loss
1080P auto

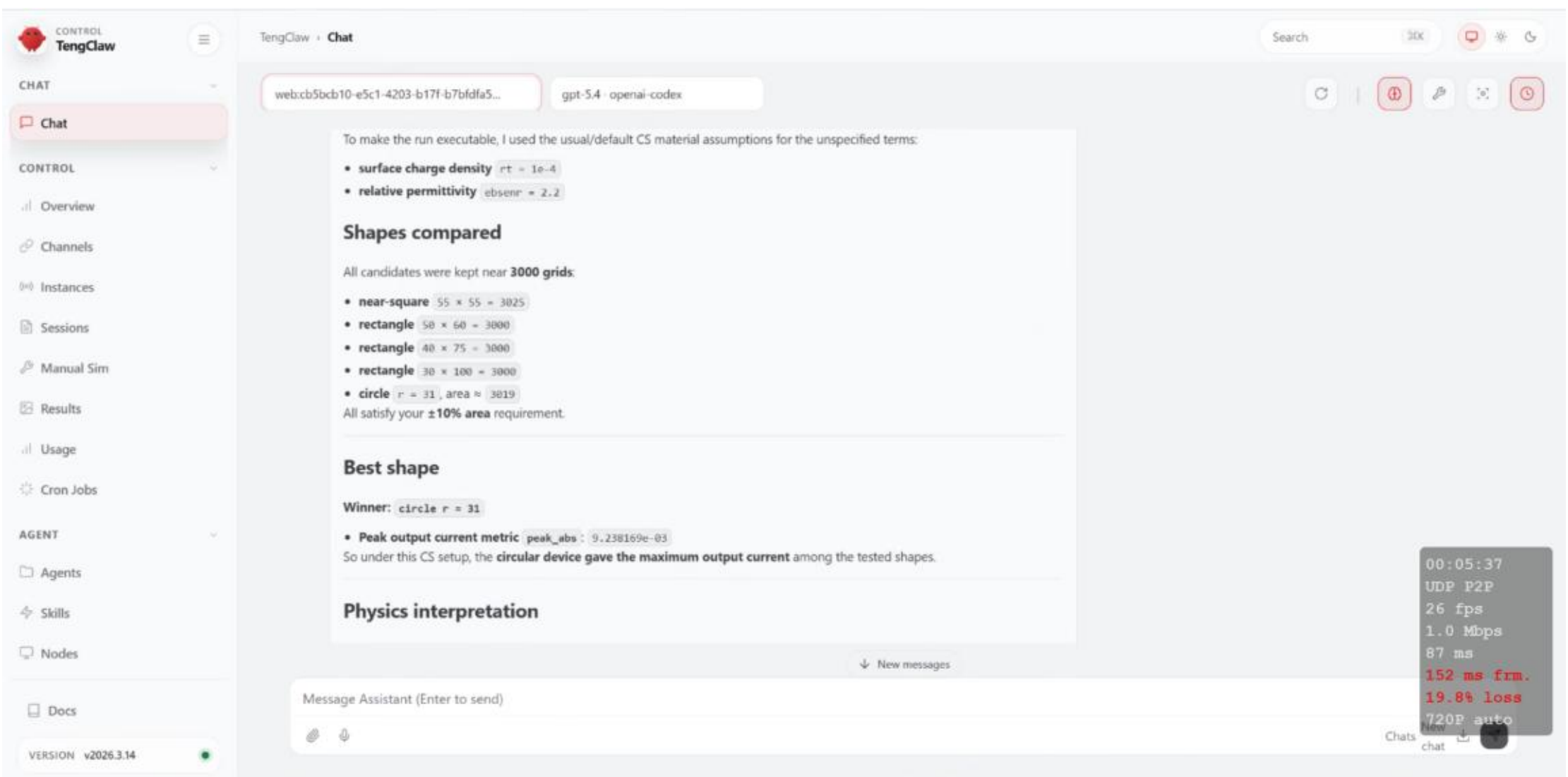

CONTROL
TengClaw
CHAT
Chat
CONTROL
Overview
Channels
Instances
Sessions
Manual Sim
Results
Usage
Cron Jobs
AGENT
Agents
Skills
Nodes
Docs
VERSION v2026.3.14
TengClaw › Chat
Search
web:cb5bcb10-e5c1-4203-b17f-b7bfdfa5...
gpt-5.4 · openai-codex
To make the run executable, I used the usual/default CS material assumptions for the unspecified terms:
surface charge density rt = 1e-4
relative permittivity ebsenr = 2.2
Shapes compared
All candidates were kept near 3000 grids:
near-square 55 × 55 = 3025
rectangle 50 × 60 = 3000
rectangle 40 × 75 = 3000
rectangle 30 × 100 = 3000
circle r = 31, area ≈ 3019
All satisfy your ±10% area requirement.
Best shape
Winner: circle r = 31
Peak output current metric peak_abs: 9.238169e-03
So under this CS setup, the circular device gave the maximum output current among the tested shapes.
Physics interpretation
New messages
Message Assistant (Enter to send)
00:05:37
UDP P2P
26 fps
1.0 Mbps
87 ms
152 ms frm.
19.8% loss
720P auto

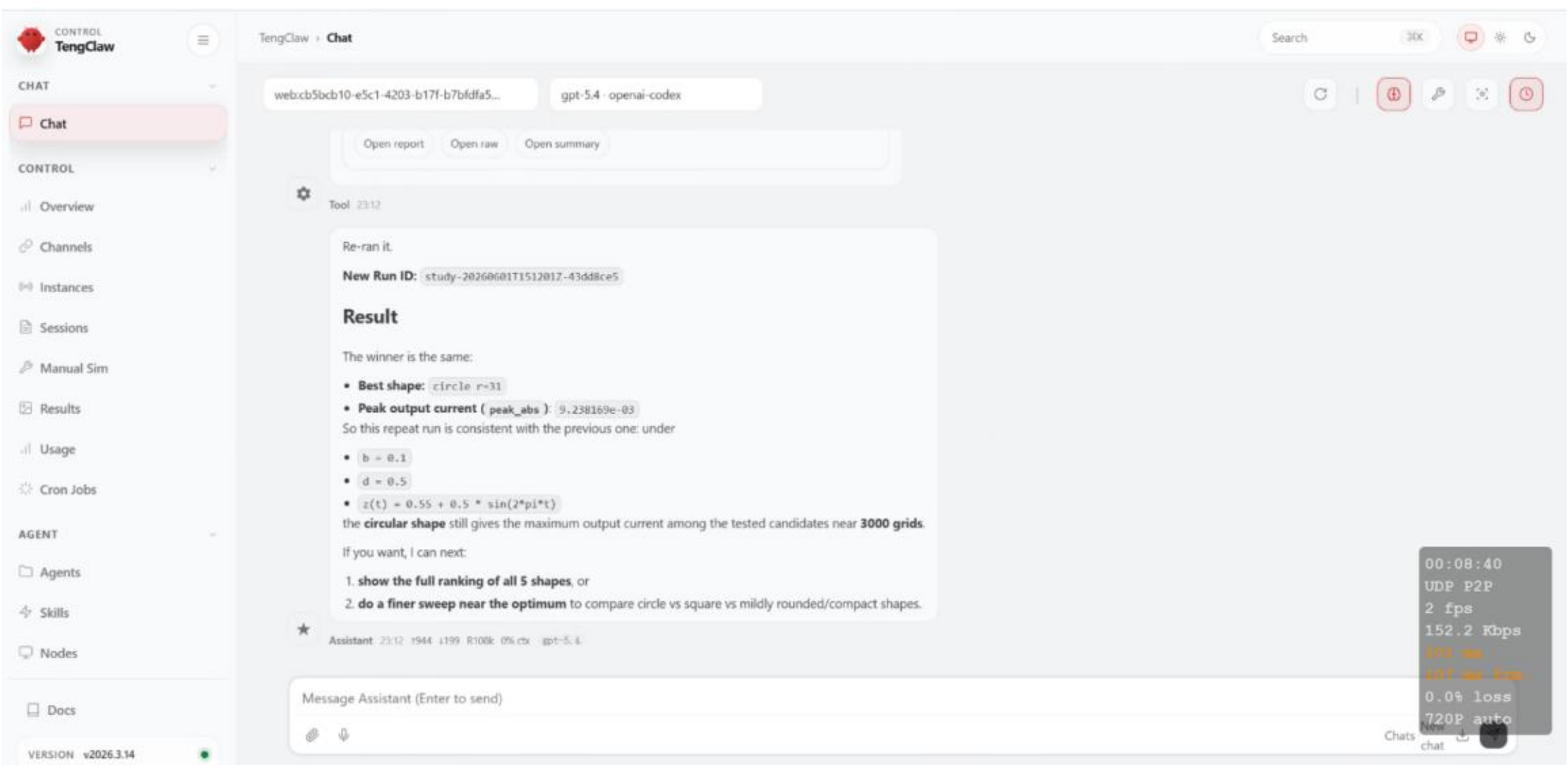

CONTROL
TengClaw
CHAT
Chat
CONTROL
Overview
Channels
Instances
Sessions
Manual Sim
Results
Usage
Cron Jobs
AGENT
Agents
Skills
Nodes
Docs
VERSION v2026.3.14
TengClaw › Chat
Search
web:cb5bcb10-e5c1-4203-b17f-b7bfdfa5...
gpt-5.4 · openai-codex
Open report
Open raw
Open summary
Tool 23:12
Re-ran it.
New Run ID: study-20260601T151201 7-43dd8ce5
Result
The winner is the same:
Best shape: circle r=31
Peak output current ( peak_abs ): 9.238169e-03
So this repeat run is consistent with the previous one: under
b = 0.1
d = 0.5
z(t) = 0.55 + 0.5 * sin(2*pi*t)
the circular shape still gives the maximum output current among the tested candidates near 3000 grids.
If you want, I can next:
1. show the full ranking of all 5 shapes, or
2. do a finer sweep near the optimum to compare circle vs square vs mildly rounded/compact shapes.
Assistant 23:12
Message Assistant (Enter to send)
00:08:40
UDP P2P
2 fps
152.2 Kbps
0.0% loss
720P auto

**Representative case 5: parameter exploration and optimization**

Starting from a baseline run, the user may ask how performance changes with dielectric thickness, relative permittivity, gap distance, vibration amplitude, frequency, device length or grid resolution. TENGCLAW interprets this request as a parameter sweep or optimization task only after the target quantity and scan range are defined. For example, maximizing peak current, maximizing transferred charge and minimizing field concentration are different scientific objectives and may lead to different optimal designs.

After the objective and range are clarified, the system executes the evaluation series, returns trend plots or comparison tables, identifies the best candidate within the requested domain and stores the sweep as reusable research state. This allows the user to inspect the reason for an optimum rather than receiving only a single numerical recommendation.

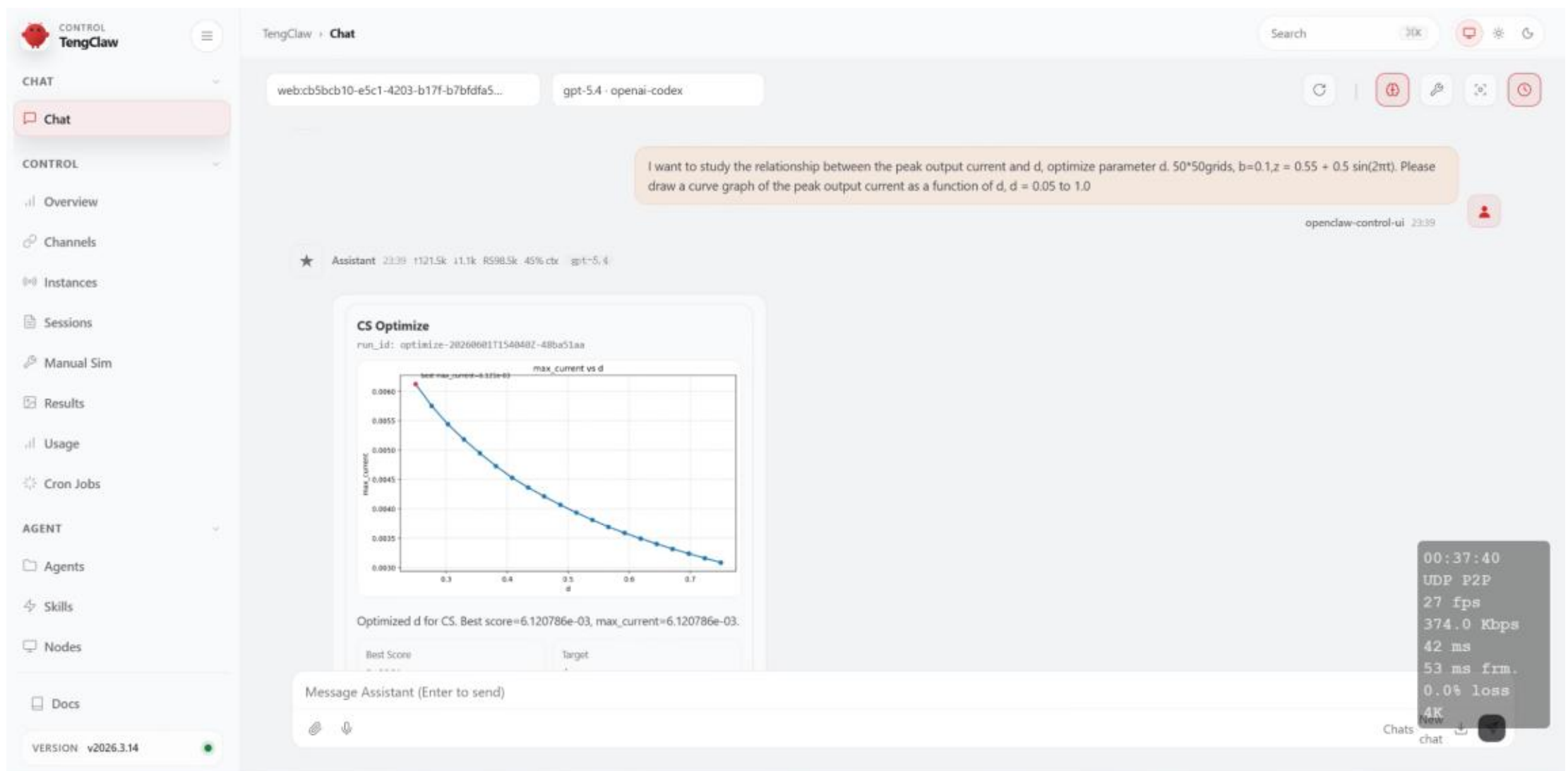

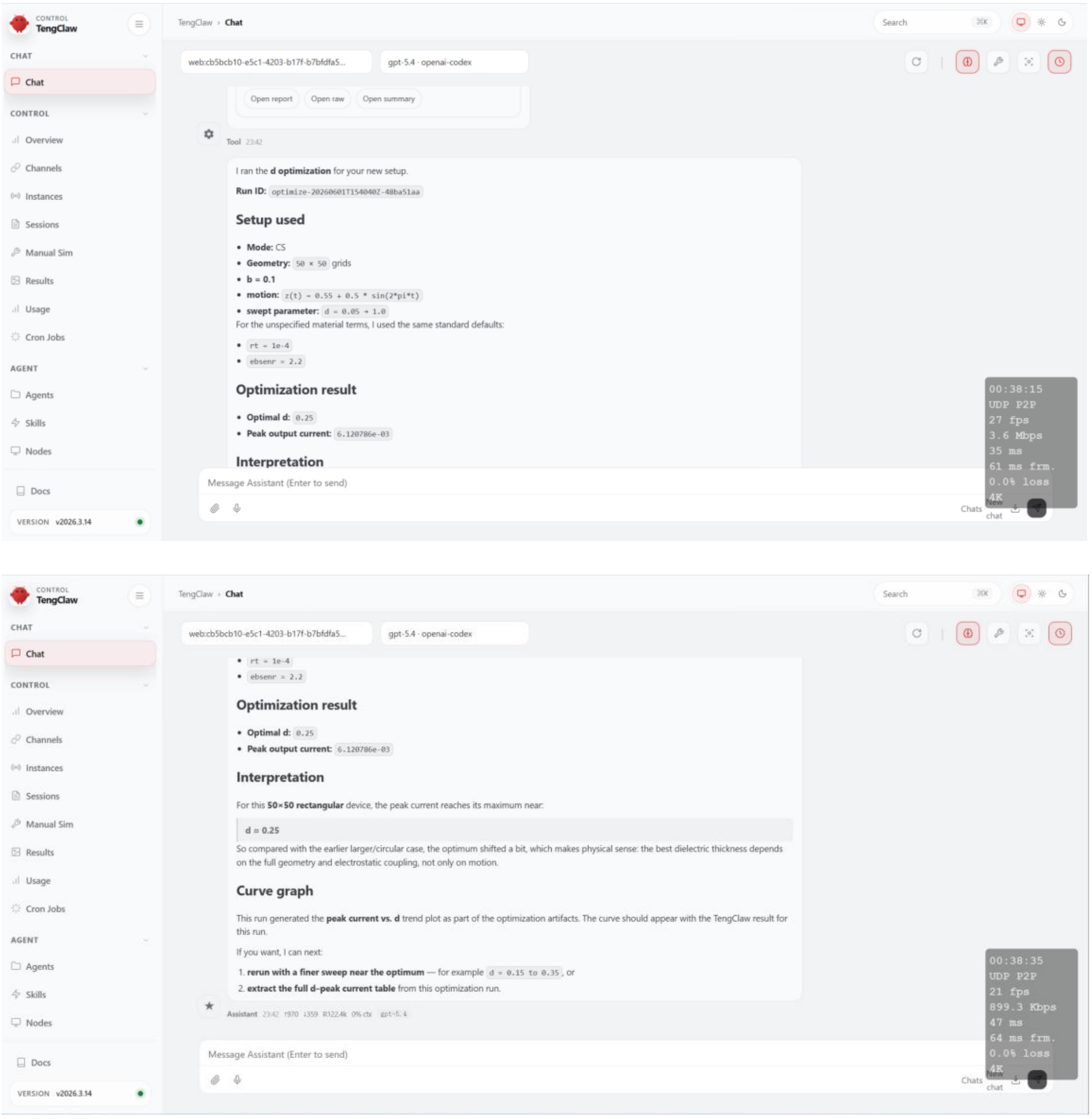

**Representative case 6: report generation and history-aware continuation**

Because each run is stored together with its interpreted intent, model branch, artifacts and conclusion, TENGCLAW can generate reports after execution. A user can ask for a concise report on the previous dynamic-output study, reopen a report associated with a baseline run or continue from the case that produced the highest current in a recent comparison. The system resolves these requests through run and graph state rather than relying only on conversational memory.

This capability is useful for long TENG design sessions, in which several simulations, field maps, parameter sweeps and comparative studies may be performed before the researcher decides which results should be summarized for analysis or manuscript preparation. In this sense, report generation is not an isolated export function. It is part of the persistent scientific workflow.

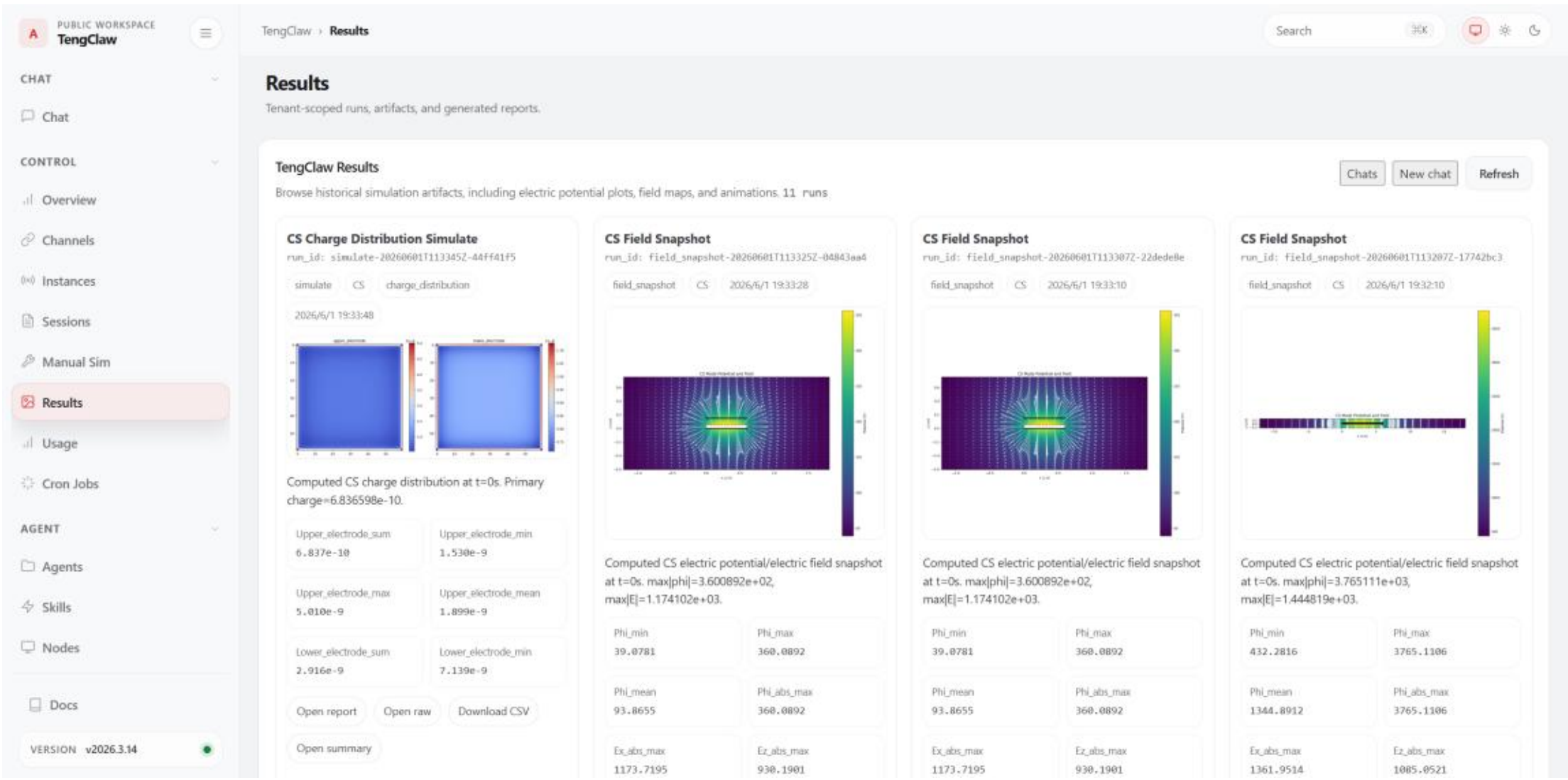

**Representative case 7: unsupported requests and deployment-dependent behavior**

If the user requests a geometry, working mode, observable or visualization that lies outside stable support, TENGCLAW does not silently fabricate a result. The deterministic critic returns an unsupported or approximate verdict, depending on whether a controlled fallback exists. In internal research mode, some extension-eligible tasks may be formulated as experimental proposals that require confirmation and validation before registration. In the public browser deployment, unsupported requests are reported as support boundaries and are not converted into automatically generated solver modules.

This conservative behavior is intentional. It prevents fluent language from being mistaken for validated physical capability and preserves the distinction between stable computation, approximate execution and experimental extension.

Taken together, these representative cases illustrate the intended use pattern of TENGCLAW. The user begins with a scientific question, the system makes the required physical commitments explicit, the deterministic critic determines whether the task is complete and supported, the source-driven hierarchy selects the admissible model branch, and the browser workspace returns artifacts that can be reopened, compared and extended. The interface is therefore not only a convenient display layer. It is the practical workspace through which the source-driven TENG theory becomes an executable, traceable and reusable modelling workflow.

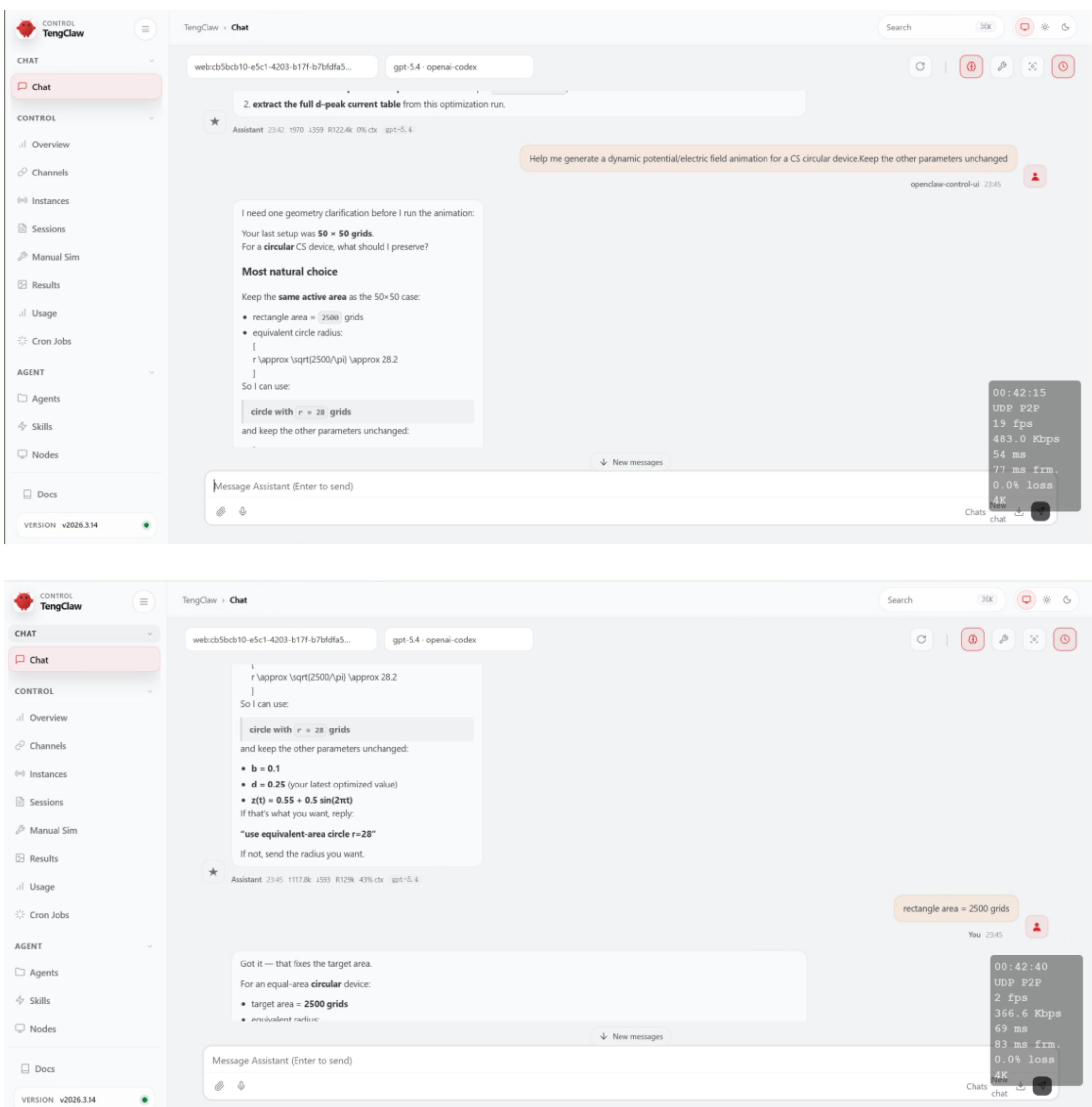

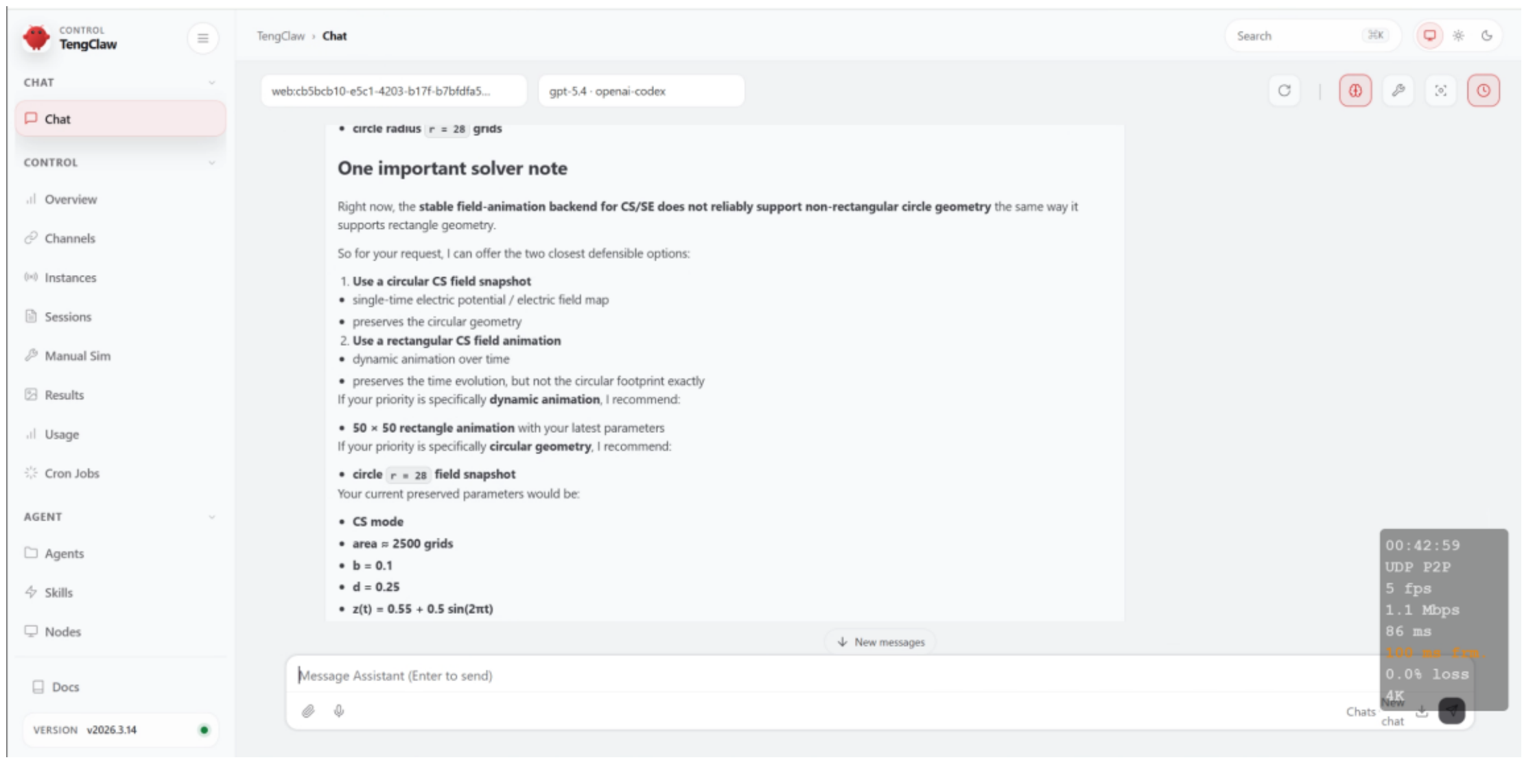

TengClaw
Chat
Overview
Channels
Instances
Sessions
Manual Sim
Results
Usage
Cron Jobs
Agents
Skills
Nodes
Docs
VERSION v2026.3.14
TengClaw › Chat
web:cb5bcb10-e5c1-4203-b17f-b7bfdfa5...
gpt-5.4 · openai-codex
One important solver note
Right now, the stable field-animation backend for CS/SE does not reliably support non-rectangular circle geometry the same way it supports rectangle geometry.
So for your request, I can offer the two closest defensible options:
1. Use a circular CS field snapshot
single-time electric potential / electric field map
preserves the circular geometry
2. Use a rectangular CS field animation
dynamic animation over time
preserves the time evolution, but not the circular footprint exactly
If your priority is specifically dynamic animation, I recommend:
50 × 50 rectangle animation with your latest parameters
If your priority is specifically circular geometry, I recommend:
circle r = 28 field snapshot
Your current preserved parameters would be:
CS mode
area ≈ 2500 grids
b = 0.1
d = 0.25
z(t) = 0.55 + 0.5 sin(2πt)
New messages
Message Assistant (Enter to send)
00:42:59
UDP P2P
5 fps
1.1 Mbps
86 ms
0.0% loss
4K

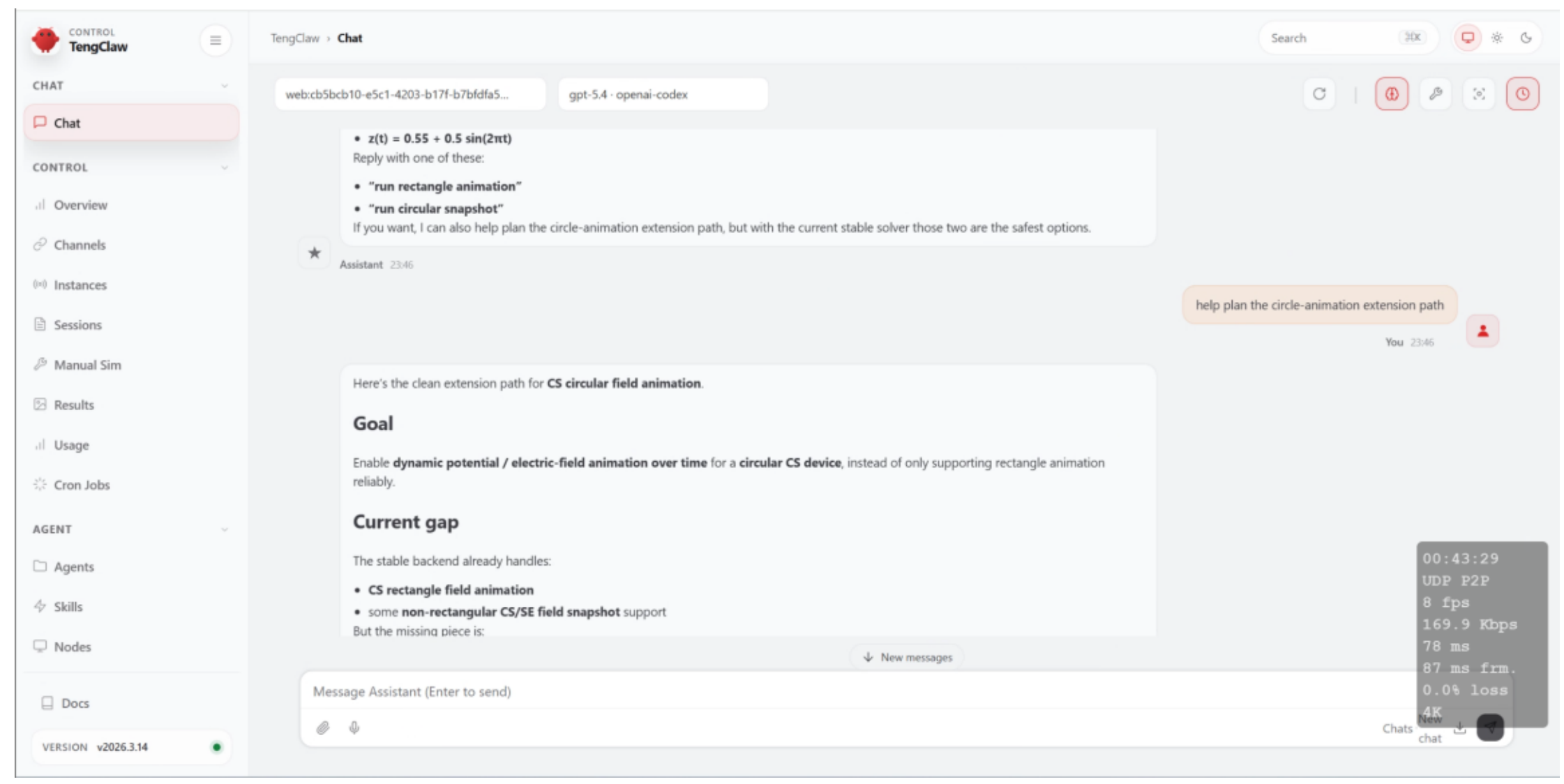

TengClaw › Chat
z(t) = 0.55 + 0.5 sin(2πt)
Reply with one of these:
"run rectangle animation"
"run circular snapshot"
If you want, I can also help plan the circle-animation extension path, but with the current stable solver those two are the safest options.
Assistant 23:46
help plan the circle-animation extension path
You 23:46
Here's the clean extension path for CS circular field animation.
Goal
Enable dynamic potential / electric-field animation over time for a circular CS device, instead of only supporting rectangle animation reliably.
Current gap
The stable backend already handles:
CS rectangle field animation
some non-rectangular CS/SE field snapshot support
But the missing piece is:
New messages
Message Assistant (Enter to send)
00:43:29
UDP P2P
8 fps
169.9 Kbps
78 ms
87 ms frm.
0.0% loss
4K

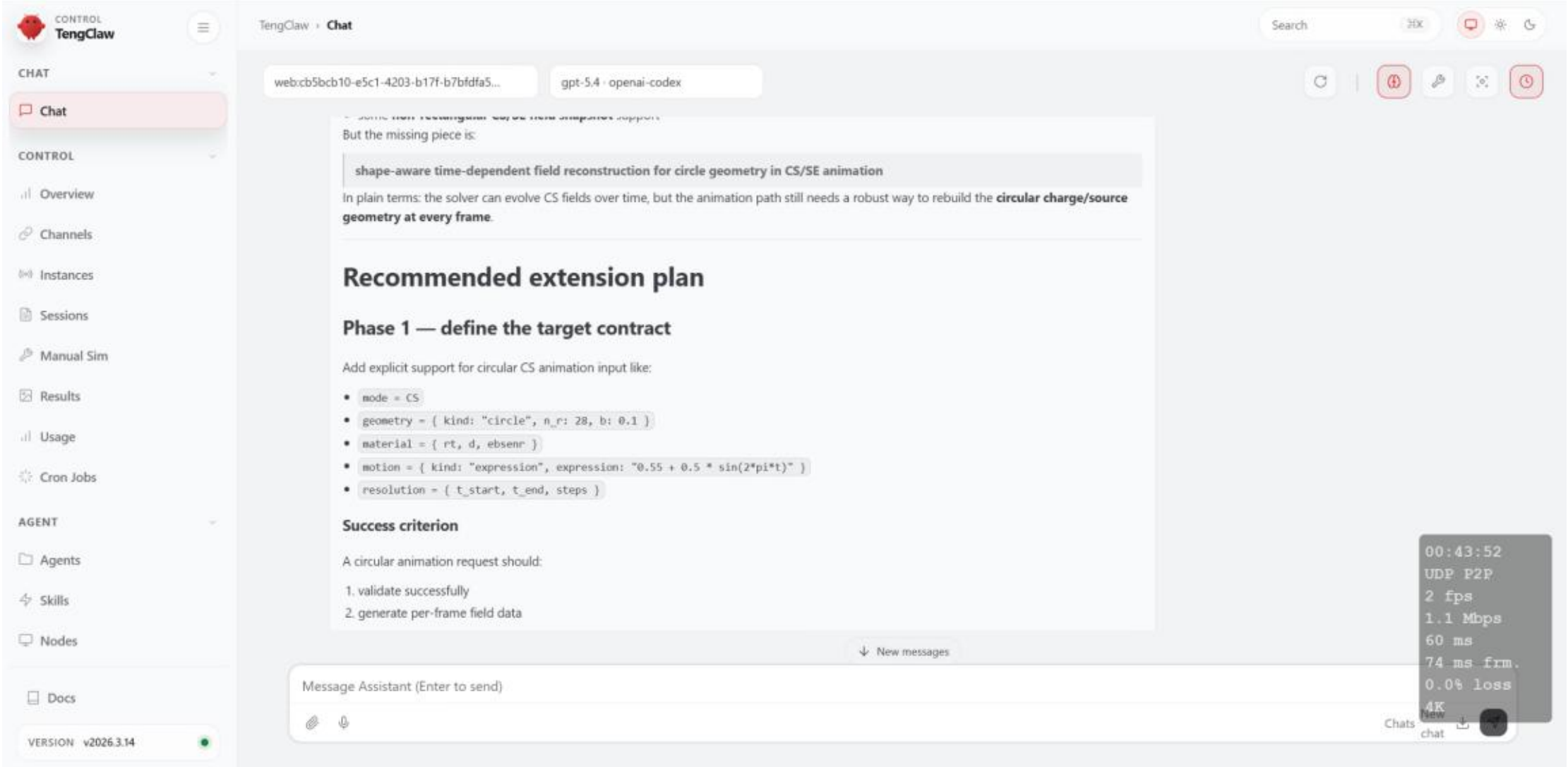

TengClaw › Chat
But the missing piece is:
shape-aware time-dependent field reconstruction for circle geometry in CS/SE animation
In plain terms: the solver can evolve CS fields over time, but the animation path still needs a robust way to rebuild the circular charge/source geometry at every frame.
Recommended extension plan
Phase 1 — define the target contract
Add explicit support for circular CS animation input like:
mode = CS
geometry = { kind: "circle", n_r: 28, b: 0.1 }
material = { rt, d, ebsenr }
motion = { kind: "expression", expression: "0.55 + 0.5 * sin(2*pi*t)" }
resolution = { t_start, t_end, steps }
Success criterion
A circular animation request should:
1. validate successfully
2. generate per-frame field data
New messages
Message Assistant (Enter to send)
00:43:52
UDP P2P
2 fps
1.1 Mbps
60 ms
74 ms frm.
0.0% loss
4K

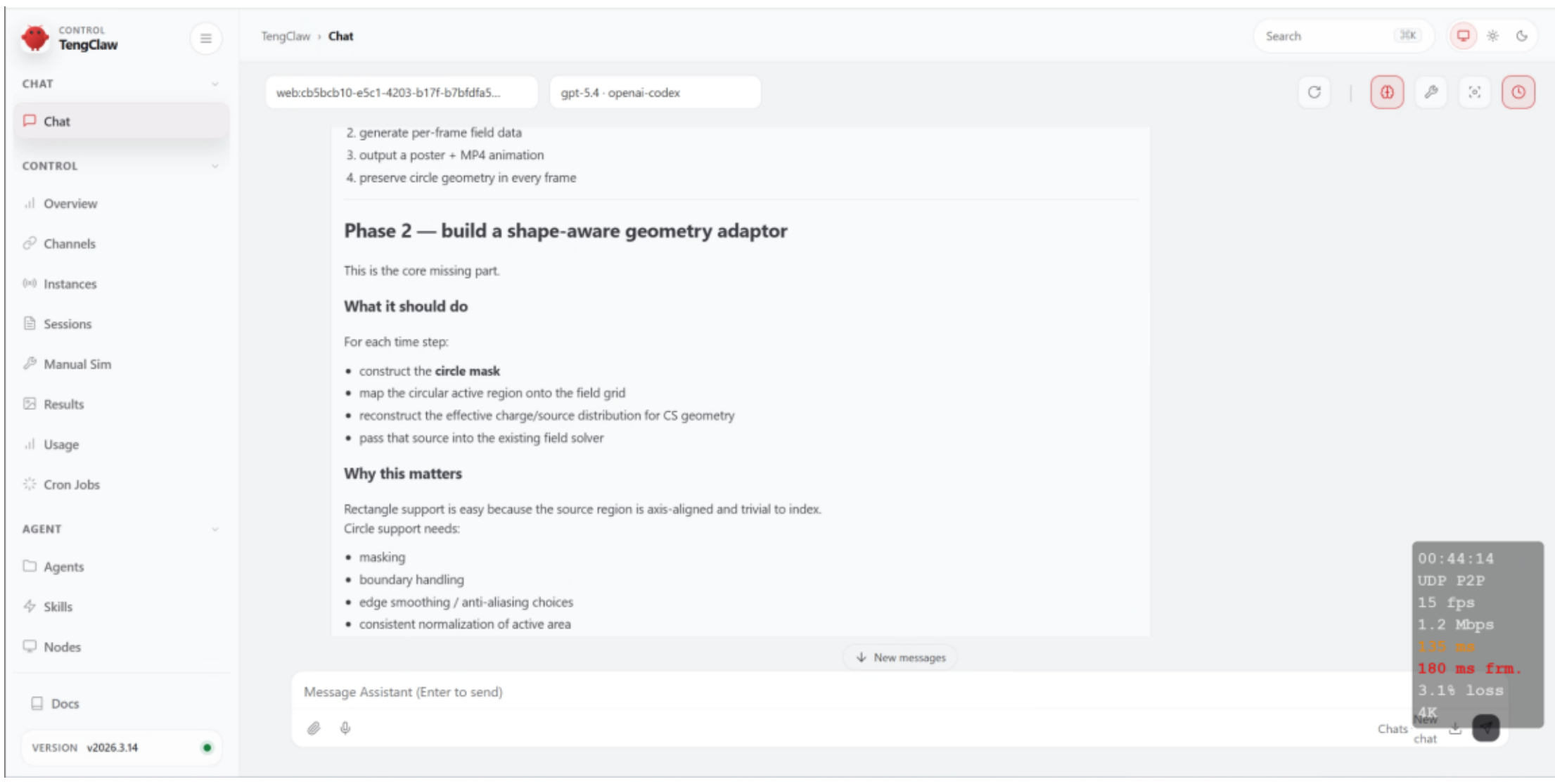

CONTROL
TengClaw
CHAT
Chat
CONTROL
Overview
Channels
Instances
Sessions
Manual Sim
Results
Usage
Cron Jobs
AGENT
Agents
Skills
Nodes
Docs
VERSION v2026.3.14
TengClaw › Chat
Search
web:cb5bcb10-e5c1-4203-b17f-b7bfdfa5...
gpt-5.4 · openai-codex
2. generate per-frame field data
3. output a poster + MP4 animation
4. preserve circle geometry in every frame
Phase 2 — build a shape-aware geometry adaptor
This is the core missing part.
What it should do
For each time step:
construct the circle mask
map the circular active region onto the field grid
reconstruct the effective charge/source distribution for CS geometry
pass that source into the existing field solver
Why this matters
Rectangle support is easy because the source region is axis-aligned and trivial to index.
Circle support needs:
masking
boundary handling
edge smoothing / anti-aliasing choices
consistent normalization of active area
New messages
Message Assistant (Enter to send)
Chats
00:44:14
UDP P2P
15 fps
1.2 Mbps
180 ms frm.
3.1% loss
4K

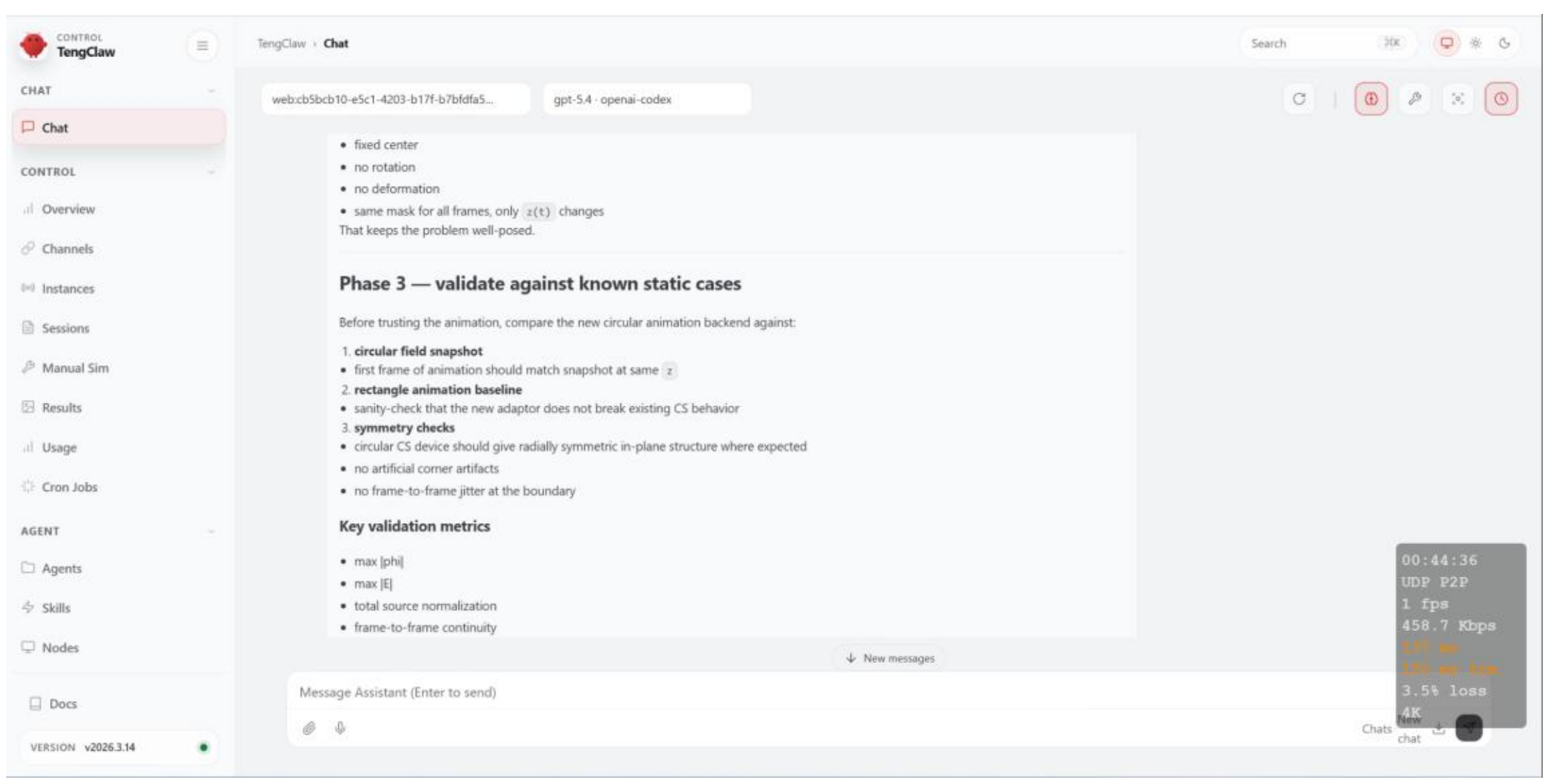

CONTROL
TengClaw
CHAT
Chat
CONTROL
Overview
Channels
Instances
Sessions
Manual Sim
Results
Usage
Cron Jobs
AGENT
Agents
Skills
Nodes
Docs
VERSION v2026.3.14
TengClaw › Chat
Search
web:cb5bcb10-e5c1-4203-b17f-b7bfdfa5...
gpt-5.4 · openai-codex
fixed center
no rotation
no deformation
same mask for all frames, only z(t) changes
That keeps the problem well-posed.
Phase 3 — validate against known static cases
Before trusting the animation, compare the new circular animation backend against:
1. circular field snapshot
first frame of animation should match snapshot at same z
2. rectangle animation baseline
sanity-check that the new adaptor does not break existing CS behavior
3. symmetry checks
circular CS device should give radially symmetric in-plane structure where expected
no artificial corner artifacts
no frame-to-frame jitter at the boundary
Key validation metrics
max |phi|
max |E|
total source normalization
frame-to-frame continuity
New messages
Message Assistant (Enter to send)
Chats
00:44:36
UDP P2P
1 fps
458.7 Kbps
3.5% loss
4K

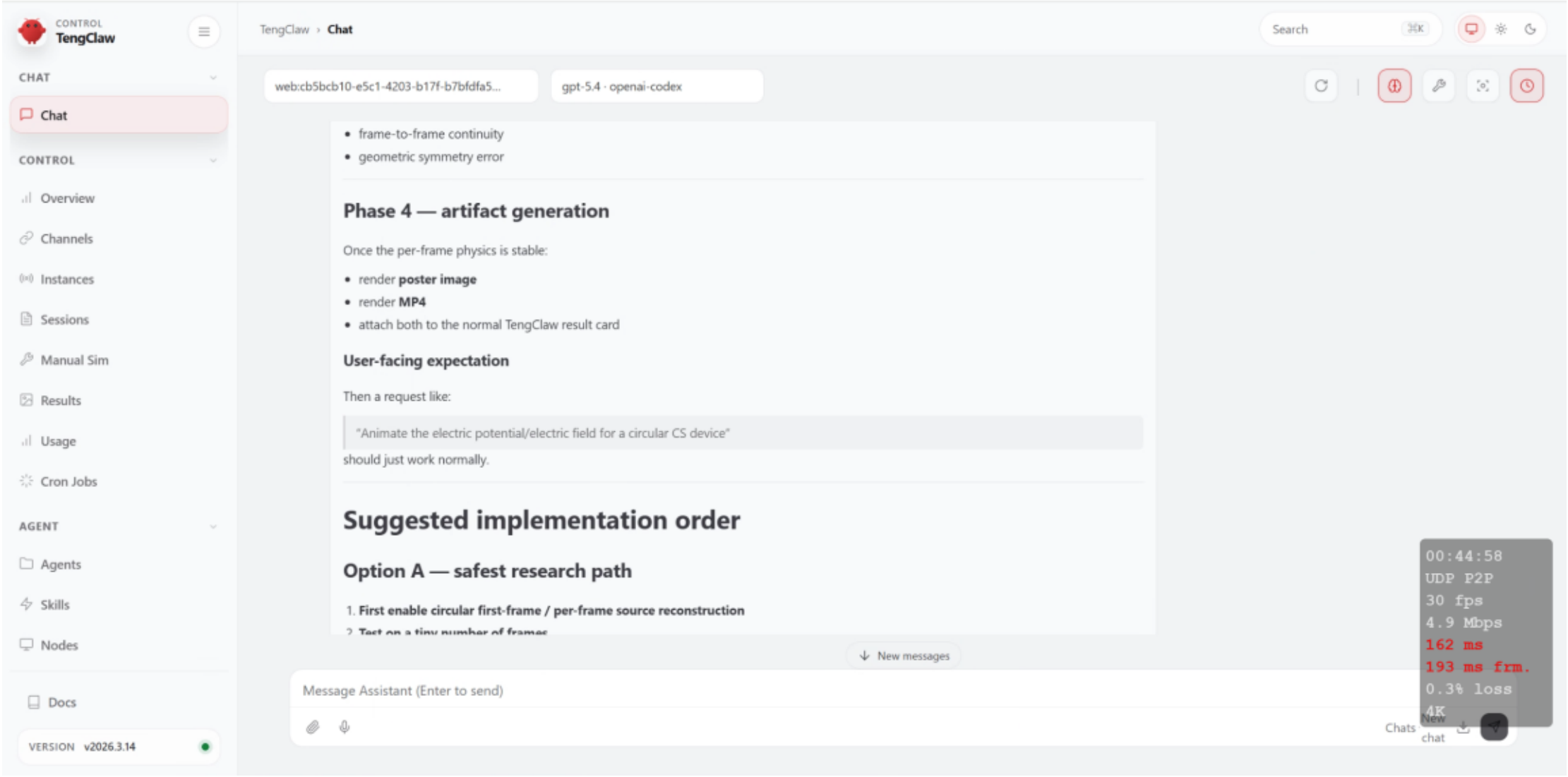

CONTROL
TengClaw
CHAT
Chat
CONTROL
Overview
Channels
Instances
Sessions
Manual Sim
Results
Usage
Cron Jobs
AGENT
Agents
Skills
Nodes
Docs
VERSION v2026.3.14
TengClaw › Chat
Search
web:cb5bcb10-e5c1-4203-b17f-b7bfdfa5...
gpt-5.4 · openai-codex
frame-to-frame continuity
geometric symmetry error
Phase 4 — artifact generation
Once the per-frame physics is stable:
render poster image
render MP4
attach both to the normal TengClaw result card
User-facing expectation
Then a request like:
"Animate the electric potential/electric field for a circular CS device"
should just work normally.
Suggested implementation order
Option A — safest research path
1. First enable circular first-frame / per-frame source reconstruction
New messages
Message Assistant (Enter to send)
Chats
00:44:58
UDP P2P
30 fps
4.9 Mbps
162 ms
193 ms frm.
0.3% loss
4K

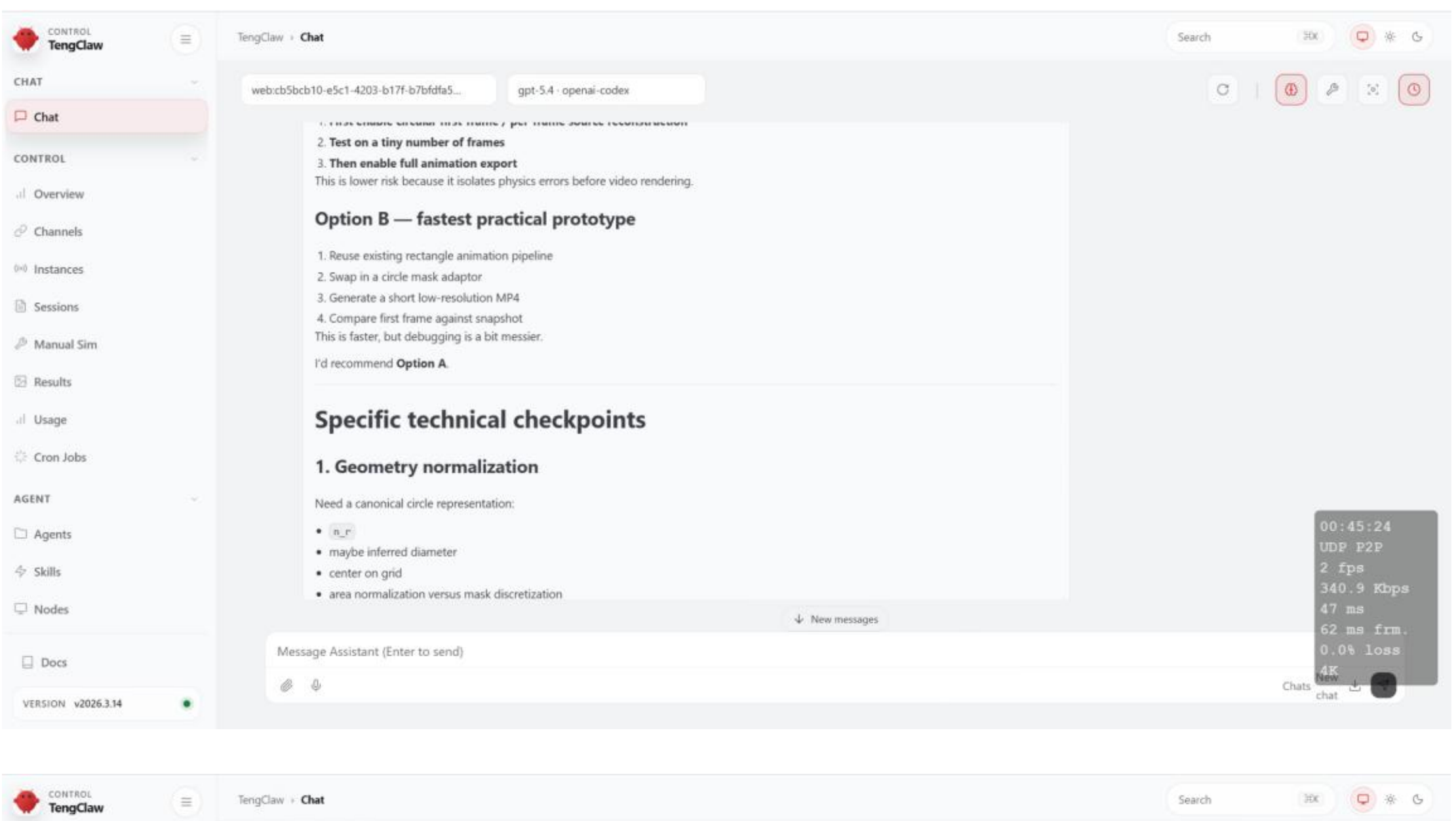

CONTROL
TengClaw
CHAT
Chat
CONTROL
Overview
Channels
Instances
Sessions
Manual Sim
Results
Usage
Cron Jobs
AGENT
Agents
Skills
Nodes
Docs
VERSION v2026.3.14
TengClaw › Chat
Search
web:cb5bcb10-e5c1-4203-b17f-b7bfdfa5...
gpt-5.4 · openai-codex
2. Test on a tiny number of frames
3. Then enable full animation export
This is lower risk because it isolates physics errors before video rendering.
Option B — fastest practical prototype
1. Reuse existing rectangle animation pipeline
2. Swap in a circle mask adaptor
3. Generate a short low-resolution MP4
4. Compare first frame against snapshot
This is faster, but debugging is a bit messier.
I'd recommend Option A.
Specific technical checkpoints
1. Geometry normalization
Need a canonical circle representation:
n_r
maybe inferred diameter
center on grid
area normalization versus mask discretization
New messages
Message Assistant (Enter to send)
Chats
00:45:24
UDP P2P
2 fps
340.9 Kbps
47 ms
62 ms frm.
0.0% loss
4K

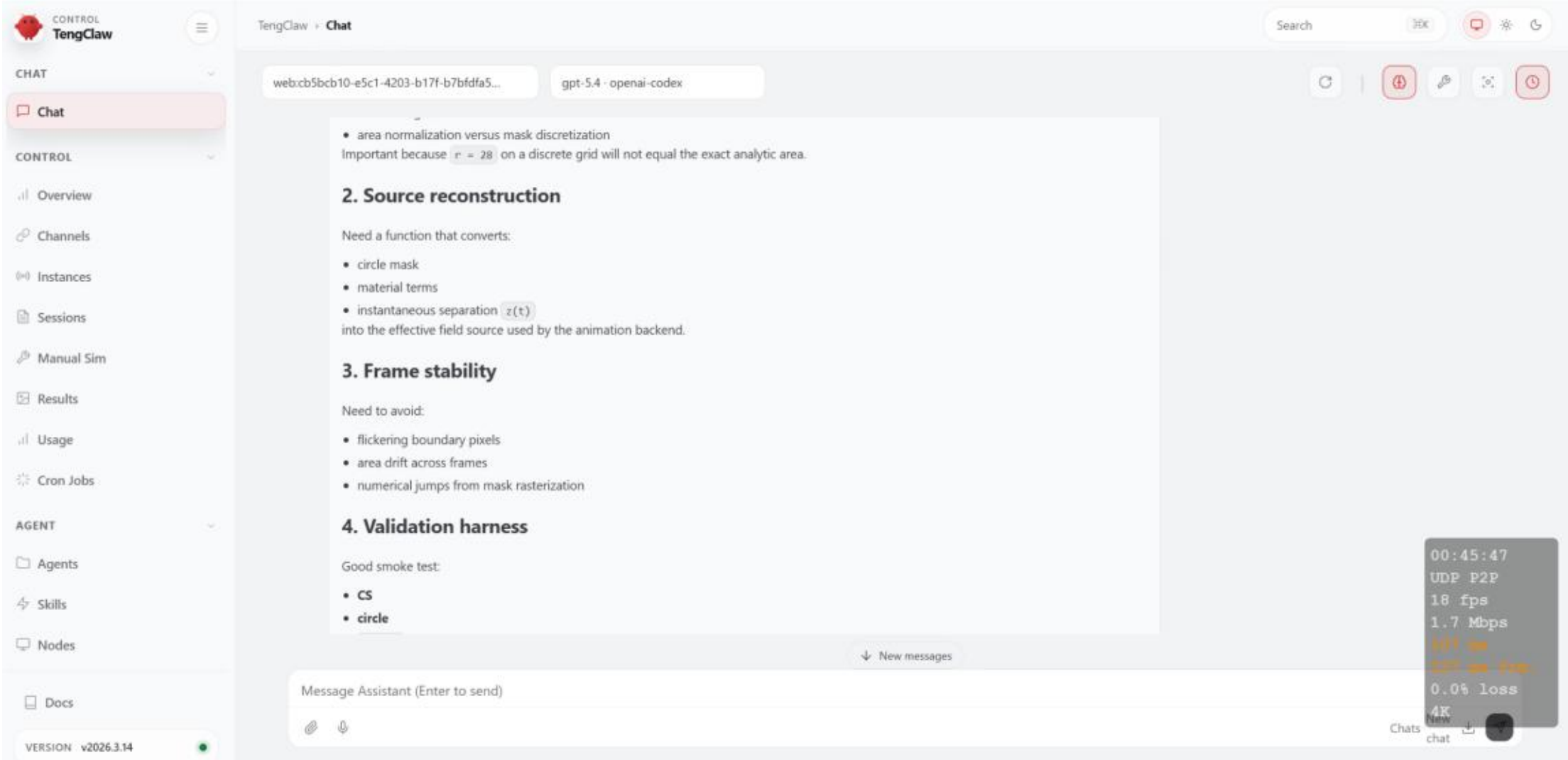

CONTROL
TengClaw
CHAT
Chat
CONTROL
Overview
Channels
Instances
Sessions
Manual Sim
Results
Usage
Cron Jobs
AGENT
Agents
Skills
Nodes
Docs
VERSION v2026.3.14
TengClaw › Chat
Search
web:cb5bcb10-e5c1-4203-b17f-b7bfdfa5...
gpt-5.4 · openai-codex
area normalization versus mask discretization
Important because r = 28 on a discrete grid will not equal the exact analytic area.
2. Source reconstruction
Need a function that converts:
circle mask
material terms
instantaneous separation z(t)
into the effective field source used by the animation backend.
3. Frame stability
Need to avoid:
flickering boundary pixels
area drift across frames
numerical jumps from mask rasterization
4. Validation harness
Good smoke test:
CS
circle
New messages
Message Assistant (Enter to send)
Chats
00:45:47
UDP P2P
18 fps
1.7 Mbps
0.0% loss
4K

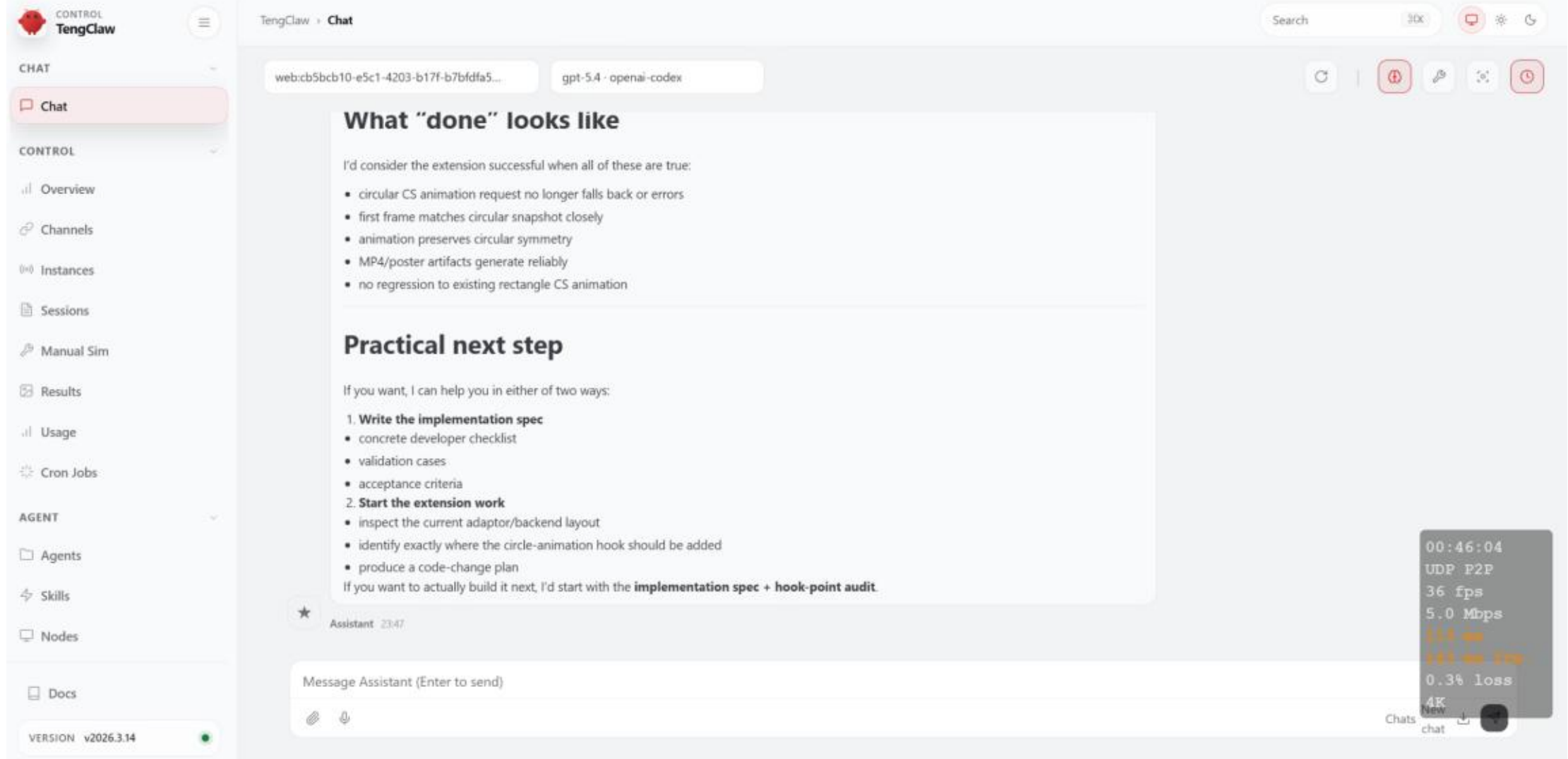

CONTROL
TengClaw
CHAT
Chat
CONTROL
Overview
Channels
Instances
Sessions
Manual Sim
Results
Usage
Cron Jobs
AGENT
Agents
Skills
Nodes
Docs
VERSION v2026.3.14
TengClaw › Chat
Search
web:cb5bcb10-e5c1-4203-b17f-b7bfdfa5...
gpt-5.4 · openai-codex
What "done" looks like
I'd consider the extension successful when all of these are true:
circular CS animation request no longer falls back or errors
first frame matches circular snapshot closely
animation preserves circular symmetry
MP4/poster artifacts generate reliably
no regression to existing rectangle CS animation
Practical next step
If you want, I can help you in either of two ways:
1. Write the implementation spec
concrete developer checklist
validation cases
acceptance criteria
2. Start the extension work
inspect the current adaptor/backend layout
identify exactly where the circle-animation hook should be added
produce a code-change plan
If you want to actually build it next, I'd start with the implementation spec + hook-point audit.
Assistant 23:47
Message Assistant (Enter to send)
Chats
00:46:04
UDP P2P
36 fps
5.0 Mbps
0.3% loss
4K

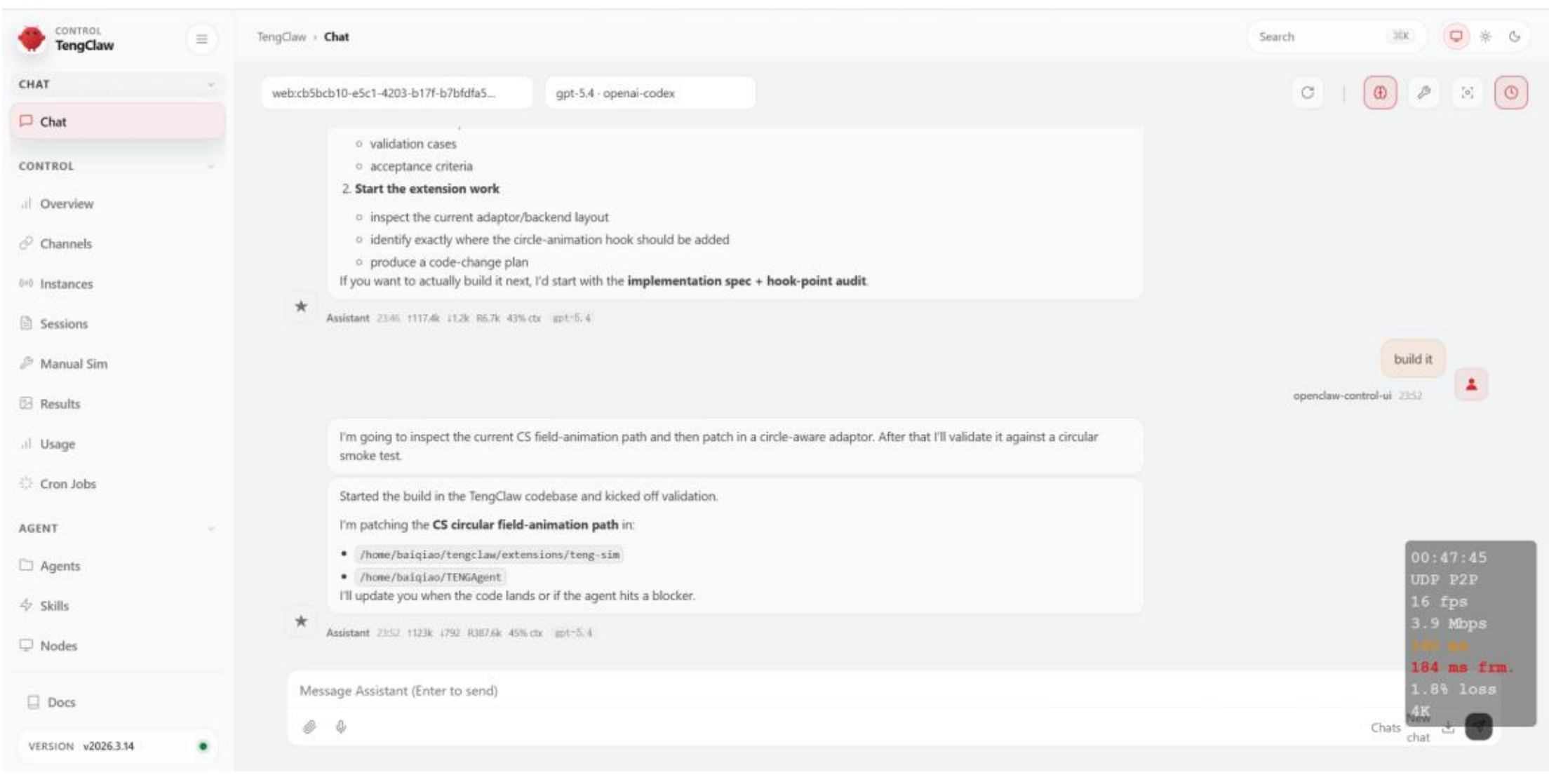
CONTROL
TengClaw
CHAT
Chat
CONTROL
Overview
Channels
Instances
Sessions
Manual Sim
Results
Usage
Cron Jobs
AGENT
Agents
Skills
Nodes
Docs
VERSION v2026.3.14
TengClaw › Chat
Search
web:cb5bcb10-e5c1-4203-b17f-b7bfdfa5...
gpt-5.4 · openai-codex
validation cases
acceptance criteria
2. Start the extension work
inspect the current adaptor/backend layout
identify exactly where the circle-animation hook should be added
produce a code-change plan
If you want to actually build it next, I'd start with the implementation spec + hook-point audit.
Assistant
build it
openclaw-control-ui
I'm going to inspect the current CS field-animation path and then patch in a circle-aware adaptor. After that I'll validate it against a circular smoke test.
Started the build in the TengClaw codebase and kicked off validation.
I'm patching the CS circular field-animation path in:
/home/baiqiao/tengclaw/extensions/teng-sim
/home/baiqiao/TENGAgent
I'll update you when the code lands or if the agent hits a blocker.
Assistant
Message Assistant (Enter to send)
00:47:45
UDP P2P
16 fps
3.9 Mbps
184 ms frm.
1.8% loss
4K

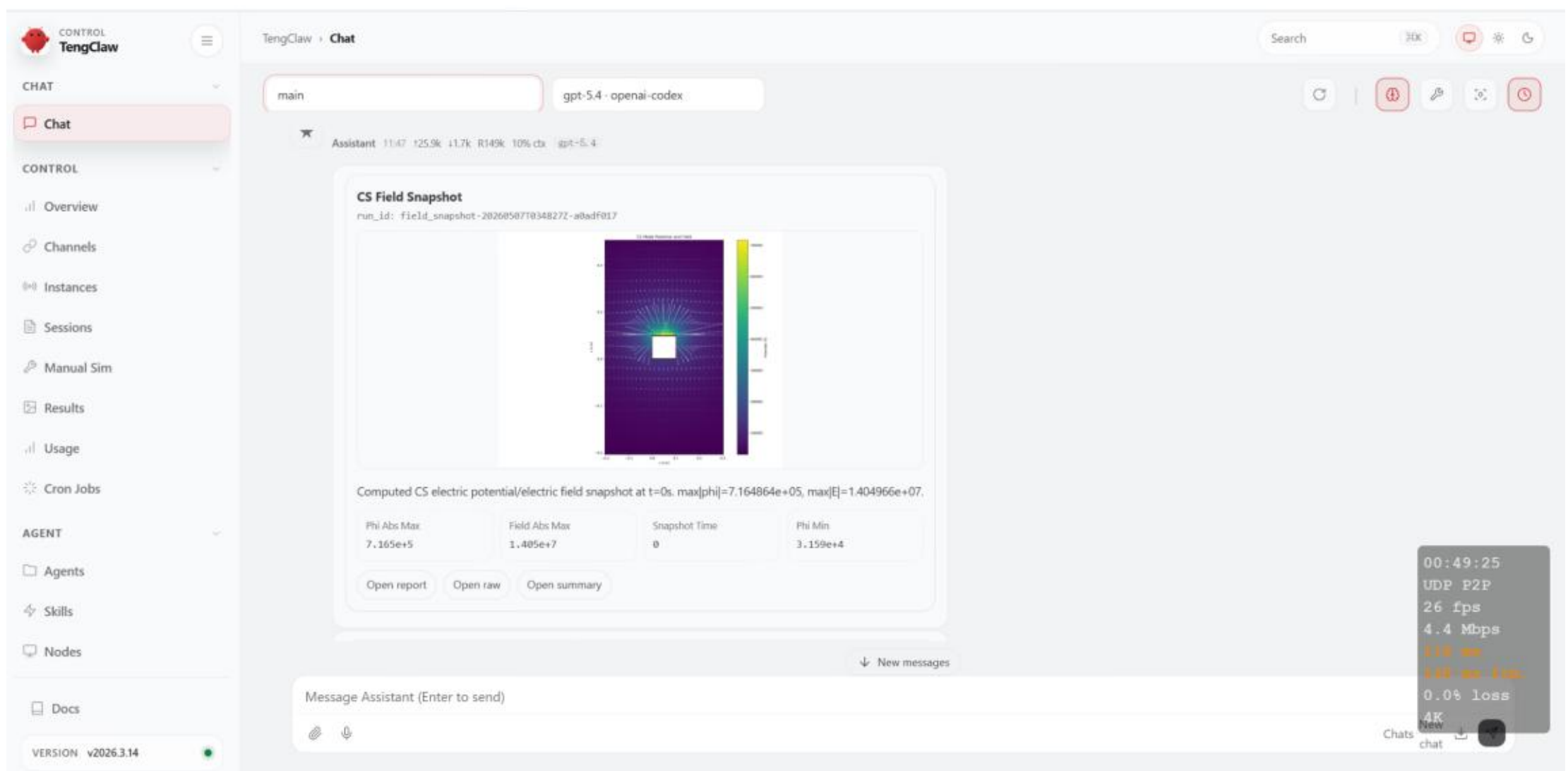
CONTROL
TengClaw
CHAT
Chat
CONTROL
Overview
Channels
Instances
Sessions
Manual Sim
Results
Usage
Cron Jobs
AGENT
Agents
Skills
Nodes
Docs
VERSION v2026.3.14
TengClaw › Chat
Search
main
gpt-5.4 · openai-codex
Assistant
CS Field Snapshot
Computed CS electric potential/electric field snapshot at t=0s. max|phi|=7.164864e+05, max|E|=1.404966e+07.
Phi Abs Max
7.165e+5
Field Abs Max
1.405e+7
Snapshot Time
0
Phi Min
3.159e+4
Open report
Open raw
Open summary
New messages
Message Assistant (Enter to send)
00:49:25
UDP P2P
26 fps
4.4 Mbps
0.0% loss
4K

## Supplementary references